\documentclass[lettersize,journal]{IEEEtran}
\usepackage{amsmath,amsfonts}
\usepackage{algorithmic}
\usepackage{algorithm}
\usepackage{array}
\usepackage[caption=false,font=normalsize,labelfont=sf,textfont=sf]{subfig}
\usepackage{textcomp}
\usepackage{stfloats}
\usepackage{url}
\usepackage{verbatim}
\usepackage{graphicx}
\usepackage{cite}
\usepackage{booktabs}
\usepackage{multirow}
\begin{document}

\title{Towards More Precise Automatic Analysis: A Comprehensive Survey of Deep Learning-based Multi-organ Segmentation}

\author{Xiaoyu Liu, Linhao Qu, Ziyue Xie, Jiayue Zhao, Yonghong Shi, and Zhijian Song
\thanks{X. Liu, L. Qu, Z. Xie, J. Zhao, Y. Shi and Z. Song are with the Digital Medical Research Center, School of Basic Medical Sciences, Fudan University, and also with the Shanghai Key Laboratory of Medical Imaging Computing and Computer Assisted Intervention, Shanghai, 200032, China. (e-mail: (liuxiaoyu21, zyxie22, jiayuezhao22) @m.fudan.edu.cn; (lhqu20, yonghong.shi, zjsong) @fudan.edu.cn);}
\thanks{X. Liu and L. Qu are co-first authors. Y. Shi and Z. Song are co-corresponding authors.}}


\markboth{Journal of \LaTeX\ Class Files,~Vol.~14, No.~8, August~2021}%
{Shell Xiaoyu Liu {\textit{et al.}}: Towards More Precise Automatic Analysis: A Comprehensive Survey of Deep Learning-based Multi-organ Segmentation}


\maketitle

\begin{abstract}
Accurate segmentation of multiple organs of the head, neck, chest, and abdomen from medical images is an essential step in computer-aided diagnosis, surgical navigation, and radiation therapy. In the past few years, with a data-driven feature extraction approach and end-to-end training, automatic deep learning-based multi-organ segmentation method has far outperformed traditional methods and become a new research topic. This review systematically summarizes the latest research in this field. For the first time, from the perspective of full and imperfect annotation, we comprehensively compile 161 studies on deep learning-based multi-organ segmentation in multiple regions such as the head and neck, chest, and abdomen, containing a total of 214 related references. The method based on  full annotation summarizes the existing methods from four aspects: network architecture, network dimension, network dedicated modules, and network loss function. The method based on imperfect annotation summarizes the existing methods from two aspects: weak annotation-based methods and semi annotation-based methods. We also summarize frequently used datasets for multi-organ segmentation and discuss new challenges and new research trends in this field.
\end{abstract}


\section{Introduction}
\IEEEPARstart{A}{ccurate} segmentation of multiple organs of the head and neck, chest, abdomen as well as other parts from medical images is crucial in computer-aided diagnosis, surgical navigation, and radiotherapy \cite{1,2}. For example, radiotherapy, which targets tumour masses and microscopic areas with high risk of tumour proliferation, is a common treatment option for cancer patients \cite{3}. However, radiotherapy can bring great risk to the normal organs around the tumour, which are known as organs at risk (OARs). Thus, accurate segmentation of tumour contours and OARs is necessary \cite{4,5}.

The early segmentation process relied heavily on manual labelling by physicians, which is a labour-intensive and time-consuming process. For example, a trained specialist may spend more than four hours manually labelling a case, which not only places a heavy burden on the healthcare system but also likely causes a delay in the radiotherapy for a patient. Moreover, different physicians or hospitals will have different results of labelling \cite{6,7,8,9}. Therefore, accurate automatic multi-organ segmentation method is urgently needed in clinical practice.

Multi-organ segmentation is a challenging task. First, the contour of the anatomical structure in image is highly variable, which is difficult to expressed by a unified mathematical rule. Second, the boundaries between different organs or tissue regions in an image are often blurred due to image noise and low intensity contrast, and these boundaries are difficult to identify using techniques of traditional digital image processing. Third, the use of different scanners, scanning protocols, and contrast agents will lead to different intensity distributions of organs in the obtained images, which poses a great challenge to the generalizability of the model. Finally, considering safety and ethical issues, many hospitals do not disclose their datasets. Many segmentation methods are trained and validated on private datasets, making it difficult to compare different methods. Therefore, designing accurate and robust multi-organ segmentation models is a very difficult and expensive task.

Traditional methods \cite{10,11,12,13} usually utilize manually extracted image features for image segmentation, such as the threshold \cite{14} method, graph cut \cite{15} method, and region growth \cite{16} method. Limited by a large number of manually extracted image features and the selection of non-robust thresholds or seeds, the segmentation results of these methods are usually unstable, and often yield only a rough segmentation result or only apply to specific organs. Knowledge-based methods can obtain anatomical information of different organs from labelled datasets, reduce the burden of manual feature extraction, and improve the robustness and accuracy of multi-organ segmentation, which commonly include multi-atlas label fusion \cite{17,18} and statistical shape models \cite{19,20}. The method based on multi-atlas label fusion-based uses image alignment to align predefined structural contours to the image to be segmented, and this method typically includes multiple steps. Therefore, the performance of this method may be influenced by various relevant factors involved in each step. The atlas-based method is still very popular, but due to the use of fixed atlases, it is difficult to handle the anatomical variation of organs between patients. In addition, it is computationally intensive and takes a long time to complete an alignment task. The statistical shape model uses the positional relationships between different organs, and uses the shape of the organs in the statistical space as a constraint to regularize the segmentation results. However, the accuracy of this approach is largely dependent on the reliability and extensibility of the shape model, and  the model based on normal anatomical structures has  very limited effect  in the segmentation of irregular structures.

\begin{figure}[ht!]
    \centering
    \includegraphics[width=\columnwidth]{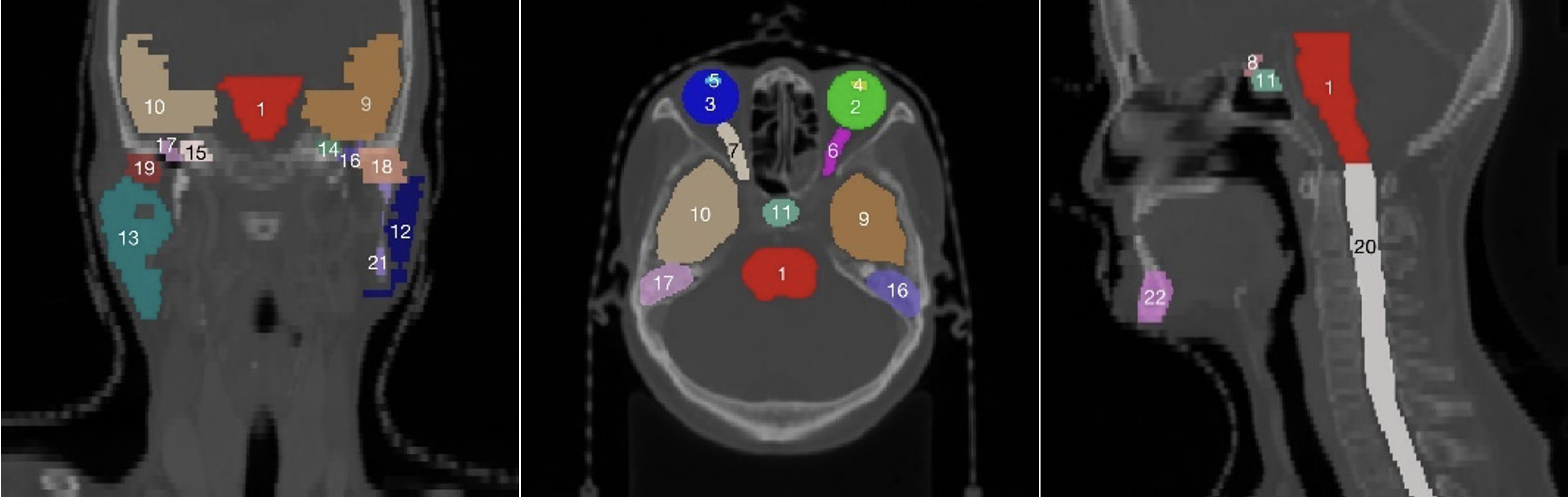}
    \caption{Schematic diagram of the organs of the head and neck, where the numbers are arranged in order: (1) brainstem, (2) left eye, (3) right eye, (4) left lens, (5) right lens, (6) left optic nerve, (7) right optic nerve, (8) Optical chiasm, (9) left temporal lobe, (10) right temporal lobe, (11) pituitary gland, (12) left parotid gland, (13) right parotid gland, (14) left temporal bone rock, (15) right temporal bone rock, (16) left temporal bone, (17) right temporal bone, (18) left mandibular condyle, (19) right mandibular condyle, (20) spinal cord, (21) left mandible, (22) right mandible.}
    \label{fig1}
\end{figure}

\begin{figure}[ht!]
    \centering
    \includegraphics[width=\columnwidth]{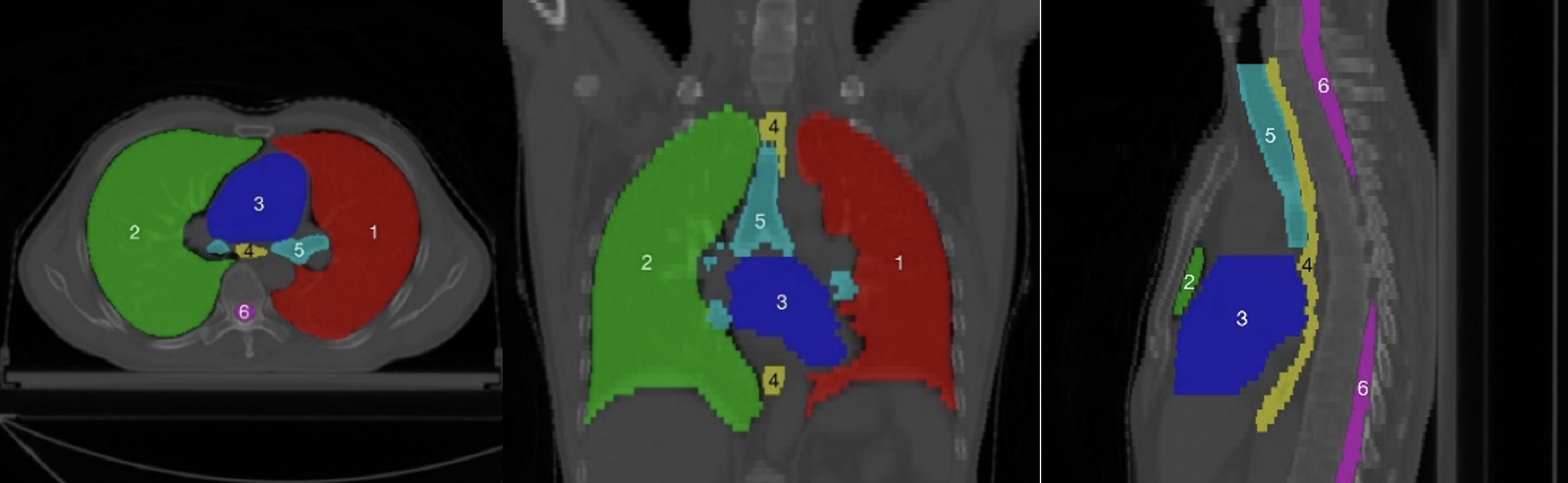}
    \caption{Schematic diagram of the thoracic organs, where the numbers are arranged in order: (1) left lung, (2) right lung, (3) heart, (4) esophagus, (5) trachea, and (6) spinal cord.}
    \label{fig2}
\end{figure}

\begin{figure}[ht!]
    \centering
    \includegraphics[width=\columnwidth]{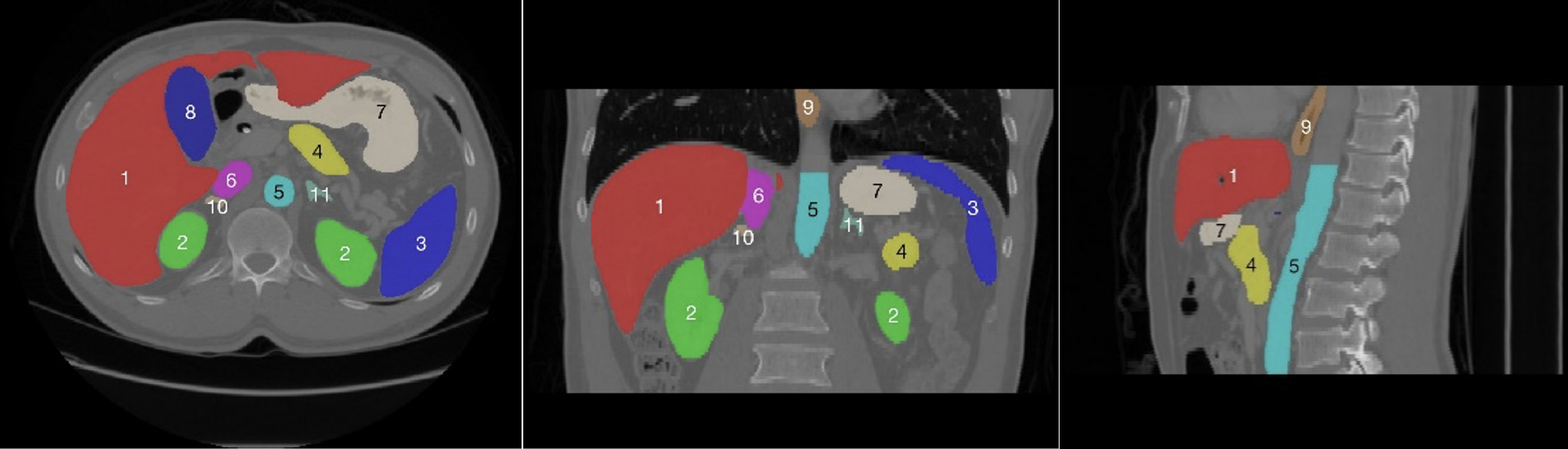}
    \caption{Schematic diagram of the abdominal organs, where the numbers are arranged in order: (1) liver, (2) kidney, (3) spleen, (4) pancreas, (5) aorta, (6) inferior vena cava, (7) stomach, (8) gallbladder, (9) esophagus, (10) right adrenal gland, (11) left adrenal gland, and (12) celiac artery.}
    \label{fig3}
\end{figure}

\begin{figure}[ht!]
    \centering
    \includegraphics[width=\columnwidth]{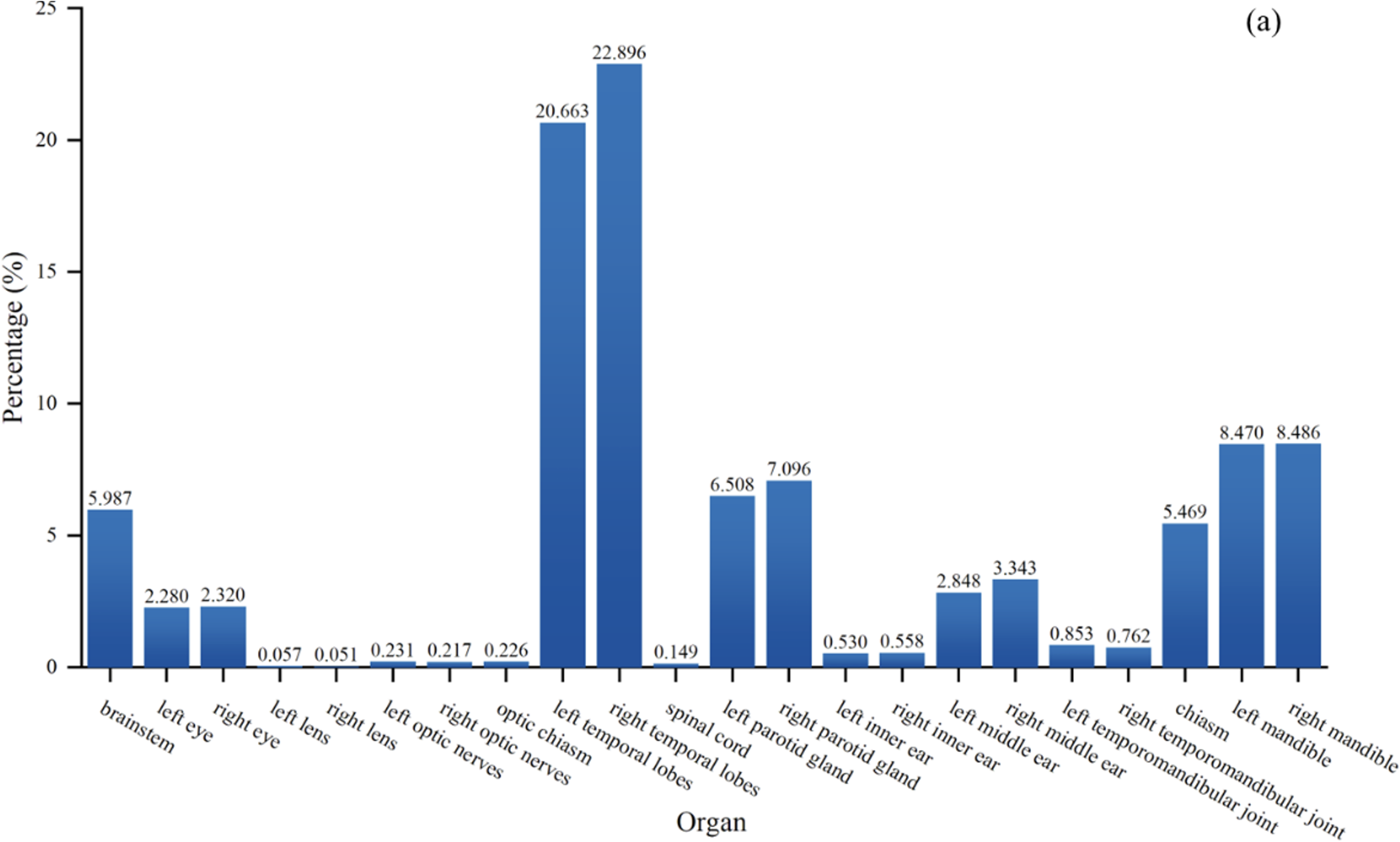}\\ 
    \includegraphics[width=\columnwidth]{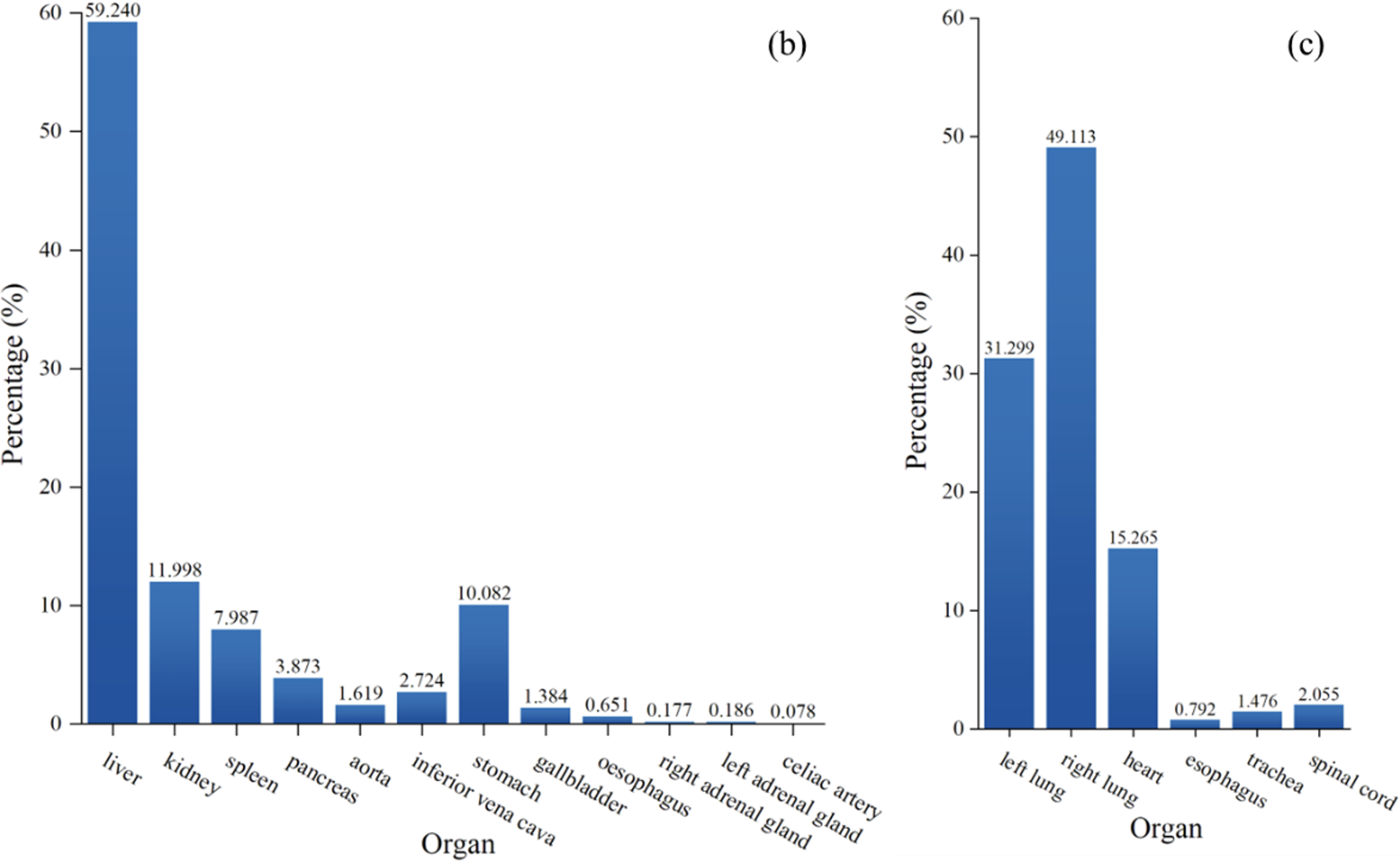}
    \caption{Illustration of the percentage of voxels in each organ of the head and neck (a), chest (b), and abdomen (c), respectively.}
    \label{fig4}
\end{figure}

Using data-driven feature extraction approach and end-to-end training, the methods based on deep learning (DL) have been widely studied in the fields of image classification \cite{21}, object detection \cite{22} and image segmentation \cite{23,24}, image fusion \cite{25}, image registration \cite{26}, etc. The segmentation method based on deep learning has become a mainstream method in the field of medical image processing. However, there are two main difficulties in multi-organ deep learning segmentation tasks. First, as shown in the head and neck in Fig. \ref{fig1}, the abdomen in Fig. \ref{fig2}, the chest in Fig. \ref{fig3}, and the statistics of the multi-organ size in each part in Fig. \ref{fig4}, there are very large differences between the organs sizes, and the serious imbalances of different organs sizes will lead to a poor segmentation performance of the trained segmentation network for small organs. Second, due to the imaging principle of CT technology and the complex anatomical structure of the human body, the contrast between organs and their surrounding tissues is often low, which leads to the inaccurate segmentation of organ boundaries by segmentation networks. Therefore, it has become a new hot research topic to develop deep multi-organ segmentation methods that can accurately segment small and large organs at the same time.

Recently, a large number of deep learning-based multi-organ segmentation methods with significantly improved performance have been proposed \cite{27}. Fu {\it{et al.}} \cite{28} systematically reviewed the medical image multi-organ segmentation methods based on deep learning by 2020 according to the network architecture. However, with the rapid development of deep learning technology, more representative new techniques and methods have been proposed, such as transformer-based multi-organ segmentation methods and imperfect annotation-based methods. A more comprehensive review and summary of these techniques and methods are very important for the development of this field.

\begin{figure}[ht!]
    \centering
    \includegraphics[width=\columnwidth]{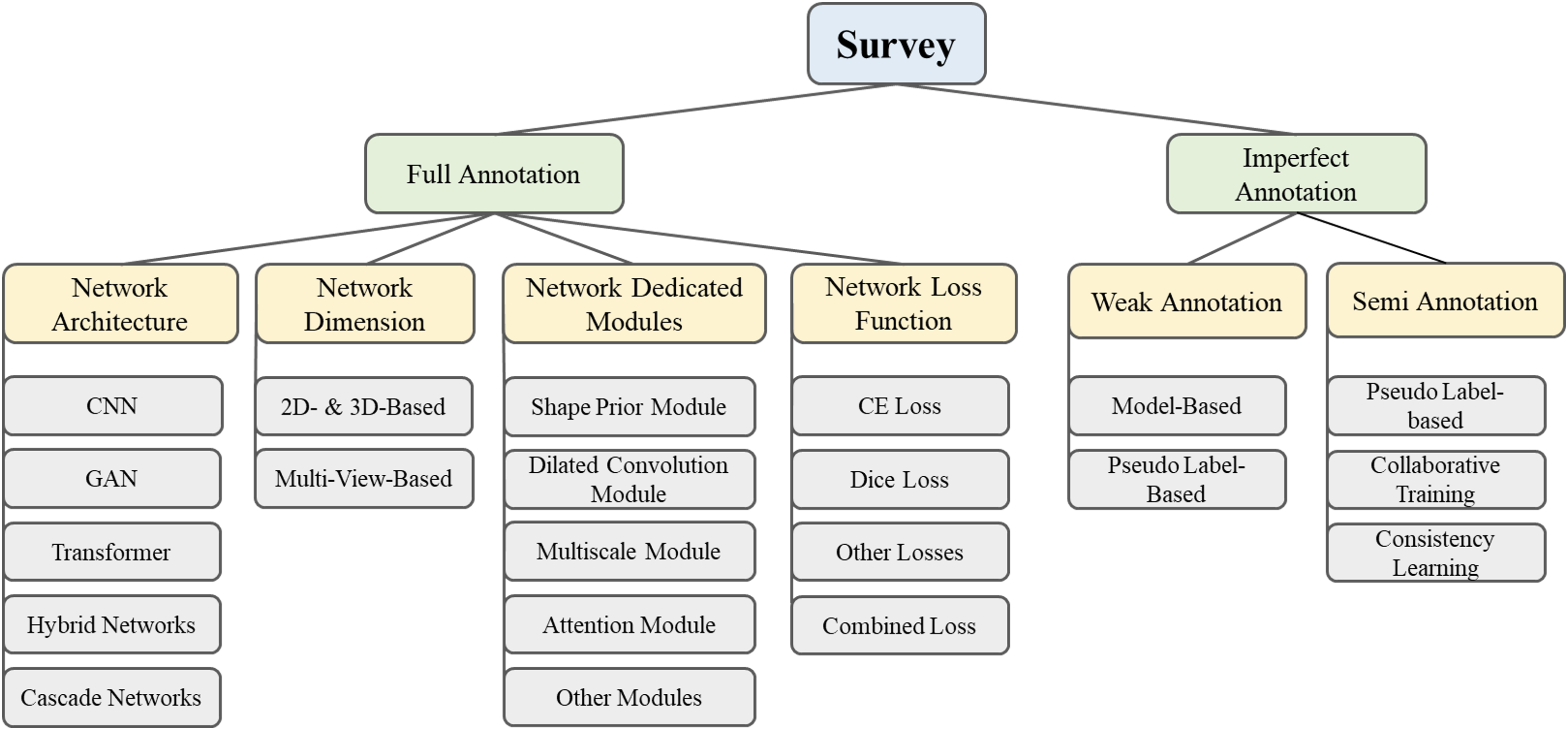}
    \caption{Framework diagram of the overview.}
    \label{fig5}
\end{figure}

This paper reviews deep learning-based multi-organ segmentation method of the head, neck, chest and abdomen published from 2016 to 2022. On Google Scholar, a search using the keywords `Multi Organ Segmentation' and `Deep Learning' yielded an initial 287 articles, 73 articles were removed according to abstract and keywords, and 161 highly relevant studies containing a total of 214 relevant references were obtained. Fig. \ref{fig5} summarizes all current state-of-the-art deep learning-based multi-organ segmentation methods according to full annotation and imperfect annotation architectures. In full annotation-based methods, we summarize the existing methods in four aspects: network architecture, network dimension, network dedicated modules, and network loss function. In imperfect annotation-based methods, we summarize the existing methods in two aspects, weak annotation and semi annotation, to investigate their innovation, contribution, and challenges.

This article is organized as follows. Section \ref{sec2} expounds the mathematical definition of multi-organ segmentation and the corresponding evaluation metrics. Section \ref{sec3} summarizes the multi-organ segmentation datasets. Section \ref{sec4} describes the literature based on full annotation-based methods, involving four parts: network architecture (section \ref{sec4}.A), network dimension (section \ref{sec4}.B), network dedicated modules (section \ref{sec4}.C), and network loss function (section \ref{sec4}.D). Section V analyzes the articles based on imperfect annotation methods, including two parts: weak annotation-based methods (section \ref{sec5}.A) and semi annotation-based methods (section \ref{sec5}.B). We discuss the existing methods and their future outlooks in section \ref{sec6}, and conclude the whole paper in section \ref{sec7}.

\section{Definition and Evaluation Metrics}
\label{sec2}
Let $\boldsymbol{X}$ represent the union of multiple organ regions in the input images, $\boldsymbol{G}$ represent the union of ground truth labels of multiple organs in the input images, $\boldsymbol{P}$ represent the union of predicted labels of multiple organs in the output images, $\boldsymbol{x_i^c \in X, g_i^c \in G, p_i^c \in P, i=1, \cdots N}$, and $\boldsymbol{c=1, \cdots C}$, where $\boldsymbol{N}$ represents the number of pixel in the image, $\boldsymbol{C}$ represents the number of categories to which the pixels belong, $\boldsymbol{f}$ represents the neural network, and $\boldsymbol{\theta}$ represents the parameters of the neural network optimization, where $\boldsymbol{P=f(X; \theta)}$.

The loss function represents the gap between the predicted and true values. In the multi-organ segmentation task, common loss functions include the cross-entropy loss and Dice loss. Section \ref{sec4_4} provides specific details about the loss function.

\begin{figure}[ht!]
    \centering
    \includegraphics[width=0.8\columnwidth]{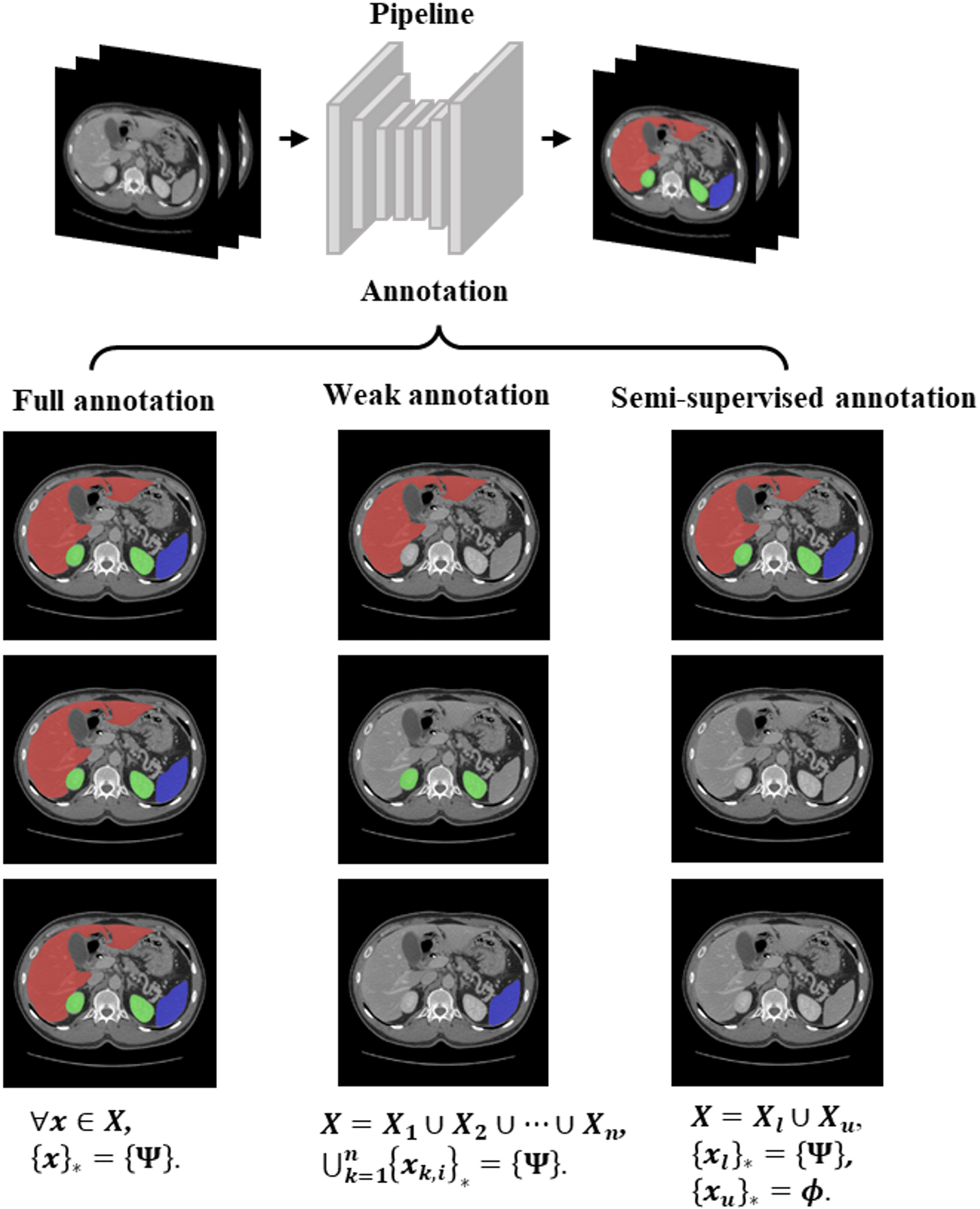}
    \caption{General overview of the learning paradigms reviewed in this paper.}
    \label{fig6}
\end{figure}

Given a multi-organ segmentation task, $\left\{\boldsymbol{\Psi}\right\}$ represents the represents the class set of organs to be segmented. $\left\{\boldsymbol{x}\right\}_\ast$ represents the set of organs annotated in $\boldsymbol{x}$. According to the available annotations, multi-organ segmentation can be implemented according to three learning paradigms: full annotation-based learning, weak annotation-based learning, and semi annotation-based learning. The last two are called imperfect annotation-based methods, as shown in Fig. \ref{fig6}. Full annotation-based learning means that the labels of all organ are given, which indicates that $\forall \boldsymbol{x} \in \boldsymbol{X},\{\boldsymbol{x}\}_*=\{\boldsymbol{\Psi}\}$ . Weak annotation often means that the data come from $\boldsymbol{n}$ different datasets. However, each dataset provides the annotations of one or more organs but not all organs, which means that $\boldsymbol{X=X_1 \cup X_2 \cup \cdots \cup X_n, \forall x_{k, i} \in X_k, k=1,2, \ldots n}$, $\boldsymbol{\left\{x_{k, i}\right\}_*<\{\Psi\}, \bigcup_{k=1}^n\left\{x_{k, i}\right\}_*=\{\Psi\}}$. Here, $\boldsymbol{x_{k, i}}$ \textbf{denotes the} $\boldsymbol{i}$\textbf{-th image in} $\boldsymbol{X_k}$. Semi annotation-based methods indicate that some of the training datasets are fully labelled and others are unlabelled, $\boldsymbol{X}=\boldsymbol{X}_{\boldsymbol{l}} \cup \boldsymbol{X_u} \cdot \boldsymbol{X}_{\boldsymbol{l}}$. $\boldsymbol{X_l}$ represents the fully labelled dataset, $\boldsymbol{X_u}$ represents the unlabelled dataset, which indicates that $\forall \boldsymbol{x_l \in X}_{\boldsymbol{l}},\left\{\boldsymbol{x}_{\boldsymbol{l}}\right\}_*=\{\boldsymbol{\Psi}\}$ and $\forall \boldsymbol{x}_{\boldsymbol{u}} \in \boldsymbol{X}_{\boldsymbol{u}},\left\{\boldsymbol{x}_{\boldsymbol{u}}\right\}_*=\boldsymbol{\phi}$, and the size of $\boldsymbol{X_l}$ is far less than the one of $\boldsymbol{X_u}$.

Usually using the \textit{\textbf{Dice}} similarity coefficient (\textit{\textbf{DSC}}), 95\% \textit{\textbf{Hausdorff}} distance (\textit{\textbf{HD95}}) and mean surface distance (\textit{\textbf{MSD}}) to evaluate the performance of the segmentation methods. \textit{\textbf{DSC}} is a measure of the volume overlap between the predicted labels and ground truth labels, \textit{\textbf{HD95}} and \textit{\textbf{MSD}} are measures of the surface distance between the predicted labels and ground truth labels.
\begin{equation}\label{eq1}
\boldsymbol{D S C=\dfrac{2 \times\left|P^c \cap G^c\right|}{\left|P^c\right|+\left|G^c\right|}}
\end{equation}
\begin{equation}\label{eq2}
\boldsymbol{H D 95=\max _{95 \%}\left[d\left(P_s^c, G_s^c\right), d\left(G_s^c, P_s^c\right)\right]}
\end{equation}
\begin{equation}\label{eq3}
M S D=\frac{1}{\left|P_s^c\right|+\left|G_s^c\right|}\left(\sum_{p_s^c \in P_s^c} d\left(p_s^c, G_s^c\right)+\sum_{g_s^c \in G_s^c} d\left(g_s^c, P_s^c\right)\right)
\end{equation}
where $\boldsymbol{P}^{\boldsymbol{c}}$ and $\boldsymbol{G}^{\boldsymbol{c}}$ represent the set of predicted pixels and the set of real pixels of the $\boldsymbol{c}$ class organ, respectively; $\boldsymbol{P_s^c}$ and $\boldsymbol{G_s^c}$ represent the set of predicted pixels and the set of real pixels of the surface of the $\boldsymbol{c}$ class organ, respectively; and $\boldsymbol{d\left(p_s^c, G_s^c\right)=\min _{g_s^c \in G_s^c}^c\left\|p_s^c-g_s^c\right\|_2}$ represents the minimal distance from point $\boldsymbol{p_s^c}$ to surface $\boldsymbol{G_s^c}$. The review reports various methods based on \textit{\textbf{DSC}} values.

\section{Multi-organ Segmentation Datasets}
\label{sec3}
To obtain high-quality organ segmentation datasets, many research teams have undertaken several collaborations with medical organizations. Table \ref{tab1} summarizes the common head and neck, thorax, and abdomen datasets used for the development and validation of multi-organ segmentation method. Table I also shows that the quantity of annotated data available for deep learning studies is still very low.
\begin{table*}[ht!]
\centering 
\caption{Frequently Used Dataset for Multi-organ Segmentation}
\label{tab1}
\begin{tabular}{@{}lm{0.11\textwidth}llm{0.2\textwidth}m{0.1\textwidth}m{0.1\textwidth}l@{}}
\toprule
Year                  & Dataset                                                                                                                                          & Modality            & Part                     & Number of organs (specific organs)                                                                                                                                                                                                                                                                                                                                                      & Quantity                                                                 & Labelling status                                              & Image size                                     \\ \midrule
2015                  & MICCAI Multi-Atlas Labelling Beyond the Cranial Vault (BTCV) \cite{29}                                                                            & CT                  & Abdomen                  & 13 (spleen, right kidney, left kidney, gallbladder, esophagus, liver, stomach, aorta, inferior vena cava, portal and splenic veins, pancreas, right adrenal gland, left adrenal gland)                                                                                                                                                                                                  & 50 (30 training and 20 testing)                                          & The training set are labelled, the test set are not labelled  & 512 × 512 × {[}85$\sim$198{]}                  \\
2015                  & MICCAI head and neck Auto Segmentation Challenge (HNC) \cite{30}                                                                                  & CT                  & Head and neck            & 9 (brainstem, mandible, chiasm, left optic nerves, right optic nerves, left parotid glands, right parotid glands, left submandibular glands, right submandibular glands)                                                                                                                                                                                                                & 35 (25 training, 10 off-site tests, 5 on-site tests)                     & Labelled                                                      & 512 × 512 × {[}110$\sim$190{]}                 \\
2015                  & \begin{tabular}[c]{@{}l@{}}Synapse multi-\\ organ segmentation\\ dataset (Synapse)\end{tabular}                                                    & CT                  & Abdomen                  & 13 (spleen, right kidney, left kidney, gallbladder, esophagus, liver, stomach, aorta, inferior vena cava, portal vein and splenic vein, pancreas, right adrenal gland, left adrenal gland)                                                                                                                                                                                              & 50 (30 training, 20 testing)                                             & Labelled                                                      & 512 × 512 × {[}85$\sim$198{]}                  \\
2015                  & Public Domain Database for Computational Anatomy (PDDCA) \cite{30}                                                                                & CT                  & Head and neck            & 9 (brainstem, mandible, chiasm, left optic nerves, right optic nerves, left parotid glands, right parotid glands, left submandibular glands, right submandibular glands)                                                                                                                                                                                                                & 48 (25 training, 8 additional training, 10 off-site and 5 on-site tests) & Labelled                                                      & 512 × 512 × {[}110$\sim$190{]}                 \\
2017                  & Thoracic Auto-segmentation Challenge (AAPM) \cite{31}                                                                                             & CT                  & Thorax                   & 5 (left lung, right lung, heart, Esophagus, spinal cord)                                                                                                                                                                                                                                                                                                                                & 60 (36 training, 12 off-site tests, 12 on-site tests)                    & Labelled                                                      & 512 × 512 × {[}103$\sim$279{]}                 \\
\multirow{2}{*}{2019} & \multirow{2}{*}{\begin{tabular}[c]{@{}l@{}}Combined (CT-MR)\\ Healthy Abdominal\\ Organ Segmentation\\ (CHAOS) \cite{32}\end{tabular}}                                                          & CT                  & \multirow{2}{*}{Abdomen} & \multirow{2}{*}{\begin{tabular}[c]{@{}l@{}}4 (left kidney, right kidney, liver,\\ spleen)\end{tabular}}                                                                                                                                                                                                                                                                                                                           & 40 (20 training and 20 testing)                                          & Labelled training set and unlabeled test set                  & 512 × 512 × {[}78$\sim$294{]}                  \\
                      &                                                                                                                                                  & MR                  &                          &                                                                                                                                                                                                                                                                                                                                                                                         & 40 (20 training, 20 testing) × 3 sequences                               & Labelled training set and unlabeled test set                  & 256 × 256 × {[}26$\sim$50{]}                   \\
2019                  & SegTHOR Challenge: Segmentation of Thoracic Organs at Risk in CT Images (SegTHOR) \cite{33}                                                      & CT                  & Thorax                   & 5 (left and right lungs, heart, esophagus, spinal cord)                                                                                                                                                                                                                                                                                                                                 & 60 (36 training, 12 off-site tests, 12 on-site tests)                    & Labelled                                                      & 512 × 512 × N                                  \\
2019                  & Annotations for Body Organ Localization based on MICCAI LITS Dataset \cite{34}                                                                    & CT                  & Thorax                   & 11 (heart, left lung, right lung, liver, spleen, pancreas, left kidney, right kidney, bladder, left femoral head, right femoral head)                                                                                                                                                                                                                                                   & 201 (131 training and 70 testing)                                        & Bounding boxes labelled                                       & 512 × 512 × N                                  \\
\multirow{2}{*}{2019} & \multirow{2}{*}{\begin{tabular}[c]{@{}l@{}}Automatic Structure\\ Segmentation for\\ Radiotherapy Planning\\ Challenge 2019\\ (StructSeg)\end{tabular}} & \multirow{2}{*}{CT} & Head and neck            & 22 (left eye, right eye, left lens, right lens, left optical nerve, right optical nerve, chiasm, pituitary, brainstem, left temporal lobes, right temporal lobes, spinal cord, left parotid gland, right parotid gland, left inner ear, right inner ear, left middle ear, right middle ear, left temporomandibular joint, right temporomandibular joint, left mandible, right mandible) & 60 (50 training, 10 testing)                                             & \multirow{2}{*}{\begin{tabular}[c]{@{}l@{}}Labelled\\ training\\ set and\\ unlabeled\\ test set\end{tabular}} & \multirow{2}{*}{512 × 512 × {[}98$\sim$140{]}} \\
                      &                                                                                                                                                  &                     & Thorax                   & 6 (left lung, right lung, spinal cord, esophagus, heart, trachea)                                                                                                                                                                                                                                                                                                                       & 60 (50 training, 10 testing)                                             &                                                               &                                                \\
2020                  & OpenKBP: The open-access knowledge-based planning grand challenge and dataset \cite{35}                                                           & CT                  & Head and neck            & 7 (brainstem, spinal cord, right parotid, left parotid, larynx, esophagus, mandible)                                                                                                                                                                                                                                                                                                    & 340 (200 training, 40 validating, 100 testing)                           & Labelled                                                      & 128×128×128                                    \\
2021                  & Abdomenct-1k \cite{36}                                                                                                                            & CT                  & Abdomen                  & 5 (liver, right and left kidneys, spleen, pancreas)                                                                                                                                                                                                                                                                                                                                     & 1112                                                                     & Labelled                                                      & 512 × 512 × N                                  \\ \bottomrule
\end{tabular}
\end{table*}

\section{Full Annotation-based Methods}
\label{sec4}
The method based on full annotation means that all organs of the multi-organ segmentation task are fully annotated. The existing methods can be analysed from four parts: network architecture, network dimension, network dedicated modules, and network loss function. Among these methods, the network architecture part summarizes the common neural network architectures and the combination or cascade of different architectures. In the network dimension part, the existing methods are classified into 2D, 3D, and multi-view methods according to the image dimension used. The part of network dedicated modules describes modules that are commonly used in multi-organ segmentation to improve the segmentation performance, The part of network loss function summarizes how common loss functions are innovated around multi-organ segmentation.

\subsection{Network Architecture}
\label{sec4_1}
Based on the design of the network architecture, multi-organ segmentation methods can be classified according to single-stage and multistage implementations. Single-stage methods include those based on CNN (Convolutional Neural Network), GAN (Generative Adversarial Network), transformer or hybrid networks. Multistage approaches include coarse-to-fine methods, localization and segmentation methods, or other cascade approaches. Tables \ref{tab_S_1}- \ref{tab_S_3} summarize the literature related to single-stage methods for the segmentation of multi-organ in the head and neck, abdomen and chest based on DSC metrics. Since there are too many organs in the head and neck as well as abdomen, this paper mainly reports 9 organs in the head and neck and 7 organs in the abdomen. Tables  \ref{tab_S_10}- \ref{tab_S_11} in the supplementary materials summarize the DSC values of other organs.

\subsubsection{CNN-Based Methods}
Convolutional Neural Network (CNN) is a feedforward neural network which can automatically extract deep features of the image. Multiple neurons are connected to each neuron in next layer, where each layer can perform complex tasks such as convolution, pooling, or loss computation \cite{92}. CNNs have been successfully applied to medical images, such as brain \cite{93,94} and pancreas \cite{55} segmentation tasks.

\paragraph{Early CNN-Based Methods}
Earlier CNN-based methods mainly used convolutional layers to extract features and then went through pooling layers and fully connected layers to obtain the final prediction results. Ibragimov and Xing \cite{37} used deep learning methods to segment OARs in head and neck CT images for the first time, training 13 CNNs for 13 OARs, and showed that the CNNs outperformed or were comparable to advanced algorithms in segmentation accuracy for organs such as the spinal cord, mandible, larynx, pharynx, eye, and optic nerve, but performed poorly in the segmentation of organs such as the parotid gland, submandibular gland, and optical chiasm. Fritscher {\it{et al.}} \cite{38} combined the shape location as well as the intensity with CNN for segmentation of the parotid gland, submandibular gland and optic nerve. Moeskops {\it{et al.}} \cite{95} investigated whether a single CNN can be used to segment six tissues in brain MR images, pectoral muscles in breast MR images, and coronary arteries in heart CTA images. The results showed that a single CNN can segment multiple organs not only on a single modality but also on multiple modalities.

\paragraph{FCN-Based Methods}
Early CNN-based methods made some improvements in segmentation accuracy compared to traditional methods. However, CNN involves multiple identical computations of overlapping voxels during the convolution operation, which may cause some performance loss. Moreover, the spatial information of the image is lost when the convolutional features are input into the final fully connected network layer. Thus, Shelhamer {\it{et al.}} \cite{96} proposed the Fully Convolutional Network (FCN), which enables end-to-end segmentation by using transposed convolutional layers that allow the size of the predicted image to match the size of the input image. Wang {\it{et al.}} \cite{97} used FCN combined with a new sample selection strategy to segment 16 organs in the abdomen, and Trullo {\it{et al.}} [83] used a variant of FCN, SharpMask \cite{98}, to segment the esophagus, heart, trachea, and aorta in the thorax, which showed the segmentation results of all four organs were improved compared with the normal FCN.

\paragraph{U-Net-Based Methods}
Based on FCN, Ronneberger {\it{et al.}} \cite{99} proposed a classical U-Net architecture, which is consisted of an encoder for the down sampling layer and a decoder for the up-sampling layer, and connects them layer by layer with skip connections, so that the features extracted from the down sampling layer can be directly transmitted to the up-sampling layer to fuse multiscale features for segmentation. U-Net has become one of the most commonly used architectures in the field of multi-organ segmentation \cite{39,40,42,48,61,69,78}. Roth {\it{et al.}} \cite{61} applied the U-Net architecture to segment the abdominal aorta, portal vein, liver, spleen, stomach, gallbladder, and pancreas. The advanced segmentation performance of multiple organs was achieved with an average Dice value of 0.893 for seven organs. Lambert {\it{et al.}} \cite{86} proposed a simplified U-Net for segmenting the heart, trachea, aorta, and esophagus of the chest. The results showed that adding dropout and using bilinear interpolation can significantly improve the segmentation performance of the heart, aorta, and esophagus compared with the ordinary U-Net. In addition to U-Net, V-Net \cite{100} proposes 3D image segmentation method based on volumetric, fully convolutional neural network. This method can directly process 3D medical data by introducing residual connections and using convolutional layers instead of pooling layers in the original U-Net. Gibson {\it{et al.}} \cite{60} used dense V-Networks to segment the pancreas, esophagus, stomach, liver, spleen, gallbladder, left kidney, and duodenum of the abdomen. Xu {\it{et al.}} \cite{77} proposed a new probabilistic V-Net model which combines a conditional variational autoencoder (cVAE) and hierarchical spatial feature transform (HSPT) for abdominal multi-organ segmentation. nnU-Net [101] is a novel framework based on U-Net architecture with the addition of adaptive pre-processing, data enhancement, and postprocessing techniques, and has shown state-of-the-art results on many publicly available datasets for different biomedical segmentation challenges \cite{50,102,103,104}. Podobnik {\it{et al.}} \cite{50} reported the results of segmentation of 31 OARs of the head and neck using the nnU-Net architecture combined with CT and MR images.

\subsubsection{GAN-Based Methods}
A typical Generative Adversarial Network (GAN) \cite{105} includes a pair of competitive networks, which are generators and discriminators. The generator attempts to deceive the discriminator by generating the artificial data, and the discriminator strives to discriminate the artificial data without being deceived by the generator; after alternate optimization training, the performance of both networks can eventually be improved. In recent years, many GAN-based multi-organ segmentation methods have been proposed and achieved high segmentation accuracy \cite{41,62,84,106,107,108,109}.

Dong {\it{et al.}} \cite{84} jointly trained GAN with a set of U-Nets as a generator and a set of FCNs as a discriminator for segmenting the left lung, right lung, spinal cord, esophagus and heart from chest CT images. The results showed that the segmentation performance of most of the organs were improved with the help of adversarial networks, and the average DSC values of the above five OARs were finally obtained as 0.970, 0.970, 0.900, 0.750 and 0.870. Tong {\it{et al.}} \cite{41} proposed a Shape-constraint GAN for automatic head and neck OARs segmentation (SC-GAN) from CT and low-field MRI images. It uses DenseNet, a deep supervised fully convolutional network to segment organs for prediction, and uses a CNN as discriminator network to correct the error of prediction. The results show that the combination of GAN and DenseNet can further improve the segmentation performance of CNN based on the original shape constraints.

GAN can improve accuracy with its adversarial losses. However, the training of GAN network is difficult and time-consuming since the generator needs to achieve Nash equilibrium with the discriminator. And its adversarial loss as a shape modifier can only achieve higher segmentation accuracy when segmenting organs with regular and unique shapes (such as liver and heart), but may not work well for irregular or tubular structures (such as pancreas and aorta).

\subsubsection{Transformer-Based Methods}
CNN-based methods can perform well for segmenting multiple organs in many tasks, but the inherent shortcomings of the perceptual field of the convolutional layers lead to the inability of CNNs to model global relationships, hindering the performance of the models. The self-attentive mechanism of the transformer \cite{110} can solve the long-term dependency problem well, achieving better results than CNNs in many tasks such as natural language processing (NLP) or computer vision \cite{111}. The performance of the medical image segmentation networks using transformer is also close or even better than the one of current state-of-the-art methods \cite{112,113,114,115}.

Cao {\it{et al.}} \cite{71} integrated the transformer with a U-shaped architecture to explore the potential of the pure transformer model for abdominal multi-organ segmentation. The results showed that the method has good segmentation accuracy. However, the method needs to initialize the network encoder and decoder using the training weights of the Swin transformer on ImageNet. Huang {\it{et al.}} \cite{75} introduced an efficient and powerful medical image segmentation architecture, MISSFormer, where the proposed enhanced mixed block can effectively overcome the feature recognition limitation problem caused by convolution. Moreover, compared with Swin-UNet, this model does not require pre-training on large-scale datasets to achieve comparable segmentation results.

Transformer-based approaches can capture long-range dependencies and achieve better performance than CNNs in many tasks. However, multi-organ segmentation problem involves the segmentation of many tiny organs, and the pure transformer network focuses on the global context modelling. This leads to the lack of detailed localization information of low-resolution features. Thus, a coarser segmentation result is usually obtained.

\subsubsection{Hybrid Networks}
CNN convolution operation can extract local features well, but it is difficult to obtain global features. The self-attentive mechanism of the transformer can effectively capture feature dependencies over long distances, but it loses local feature details, which may obtain poor results for the segmentation accuracy of small organs. Therefore, some researchers have combined the CNN and transformer to overcome the limitations of both architectures \cite{82,113,116,117,118,119}.

Suo {\it{et al.}} \cite{76} proposed an intra-scale and inter-scale collaborative learning network (I2-Net) by combining features extracted by the CNN and transformer to segment multiple organs of the abdomen, which improved the segmentation performance of small and medium-sized organs by 4.19\% and 1.83\%-3.8\%, respectively. Kan {\it{et al.}} \cite{51} proposed ITUnet, which adds the features extracted by the transformer to the output of each block of the CNN-based encoder, which can obtain segmentation results provided by both the local and global information of the image. ITUnet has better accuracy and robustness than other methods, especially on difficult organs such as the lens. Chen {\it{et al.}} \cite{72} proposed a new network architecture, TransUNet, which uses a transformer to further encode CNN encoders to build stronger encoders and report competitive results for multi-organ segmentation of the head and neck. Hatamizadeh {\it{et al.}} \cite{66} proposed a new architecture U-net transformer (UNETR) using a transformer as an encoder and the CNN as a decoder, which achieves better segmentation accuracy by capturing global and local dependencies.

In addition to the methods combining CNN and transformer, there are some other hybrid frames. For example, Chen {\it{et al.}} \cite{120} combined U-Net and long short-term memory (LSTM) to realize the segmentation of five organs in the chest, and the DSC values of all five organs were above 0.8. Chakravarty {\it{et al.}} \cite{121} proposed a hybrid architecture combining CNN and RNN to segment the optic disc, nucleus, and left atrium. The hybrid architecture-based approach can combine and utilize the advantages of the two architectures for the accurate segmentation of small and medium-sized organs, which is a key research direction for the future.

\subsubsection{Cascade Networks}
Due to most organs occupy only a small volume in images, the segmentation models are easy to segment large organs and ignore small organs, which prompted researchers to propose cascade multistage methods. Multistage methods can be divided into two main categories, depending on the information provided by the primary network to the secondary network. The first category is called coarse-to-fine multi-organ segmentation method, where the first network performs coarse segmentation, and its results are passed to another network to achieve fine segmentation. The second category is called multi-organ segmentation method based on localization and segmentation, where candidate boxes for the location of each organ are identified by registration methods or localization networks, and then input into the second-level network for fine segmentation. In addition, the first network can provide other information, such as the shape location or proportion, to better guide the segmentation of the second network. Tables \ref{tab_S_4}-\ref{tab_S_9} summarize the relevant literature of the cascade methods for head and neck, chest and abdomen based on DSC metrics, and tables \ref{tab_S_7}-\ref{tab_S_8} in the supplementary materials summarize the DSC metrics of other organs.

\begin{figure}[ht!]
    \centering
    \includegraphics[width=\columnwidth]{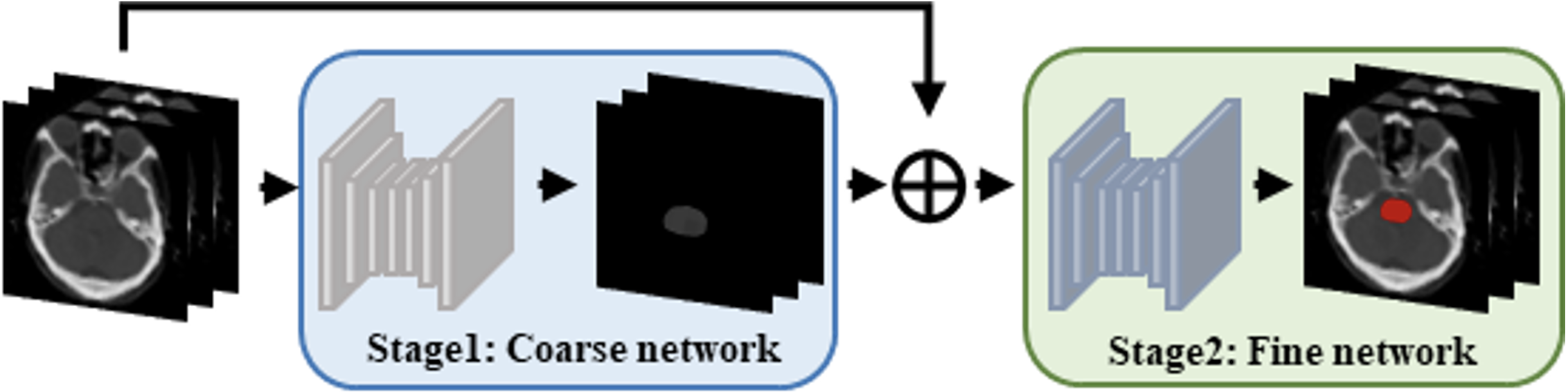}
    \caption{Coarse-to-fine-based segmentation method.}
    \label{fig7}
\end{figure}

\paragraph{Coarse-to-Fine-Based Methods}
The coarse-to-fine-based methods first inputs the original image and its corresponding labels into the first network. After training, the first-level network obtains the coarse segmentation probability map, which will be multiplied by the original image, and the results will be input into the second network to refine the rough segmentation. This process is shown in Fig. \ref{fig7}. In recent years, a number of coarse-to-fine methods have been proposed for multi-organ segmentation \cite{123,124,125,126,129,130,133,152,153,154}, and the references are shown in Tables \ref{tab_S_5}-\ref{tab_S_7}.

Trullo {\it{et al.}} \cite{83} proposed two synergistic depth architectures to jointly segment all organs, including the esophagus, heart, aorta, and trachea. Probabilistic maps obtained in the first stage were passed to the second stage to learn anatomical constraints, and then four networks were trained for four structures in the second stage to distinguish the background from each target organ in separate refinements. Zhang {\it{et al.}} \cite{129} proposed a new cascaded network model with Block Level Skip Connections (BLSC) between two cascaded networks. This architecture enabled the second-stage network to capture the features learned by each block in the first-stage network and accelerated the convergence of the second-stage network. Xie {\it{et al.}} \cite{130} proposed a new framework called the Recurrent Saliency Transformation Network (RSTN). This framework enabled coarse scale segmentation masks to be passed to the fine stage as spatial weights, while gradients can be backpropagated from the loss layer to the whole network, so as to realize the joint optimization of the two stages, thus improving the segmentation accuracy of small targets. Ma {\it{et al.}} \cite{125} proposed a new end-to-end coarse-to-fine segmentation model to automatically segment multiple OARs in head and neck CT images. This model used a predetermined threshold to classify the initial results of the coarse stage into large and small OARs, and then designed different modules to refine the segmentation results.

This coarse-to-fine approach effectively reduces the complexity of the background and enhances the discriminative information of the target structures. Compared with the single-stage approach, this coarse-to-fine-based method improves the segmentation results for small organs, but there are limitations in memory and training time because at least two networks need to be trained.

\begin{figure}[ht!]
    \centering
    \includegraphics[width=\columnwidth]{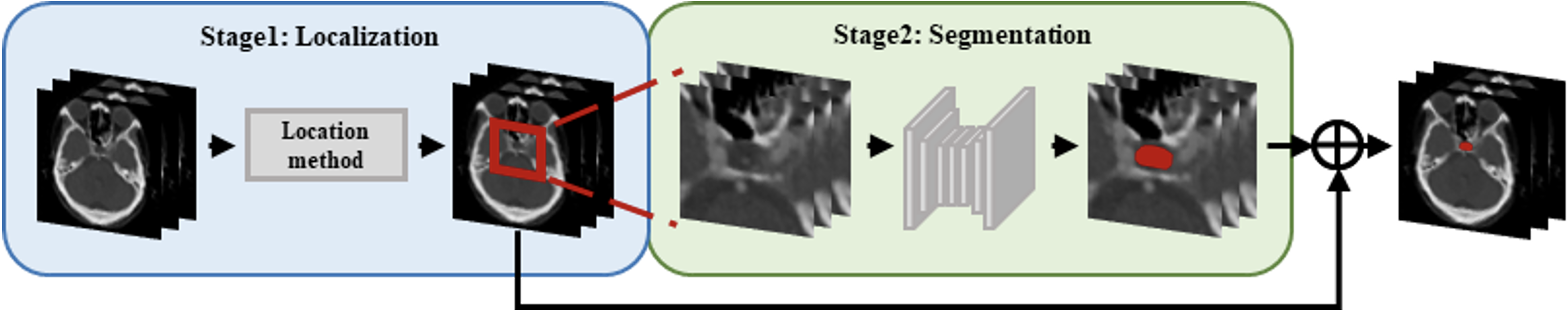}
    \caption{Localization and segmentation based method.}
    \label{fig8}
\end{figure}

\paragraph{Localization and Segmentation Based Methods}
The localization and segmentation methods are also multistage cascade methods. Here, the first-level network provides location information, returns a candidate frame, and crops the region of interest of the image according to the location information, and uses it as the input of the second network. In this way, when the second network performs segmentation, one organ can be targeted, excluding the interference of other organs or background noise and improving the segmentation accuracy. The process is shown in Fig. \ref{fig8}. The organ location in the first stage can be obtained through registration or localization network. The relevant literature of multi-stage method based on location and segmentation are listed in Tables \ref{tab_S_7}-\ref{tab_S_9}, and the DSC values of other organs are listed in tables \ref{tab_S_14}-\ref{tab_S_15}.

Wang {\it{et al.}} \cite{137}, Men {\it{et al.}} \cite{138}, Lei {\it{et al.}} \cite{144}, Francis {\it{et al.}} \cite{151}, Tang {\it{et al.}} \cite{139} proposed decomposing OARs segmentation into two stages of localization and segmentation. The first stage localizes the target OARs using the bounding box, the second stage segments the target OARs within the bounding box, and both stages use neural networks. Among them, Wang {\it{et al.}} \cite{137} and Francis {\it{et al.}} \cite{151} used a 3D U-net in both stages. Lei {\it{et al.}} \cite{144} used Faster RCNN to automatically locate the ROI of organs in the first stage. Korte {\it{et al.}} \cite{146} demonstrated that the CNN is a suitable method for automatically segmenting parotid and submandibular glands in MRI images of HNC patients. The segmentation accuracy of the parotid and submandibular glands can be improved by cascading localizing CNNs, cropping and segmenting high-resolution CNNs. FocusNet \cite{108,142} presented a novel deep neural network to solve the class imbalance problem in the segmentation of head and neck OARs. The small organs are first localized by the organ localization network. Then, combined with the high-resolution information of each small organ, multiscale features are input to the segmentation network together to accurately segment the small organs.

The organ localization by Larsson {\it{et al.}} \cite{148}, Zhao {\it{et al.}} \cite{149}, Ren {\it{et al.}} \cite{122} and Huang {\it{et al.}} \cite{145} was obtained with registration method followed by the application of convolutional neural networks for segmentation. Among them, Ren {\it{et al.}} \cite{122} designed interleaved cascades of 3D-CNNs to segment each structure of interest. Since adjacent tissues are usually highly correlated from a physiological and anatomical perspective, using the initial segmentation results of a specific tissue can help refine the segmentation of other neighbouring tissues. Zhao {\it{et al.}} \cite{149} proposed a new flexible knowledge-assisted convolutional neural network which combine deep learning and traditional methods to improve the segmentation accuracy in the second stage.

The vast majority of approaches require to determine the target areas prior to segmentation network training by different localization methods. For example, Ren {\it{et al.}} \cite{122} localized organ regions through a multi-atlas-based method. Wang {\it{et al.}} \cite{137} used separate CNNs to localize candidate areas. That is, their target organ region localization is constructed independently of organ segmentation, which will hinder the transmission of information between these two related learning tasks. On this basis, Liang {\it{et al.}} \cite{141} proposed a multi-organ segmentation framework based on multi view spatial aggregation, which combines the learning of the organ localization subnetwork and the segmentation subnetwork to reduce the influence of background regions and neighbouring similar structures in the input data. Additionally, the proposed fine-grained representation based on ROIs can improve the segmentation accuracy of organs with different sizes, especially the segmentation results of small organs.

The type of multistage method improves the organ segmentation accuracy, especially for small organs, which largely reduces the interference of the background. However, this two-step process has certain requirements for memory and training time, and the segmentation accuracy also depends largely on the regional localization accuracy. Better localization of organs and improvement of segmentation accuracy are still directions to be investigated in the future.

\paragraph{	Other Cascade Methods}
In addition to probability maps and localization information, the first network can also provide other types of information, such as scale information and shape priors. For example, Tong {\it{et al.}} \cite{155} combines the FCNN and a shape representation model (SRM) for head and neck OARs segmentation. The first-level network is the SRM for learning highly representative shape features in head and neck organs. The direct comparison of the FCNN with and without SRM shows that the SRM significantly improves the segmentation accuracy of nine organs with different sizes, morphological complexity, and different CT contrasts. Roth {\it{et al.}} \cite{127} proposed a multiscale 3D FCN approach which is accomplished by two cascaded FCNs, where low-resolution 3D FCN predictions are upsampled, cropped, and connected to higher-resolution 3D FCN inputs. In this case, the primary network provides scale information to the secondary network. And the method uses the scale space pyramid with automatic context to perform high-resolution semantic image segmentation, while considering large contextual information from the lower resolution levels.

\subsection{Network Dimension}
\label{sec4_2}
\begin{figure}[ht!]
    \centering
    \includegraphics[width=\columnwidth]{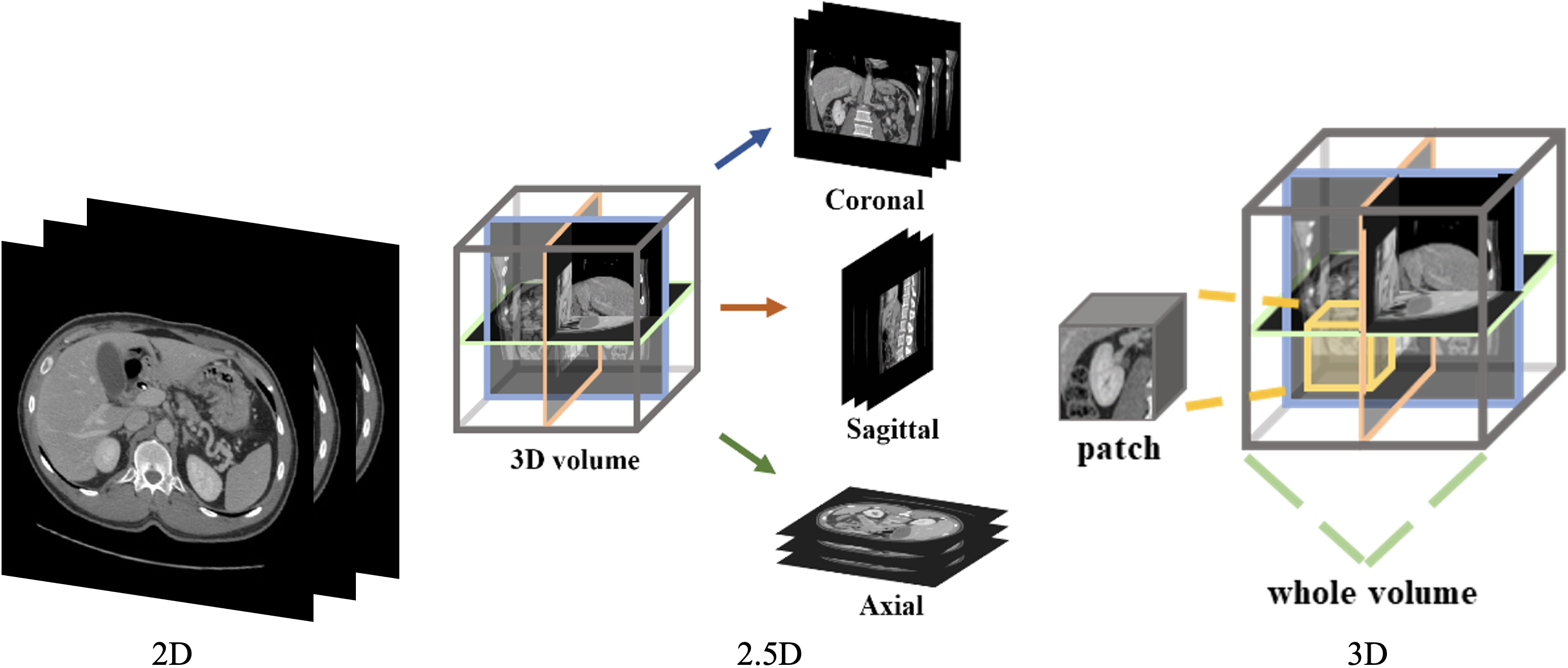}
    \caption{Different network dimensions.}
    \label{fig9}
\end{figure}

Considering the dimensionality of input images and convolutional kernels, multi-organ segmentation neural networks can be classified into 2D, 2.5D and 3D architectures, as shown in Fig. \ref{fig9}, and the differences between the three architectures will be discussed in follows.

\subsubsection{2D- \& 3D-Based Methods}
The input of the 2D multi-organ segmentation neural network is slices from a three-dimensional medical image, and the convolution kernel is also two-dimensional. Men {\it{et al.}} \cite{58}, Trullo {\it{et al.}} \cite{83}, Gibson {\it{et al.}} \cite{60}, Chen {\it{et al.}} \cite{44}, Zhang {\it{et al.}} \cite{48}, Chen {\it{et al.}} \cite{67} used 2D networks for multi-organ segmentation. 2D architectures can reduce the GPU memory burden, but CT or MRI images are inherently 3D. Moreover, slicing images into 2D tends to ignore the rich information in the entire image voxel, so 2D models are insufficient for analysing the complex 3D structures in medical images.

3D multi-organ segmentation neural network architectures use 3D convolutional kernels, which can directly extract feature information from 3D medical images. Roth {\it{et al.}} \cite{61}, Zhu {\it{et al.}} \cite{39}, Gou {\it{et al.}} \cite{42}, and Jain {\it{et al.}} \cite{156} used 3D architectures for multi-organ segmentation. However, due to GPU memory limitations, 3D architectures may face computationally intensive and memory shortage problems, so the majority of 3D network methods use sliding windows acting on patches. Zhu {\it{et al.}} \cite{39} proposed a deep learning model called AnatomyNet, which receives full-volume head and neck CT images as the inputs and generates masks of all organs to be segmented at once. AnatomyNet only uses a down sampling layer in the first encoding block to consider the trade-off between GPU memory usage and network learning capability, which can occupy less GPU memory than other network structures while preserving information about small anatomical structures.

\subsubsection{Multi-View-Based Methods}
In medical image segmentation, it is crucial to make good use of the spatial information between medical image slices. Directly input 3D images into the network, the 3D images will occupy huge memory, or convert 3D images to 2D images, the spatial information between medical image slices will be directly discarded. Thus, the idea of multiple views has appeared, which means using 2.5D neural networks with multiple 2D slices and combining 2D convolution and 3D convolution.

The 2.5D multi-organ segmentation neural network architecture still uses 2D convolutional kernels, but the input of the network is multiple slices, either a stack of adjacent slices using interslice information \cite{64,89}, or slices along three orthogonal directions (axial, coronal, and sagittal) \cite{37,38,124,143}. This 2.5D approach saves computational resources and makes good use of spatial information. It is also widely used in semi supervised-based methods, which are reviewed in Section \ref{sec5_2}. Zhou {\it{et al.}} \cite{157} segmented each 2D slice using the FCN by sampling a 3D CT case on three orthogonally oriented slices (2D images) and then assembled the segmented output (i.e., 2D slice results) back into 3D. Chen {\it{et al.}} \cite{67} developed a multi-view training method at the ratio of 4:1:1 on different views (axial, coronal, and sagittal) and applied a majority voting strategy to combine the three predictions into a final segmentation. The results show that the method can remove some wrong segmentation areas in the single-view output, especially for the small intestine and duodenum. Wang {\it{et al.}} \cite{128} used a statistical fusion approach to combine segmentation results from three views and relate the structural similarity of 2D views to the original 3D image. Liang {\it{et al.}} \cite{143} performed context-based iterative refinement training on each of the three views and aggregated all the predicted probability maps of the three orthogonal views in the last iteration to obtain the final segmentation results. Experiments show that this multi-view framework outperforms the segmentation results of the three separate views.

Tang {\it{et al.}} \cite{68} proposed a new framework for combining 3D and 2D models, which implements segmentation through high-resolution 2D convolution and extracting spatial contextual information through low-resolution 3D convolution. The corresponding 3D features used to guide 2D segmentation are controlled by a self-attentive mechanism, and the results show that this method consistently outperforms existing 2D and 3D models. Chen {\it{et al.}} \cite{44} proposed a hybrid convolutional neural network, OrganNet2.5D, which can make full use of 3D image information to process different planar and depth image resolutions. OrganNet2.5D integrates 2D convolution and 3D convolution to extract both clear underlying edge features and rich high-level semantic features.

Some current studies only deal with 2D image, which avoids memory and computation problems but does not make full use of 3D image information. 2.5D methods can make better use of information from multiple views and improve single-view segmentation compared to 2D networks, but the spatial contextual information they can extract is still limited. Moreover, the current 2.5D methods using in multi-organ segmentation are the aggregation of three perspectives at the outcome level, and the intermediate processes are independent of each other; better use of the intermediate learning process is also the direction to be investigated \cite{158,159,160}. Some studies have performed 3D convolution, but local patches need to be processed. For example, Networks that process full-volume 3D CT images, similar to AnatomyNet, use only a down sampling layer to preserve information about small anatomical structures, so the receptive field of these networks is limited. To solve this problem, DenseASPP with four expansion rates (3,6,12,18) is introduced into FocusNet \cite{142}; however, when the expansion rates of the cascaded expanded convolution have a common factor relationship, grid problems affecting the segmentation accuracy may occur. Pure 3D networks also face the problem of increased parameter and computational burden, which limits the depth and performance of the network. Therefore, considering the memory and computational burden, better combination of multi-view information for more accurate multi-organ segmentation is still the future research direction.

\subsection{Network Dedicated Modules}
\label{sec4_3}
The network architecture is very important to improve the multi-organ segmentation accuracy, but its design process is complex. In multi-organ segmentation tasks, there are many special mechanisms to improve the accuracy of organ segmentation, such as the dilation convolution module, feature pyramid module, and attention module. They improve multi-organ segmentation accuracy by increasing the perceptual field, aggregating features of different scales, and focusing the network on the segmented region. Cheng {\it{et al.}} \cite{184} studied the performance improvement of each module of the network compared with the basic U-Net network in the head and neck segmentation task.

\subsubsection{Shape Prior Module}
Shape prior is more suitable for medical images than natural images because the spatial relationships between internal structures in medical images are relatively fixed. Therefore, considering anatomical priors in a multi-organ segmentation task will significantly improve the performance of multi-organ segmentation.

The current methods using anatomical priors fall into two main categories. One category is based on the idea of statistics, which calculates the average distribution of organs in a fully labelled dataset so that the prediction results can be as close as possible to the average distribution of organs \cite{38,55,84,161,162}. The other is to train a shape representation model, which pretrains the shape representation model using the annotation of the training dataset, and then uses it as a regularization term to constrain the predictions of the segmentation network during training \cite{41,155}. It has also been shown that generative models can learn anatomical priors \cite{163}. Therefore, it is a future research direction to consider using generative models (e.g., diffusion models, which are popular in the last two years \cite{164,165}) to better obtain anatomical prior knowledge to improve segmentation performance.

\subsubsection{Dilated Convolutional Module}
In traditional CNNs, down sampling and pooling operation are usually used several times to reduce the computation and expand the field of perception, which will lose the spatial information and make image reconstruction difficult. Dilated convolution (also known as ``Atrous'') introduces another parameter to the convolution layer, namely, the expansion rate, which can expand the field of perception to extract features across a larger spatial range without increasing the computational cost. Dilated convolution is a commonly used method in multi-organ segmentation tasks \cite{54,55,69,88,89} that increases the size of the sampling space, allowing the neural network to extract features in a larger receptive field that captures multiscale contextual information. These contextual features can capture finer structural information, which is important for pinpointing organ location. Gibson {\it{et al.}} \cite{55} used CNN networks with dilated convolution to accurately segment the liver, pancreas, stomach, and esophagus from abdominal CT. Men {\it{et al.}} \cite{58} proposed a new method based on deep extended convolutional neural network (DDCNN) for fast and consistent automatic segmentation of clinical target volumes (CTVs) and OARs. Vesal {\it{et al.}} \cite{88} introduced dilated convolution to 2D U-Net for segmenting the esophagus, heart, aorta, and thoracic trachea.

\subsubsection{Multiscale Module}
Neural networks extract the features of the target layer by layer. The lower layer networks have smaller perceptual fields and stronger representation of geometric detail information, and they have higher resolution but weaker representation of semantic information. The higher layer networks have larger perceptual fields and stronger representation of semantic information, but they have lower resolution of feature maps and weaker representation of geometric information, leading to the information loss of small targets. Common multiscale fusion modules include bottom-up, top-down, and laterally connected feature pyramids (FPNs) \cite{166}, spatial pooling pyramids (ASPPs) \cite{167} combining dilated convolution and multiscale fusion, and others. In multi-organ segmentation tasks, multiscale feature fusion has been widely used in multi-organ segmentation due to the different sizes of the organs of interest. Jia and Wei \cite{69} introduced the feature pyramid into the multi-organ segmentation network using two opposite feature pyramids, top-down and bottom-up forms, which can effectively handle multiscale changes and improve the segmentation accuracy of small targets. Shi {\it{et al.}} \cite{89} used the pyramidal structure of lateral connections between encoders and decoders to capture contextual information at multiple scales. Srivastava {\it{et al.}} \cite{49} proposed a new segmentation architecture named OARFocalFuseNet, which uses a focal modulation scheme to aggregate multiscale contexts in a specific resolution stream when performing multiscale fusion.

\subsubsection{Attention Module}
The attention module can highlight important features by dynamically weighting them. This novel attention mechanism allows exploring the inherent self-attentiveness of the network and is essential for multi-organ segmentation tasks \cite{62,168}. Common attention mechanisms include channel attention, spatial attention, and self-attention.

Squeeze-and-excitation (SE) module \cite{169} is a typical channel attention module which can focus on key parts of an image by generating a channel attention tensor. AnatomyNet \cite{39} uses 3D SE residual blocks to segment the OARs of the head and neck, enabling the extraction of 3D features directly from CT images and adaptively calibrating the mapping of residual features within each feature channel. Liu {\it{et al.}} \cite{45} proposed a new cross-layer spatial attentional map fusion network (CSAF-CNN) to segment multiple organs in the chest, which can effectively integrate the weights of different spatial attentional maps in the network, thus obtain more useful attentional maps. The average DSC of 22 organs in the head and neck was 72.50\%, which was significantly better than U-Net (63.9\%) and SE-UNet (67.9\%). Gou {\it{et al.}} \cite{42} designed a self-channel-spatial-attention neural network (SCSA-Net) for 3D head and neck OARs segmentation, which can adaptively enhance both channel and spatial features. Compared with SE-Res-Net and SE-Net, SCSA-Net improved the DSC of the optic nerve and submandibular gland by 0.06 and 0.03 and 0.05 and 0.04, respectively. Lin {\it{et al.}} \cite{70} suggested to embed the variance uncertainty into the attention architecture and proposed a variance-aware attention U-Net network to improve the attention to error-prone regions (e.g., boundary regions) in multi-organ segmentation. Compared with existing methods, the segmentation results of small organs and organs with irregular structures (e.g., duodenum, esophagus, gallbladder, and pancreas) are significantly improved. Zhang {\it{et al.}} \cite{48} proposed a new hybrid network (Weaving attention U-Net, WAU-Net) with a U-Net++ \cite{170} structure that uses CNNs to extract the underlying features, and uses axial attention blocks to efficiently model global relationships at different levels of the network, which achieve competitive performance in the head and neck multi-organ segmentation task.

\subsubsection{Other Modules}
The dense block \cite{171} can efficiently use the information of the intermediate layer, and the residual block \cite{172} can prevent gradient disappearance during backpropagation. These two modules are often embedded in the basic segmentation framework. The convolution kernel of the deformable convolution \cite{173} can adapt itself to the actual situation and better extract the input features. Heinrich {\it{et al.}} \cite{63} proposed a 3D abdominal multi-organ segmentation architecture with sparse deformable convolutions (OBELISK-Net) and showed that the combination with conventional CNNs can further improve the segmentation of small organs with large shape variations (e.g., pancreas, esophagus). The deformable convolutional block proposed by Shen {\it{et al.}} \cite{80} can handle variations in the shape and size of different organs by generating reasonable receptive fields for different organs with additional trainable offsets. The strip pooling (strip pooling) \cite{174} module can target long strip structures (e.g., esophagus and spinal cord) by using long pooling instead of traditional square pooling to avoid merging contaminated information from unrelated regions and better capture anisotropic and remote contextual information. For example, Zhang {\it{et al.}} \cite{131} used a pool of anisotropic strips with three different directional receptive fields to capture the spatial relationships between multiple organs in the abdomen. Compared to network architectures, network modules have been widely utilized because of their relatively simple design process and the relative ease of embedding them into various architectures.

\subsection{Network Loss Function}
\label{sec4_4}
As we all known, in addition to the network architecture or network modules, the segmentation accuracy also depends on the selected loss function. In multi-organ segmentation tasks, selecting a suitable loss function can reduce the class imbalance in deep learning and improve the segmentation accuracy of small organs.

Jadon \cite{175} summarized the commonly used loss functions in semantic segmentation, which are classified into distribution-based loss functions, region-based loss functions, boundary-based loss functions, and compound-based loss functions. Common loss functions used for multi-organ segmentation include CE loss \cite{176}, Dice loss \cite{177}, Tversky loss \cite{178}, focal loss \cite{179} and their combined loss functions.

\subsubsection{CE Loss}
The CE loss (cross-entropy loss function) \cite{176} is an information theoretic measure that calculates the difference between the prediction of the network and the ground truth. Men {\it{et al.}} \cite{58}, Moeskops {\it{et al.}} \cite{95}, Zhang {\it{et al.}} \cite{48} used CE loss for multi-organ segmentation. However, when the number of foreground pixels is much smaller than the background, CE loss will heavily bias the model towards the background, resulting in poor segmentation results. The weighted CE loss \cite{180} adds weight parameters to each category based on CE loss. so that it can obtain better results in the case of unbalanced sample sizes compared to the original CE loss. Since there is a significant class imbalance problem in multi-organ segmentation, i.e., a very large difference in the number of voxels in different organs, using weighted CE loss will achieve better results than using only the CE loss. Trullo {\it{et al.}} \cite{83} used a weighted CE loss to segment the heart, esophagus, trachea, and aorta in thechest image; Roth {\it{et al.}} \cite{61} applied a weighted CE loss to abdomen multi-organ segmentation.

\subsubsection{Dice Loss}
Milletari {\it{et al.}} \cite{100} proposed the Dice loss as a volume-based overlap measure, converting the voxel measure to the semantic label overlap measure, and becoming a commonly loss function in the segmentation task. Ibragimov and Xing \cite{37} used the Dice loss to segment multiple organs of the head and neck. However, the use of the Dice loss alone does not eliminate the problem that the inherent nature of neural networks is beneficial to large volume organs. Inspired by the weighted CE loss, Sudre {\it{et al.}} \cite{177} introduced the weighted Dice score (GDSC), which adaptively weighed its Dice values according to the current class size. Shen {\it{et al.}} \cite{59} investigated three different types of GDSC based on class label frequencies (uniform, simple, and square) and evaluated their effects on segmentation accuracy. Gou {\it{et al.}} \cite{42} used GDSC for head and neck multi-organ segmentation. Tappeiner {\it{et al.}} \cite{181} introduced the class adaptive Dice loss to further compensate for high imbalances based on nnU-Net, and the results showed that the method could improve the performance of class imbalance segmentation tasks.

\subsubsection{Other Losses}
The Tversky loss \cite{178} is a generalization of the Dice loss and can be optimized by adjusting the parameters to control the balance between false positives and false negatives. The focal loss \cite{179} was proposed in the field of object detection to enhance the attention on samples that are difficult to segment. Similar to the focal loss, the focal Tversky loss \cite{182} focuses on segmenting difficult samples by reducing the weights of simple sample losses. Berzoini {\it{et al.}} \cite{78} used the focal Tversky loss on smaller organs, thus balancing the indices between organs of different sizes, increasing the weights of small samples that are difficult to segment and finally solving the class imbalance problem caused by the kidney and bladder. Inspired by the exponential logarithmic loss (ELD-Loss) \cite{183}, Liu {\it{et al.}} \cite{45} introduced the top-k exponential logarithmic loss (TELD-Loss) to solve the class imbalance problem in the head and neck. The results showed that using this loss function has a strong ability to handle mislabelling.

\subsubsection{Combined Loss}
Each type of loss function has its own advantages and disadvantages. Combining multiple functions can be used for multi-organ segmentation. A more common method is the weighted sum of the Dice loss and CE loss, which attempts to solve the class imbalance problem with the Dice loss while using the CE loss for curve smoothing. Isensee {\it{et al.}} [101] proposed combining the Dice loss and CE loss to measure the overlap of voxel-like predicted outcomes and ground truth. Isler {\it{et al.}} \cite{54}, Srivastava {\it{et al.}} \cite{49}, Xu {\it{et al.}} \cite{77}, Lin {\it{et al.}} \cite{70}, and Song {\it{et al.}} \cite{73} used the weighted combination of the Dice loss and CE loss for multi-organ segmentation. When small objects are involved, using only the Dice loss leads to a lower accuracy; when the predicted region does not overlap with the labelled region, using the CE loss allows the prediction to be as close to the label as possible. Zhu {\it{et al.}} \cite{39} specifically studied different loss functions for the unbalanced head and neck region, and pointed out that the combination of the Dice loss and focal loss was superior to the ordinary Dice loss. Both Cheng {\it{et al.}} \cite{184} and Chen {\it{et al.}} \cite{44} used this combined loss function.

The conventional Dice loss is detrimental for smaller structures because a small amount of voxel misclassification leads to a large decrease in the Dice score. Applying the exponential logarithmic loss or combining the focal loss with the Dice loss can solve this problem. Using this kind of loss function does not require much adjustment to the network, however, it reduces the segmentation accuracy of the hard voxels in the region. On this basis, Lei {\it{et al.}} \cite{47} proposed a new hardness-aware loss function that can focus more on hard voxels to achieve accurate segmentation. The ultimate goal of neural network optimization is the loss function, and designing a suitable loss function so that the network can improve the segmentation accuracy of various organs is still a research direction.

\section{Imperfect Annotation-based Methods}
\label{sec5}
Currently, most of the methods in the multi-organ segmentation field are based on fully annotated methods. However, medical image data is usually hard to acquire and annotate. In particular, for multi-organ segmentation tasks, obtaining fully annotated datasets is quite difficult, which inspired the idea of using imperfect annotation. In this paper, imperfect annotations are classified into two categories. The first category is weak annotation-based methods, where weak annotation indicates that the data annotation is incomplete or imprecise in each case. For example, in multi-organ segmentation, each image has only one kind of organ annotated; each image has no pixel-level annotation but only category annotation; or the annotation is scribbled or contains noise. Another category is semi supervised-based methods, where semi supervision indicates that only a small portion of the total data is annotated and most of the remaining is unannotated. In the following, we introduce the application of these two types of methods in multi-organ segmentation.

\subsection{Weak Annotation-Based Methods}
\label{sec5_1}
In medical image segmentation, it is a difficult task to obtain the annotation of multiple organs simultaneously on the same set of images. For example, many existing single-organ datasets, such as LiTS \cite{185}, KiTS \cite{186} (p19), and pancreas datasets \cite{187}, can only provide annotations for a single organ. However, multi-organ segmentation networks cannot be effectively trained solely based on these single-organ annotated datasets. Therefore, many studies have started to explore learning unified multi-organ segmentation networks from partially labelled datasets. Based on the implementation methods, we divide the current studies into model-based approaches and pseudo label-based approaches.

\subsubsection{Model-Based Methods}
The idea of the model-based approach is to realize a unified network for multiple partially labelled organs. Chen {\it{et al.}} \cite{188} introduced a multi-branch decoder structure with a shared encoder and eight decoders to address the partial labelling problem. However, this structure is not flexible enough to be extended to new classes. Dmitriev and Kaufman \cite{189} proposed conditional CNNs for learning multi-organ segmentation models, which integrate information of organ categories into the segmentation network. Zhang and Xie {\it{et al.}} \cite{190,191} proposed the idea of DoDNet. Similar to conditional CNN, they spliced the task encoding with the features extracted by the encoder, and introduced a dynamic parameter mechanism in the segmentation head. Zhang {\it{et al.}} [103] used the leading framework nn-UNet \cite{101} as the backbone model, adding task encoding as supporting information to the decoder of nn-UNet, and combined the deep supervision mechanism to further refine the output of organs of different sizes. Wu {\it{et al.}} \cite{192} proposed TGNet composed of task-guided attention module and task-guided residual block, which can highlight task-relevant features while suppressing task-irrelevant information during feature extraction. Liu {\it{et al.}} \cite{193} first introduced incremental learning (IL) to aggregate partially labelled datasets in stages, and verified that the distribution of different partially labelled datasets misleads the process of IL. Xu and Yan \cite{194} proposed a new federated multi-encoding U-Net (Fed-MENU) method that can effectively use independent datasets with different partial labels to train a unified model for multi-organ segmentation. The model outperformed any model trained on a single dataset as well as the model trained on all datasets combined. Fang and Yan \cite{195} and Shi {\it{et al.}} \cite{196} trained uniform models on partially labelled datasets by designing new network and proposing specific loss function.

\subsubsection{Pseudo Label-Based Methods}
The pseudo label-based methods generate pseudo labels of unlabelled organs by using partial-organ segmentation models trained in partially labelled datasets, which can be converted to fully supervised methods. Zhou {\it{et al.}} \cite{161} proposed an a Prior-aware Neural Network (PaNN), which utilized prior statistics obtained from a fully labelled dataset to guide the training process based on partially labelled datasets. Huang {\it{et al.}} \cite{197} proposed a weight-averaging joint training framework, which can correct the noise in the pseudo labels, so as to learn a more robust model. Zhang {\it{et al.}} \cite{198} proposed a multi-teacher knowledge distillation framework that utilizes pseudo labels predicted by teacher models trained on partially labelled datasets to train student models for multi-organ segmentation. Lian {\it{et al.}} \cite{162} proposed a multi-organ segmentation model (PRIMP) based on single and multiple organs anatomical priors. The model first generates pseudo labels for each partially labelled dataset so as to obtain a set of multi-organ datasets with pseudo label. Then the multi-organ segmentation model is trained on this dataset, and tested on another new dataset. For the first time, this method considers the domain discrepancy between partially labelled datasets and the tested multi-organ datasets.

In addition to partial annotation, weak annotation also includes image-level annotation, sparse annotation, and noisy annotation \cite{199}. Regarding multi-organ segmentation tasks, Kanavati {\it{et al.}} \cite{200} proposed a weakly supervised organ segmentation method based on classification forests for the liver, spleen, and kidney, in which the labels are scribbled on the organs.

\subsection{Semi Supervised-Based Methods}
\label{sec5_2}
Semi supervised multi-organ segmentation methods make full use of unlabelled data to improve the segmentation performance, thus reducing the need for extensive annotation. In recent years, semi supervised learning has been widely used in medical image segmentation, such as heart segmentation \cite{201,202,203}, pancreas segmentation \cite{204}, and tumour target region segmentation \cite{205}. A detailed review of semi supervised learning in medical images was presented by Jiao {\it{et al.}} \cite{206}, who classified semi supervised medical image segmentation methods into three paradigms: pseudo label-based methods, consistency regularization-based methods, and knowledge prior-based methods. In this review, we focus on semi supervised multi-organ segmentation methods.

Ma {\it{et al.}} \cite{36} established a new benchmark for semi supervised abdominal multi-organ segmentation, which developed a method based on pseudo labelling. The teacher model was first trained on the labelled data, and generated the pseudo labels for the unlabelled data. Then, the student model was trained on both the real labelled and pseudo labelled data. Finally, the teacher model was substituted with the student model to complete the training. The results on the liver, kidney, spleen, and pancreas show that using unlabelled data can improve the performance of multi-organ segmentation.

Multi-view methods are also widely used in semi supervised multi-organ segmentation, where the model is made to learn in a collaborative training manner to extract useful information from multiple planes (e.g., sagittal, coronal, and axial planes), and then use multi-plane fusion to generate more reliable pseudo labels, and thus train better segmentation networks. Zhou {\it{et al.}} \cite{207} designed a system framework, DMPCT, for multi-organ segmentation of abdominal CT scans by fusing multi-planar information on unlabelled data during training. The framework uses a multi-planar fusion module to synthesize inferences and iteratively update pseudo labels for multiple configurations of unlabelled data. Xia {\it{et al.}} \cite{208} proposed an uncertainty-aware multi-view collaborative training (UMCT) method based on uncertainty perception, which first obtains multiple views by spatial transformations such as rotation and alignment, then trains a 3D deep segmentation network on each view, and performs joint training by implementing multi-view consistency on unlabelled data.

In addition to the collaborative training approach, multi-organ segmentation is also suitable for consistency-based learning due to the large number of prospect categories and dense distribution of organs. Consistency learning encourages consistent output through networks with different parameters. Lai {\it{et al.}} \cite{209} developed a semi supervised learning-based DLUNet for abdominal multi-organ segmentation, which consists of two lightweight U-Nets in the training phase. Moreover, regarding unlabelled data, the outputs obtained from two networks are used to supervise each other, which can improve the accuracy of these unlabelled data. It eventually achieves an average DSC of 0.8718 for 13 organs in the abdomen.

In addition, there are other semi supervised multi-organ segmentation-based methods. Lee {\it{et al.}} \cite{210} proposed a discriminator module based on human-in-the-loop quality assurance (QA) to supervise the learning of unlabelled data. They used QA scores as a loss function for unlabelled data. Raju Cheng {\it{et al.}} \cite{211} proposed a powerful semi supervised organ segmentation method, CHASe, for liver and lesion segmentation. It integrates co-training and heteromodality learning into a co-heterogeneous training framework. The framework is trained on a small single-phase dataset and can be adapted to label-free multicentre and multiphase clinical data.

\section{Discussion and Future Trends}
\label{sec6}
In this paper, a systematic review of deep learning methods for multi-organ segmentation is presented from the perspectives of both full annotation and imperfect annotation. The main innovations of the full annotation method focus on the design of network architectures, the combination of network dimensions, the innovation of network modules and the proposal of new loss functions. In terms of the network architecture design, with the development of the transformer \cite{114} architectures, better utilization of these advanced architectures for multi-organ segmentation is a promising direction, as well as the automatic search for the optimal architecture for each organ through neural network architecture search (NAS) \cite{212}. In the network dimension, optimally combining 2D and 3D architectures is a worthwhile research direction. In terms of network module, more dedicated modules need to design to improve the segmentation accuracy according to the multi-organ segmentation task. In terms of the loss functions, targeting the class imbalance, geometric prior or introducing adversarial learning loss will have great potential for designing more comprehensive and diverse loss functions.

Full annotation methods rely on fully annotated and high-quality datasets. Many imperfect annotation-based methods have been proposed for medical image segmentation in the last two years, including the aforementioned multi-organ segmentation based on weak annotation-based methods and semi annotation-based methods. However, compared to full annotation-based methods, the imperfect annotation-based methods have been less studied. It is a future research focus if imperfect annotation-based methods can be used more adequately to achieve the performance close to that of the full annotation-based methods.

Deep learning has already played a significant role in multi-organ segmentation task, but many challenges remain to be explored in the future, which are summarized in follows:

\subsection{Higher Segmentation Accuracy}
\label{sec6_1}
The current multi-organ segmentation method is more effective in solving the segmentation of large organs and organs with standard contours, such as the brainstem and mandible in the head and neck; the left and right lungs and heart in the chest; and the liver, spleen, and stomach in the abdomen. Moreover, the DSC of various methods can basically reach 0.8 or higher, while for small organs, such as the optical chiasm in the head and neck (see Fig. \ref{fig1}(8)), the left and right optic nerves (see Fig. \ref{fig1}(6 and 7)), the DSC can only reach about 0.7; irregular organs such as the pancreas in the abdomen (Fig. \ref{fig2}(4)), and long striped organs such as the spinal cord (Fig. \ref{fig2}(6)), the segmentation results are also not very satisfactory. The future research direction is to enhance the segmentation accuracy of these types of organs using more advanced automatic segmentation frameworks.

\subsection{More Comprehensive Public Datasets}
\label{sec6_2}
Currently, public datasets covering multiple organs are not sufficient. And the vast majority of methods are validated on their private datasets, making it difficult to verify the generalizability of the models. Therefore, there is a need to establish multicentre public datasets of multi-organ segmentation with large data volumes, wide coverage, and strong clinical relevance in the future.

\subsection{Better Use of Imperfect Annotations}
\label{sec6_3}
The vast majority of current methods are based on full annotation methods. Since medical image data are usually not easy to collect and annotating all the organs on the same image is a time-consuming and laborious work. Further studies can be performed to better utilize imperfect annotations \cite{213,214}, including the use of weakly annotated datasets and semi annotated datasets.

\subsection{Study of Transfer Learning Models}
\label{sec6_4}
Existing deep learning models usually trained on one part of the body, which usually tend to obtain poor results when migrated to other datasets or applied to other parts of the body. Therefore, transfer learning models need to be explored in the future. For example, Fu {\it{et al.}} \cite{65} proposed a new method called domain adaptive relational reasoning (DARR). It is used to generalize 3D multi-organ segmentation models to medical data from different domains. In addition, a very significant problem with medical images compared to other natural images is that many private datasets are not publicly available, and many hospitals only release trained models. Therefore, source free domain adaptation problem will be a very important research direction in the future. For example, Hong {\it{et al.}} \cite{81} proposed a source free unsupervised domain adaptive cross-modal approach for abdomen multi-organ segmentation.

\section{Conclusion}
\label{sec7}
In this paper, we systematically review 214 deep learning-based multi-organ segmentation studies in two broad categories, namely full annotation-based methods and imperfect annotation-based methods for multiple parts, including the head and neck, thorax and abdomen. In the fully labelled methods, we summarize the existing methods according to network architectures, network modules, network dimensions, and loss functions. In the imperfect annotation-based methods, we summarize both weak annotation-based methods and semi annotated-based methods. On this basis, we also put forward tailored solutions for some current difficulties and shortcomings in this field, and illustrate the future trends. The comprehensive survey shows that multi-organ segmentation algorithm based on deep learning is rapidly developing towards a new era of more accurate, more detailed and more automated analysis.

\section*{Acknowledgments}
This work was supported by the National Natural Science Foundation of China under grant 82072021. This work was also supported by the medical-industrial integration project of Fudan University under grant XM03211181.

\clearpage

{\appendices
\begin{table*}[ht!]
\section*{Supplementary Materials}
\centering 
\caption{DSC-Based Summary of the Literature on Multi-Organ Single-State Segmentation Methods for the Head and Neck}
\label{tab_S_1}
\resizebox{\textwidth}{!}{
\begin{tabular}{@{}llllllllllllll@{}}
\toprule
\multirow{2}{*}{Ref}                                     & \multirow{2}{*}{Backbone}                  & \multirow{2}{*}{Datasets}              & \multirow{2}{*}{Quantity} & \multirow{2}{*}{Organ type} & \multirow{2}{*}{Brainstem} & \multirow{2}{*}{Mandible} & \multicolumn{2}{l}{Parotid gland} & \multicolumn{2}{l}{Submandibular gland} & \multicolumn{2}{l}{Optic nerve} & \multirow{2}{*}{Chiasm} \\ \cmidrule(lr){8-13}
                                        &                           &                       &          &            &           &          & Left            & Right           & Left               & Right              & Left           & Right          &        \\ \midrule
Ibragimov and Xing \cite{37}             & 2.5D CNN                  & Private (CT)          & 50       & 13         & -         & 0.895    & 0.766           & 0.779           & 0.697              & 0.730              & 0.639          & 0.645          & 0.374  \\
Fritscher {\it{et al.}} \cite{38}              & 2.5D CNN                  & HNC (CT) \cite{30}     & 30       & 3          & -         & -        & 0.810           & -               & 0.650              & -                  & -              & -              & 0.520  \\
Zhu {\it{et al.}} \cite{39}                      & 3D U-Net                  & Private (CT)          & 261      & 9          & 0.867     & 0.925    & 0.881           & 0.874           & 0.814              & 0.813              & 0.721          & 0.706          & 0.532  \\
Van Rooij {\it{et al.}} \cite{40}               & 3D U-Net                  & Private (CT)          & 157      & 11         & 0.640     & -        & 0.830           & 0.830           & 0.820              & 0.810              & -              & -              & -      \\
\bottomrule
\end{tabular}}
\end{table*}

\begin{table*}[ht!]
\caption{DSC-Based Summary of the Literature on Multi-Organ Single-Stage Segmentation Methods for the Abdomen}
\label{tab_S_2}
\resizebox{\textwidth}{!}{%
%
%
}
\end{table*}

\begin{table*}[ht!]
\caption{DSC-Based Summary of the Literature on Multi-Organ Single-State Segmentation Methods for the Thorax}
\label{tab_S_3}
\resizebox{\textwidth}{!}{%
%
 \\
Vesal {\it{et al.}} \cite{88}                                                                  & 2D U-Net                                                    & SegTHOR (CT) \cite{33}                                                                 & 60                                              & 4                                             & 0.941                                                 & 0.858                                                 & 0.926                                                 & -                                             & -                                             & 0.938                                                 & -                                             \\
Shi {\it{et al.}} \cite{89}                                                                    & 2.5D U-Net                                                  & StructSeg (CT) \cite{1}                                                                & 50                                              & 5                                             & 0.941                                                 & 0.821                                                 & 0.882                                                 & 0.968                                         & 0.971                                         & -                                                     & 0.902                                         \\
Mahmood {\it{et al.}} \cite{90}                                                                & 2D U-Net                                                    & AAPM (CT) \cite{31}                                                                    & 60                                              & 5                                             & 0.880                                                 & 0.660                                                 & -                                                     & 0.970                                         & 0.970                                         & -                                                     & 0.800                                         \\
Zhang {\it{et al.}} \cite{91}                                                                 & 2D FCN                                                      & Private (CT)                                                                          & 36                                              & 6                                             & 0.860                                                 & 0.670                                                 & 0.910                                                 & 0.950                                         & 0.960                                         & -                                                     & 0.890       \\ 
\bottomrule
\end{tabular}%
}
\end{table*}

\begin{table*}[ht!]
\caption{DSC-Based Summary of the Literature on Multi-Organ Coarse-to-Fine Segmentation Methods for the Head and Neck}
\label{tab_S_4}
\resizebox{\textwidth}{!}{%
\begin{tabular}{@{}lllllllllllllll@{}}
\toprule
\multirow{2}{*}{Ref}                        & \multirow{2}{*}{Coarse}     & \multirow{2}{*}{Fine}     & \multirow{2}{*}{Datasets}          & \multirow{2}{*}{Quantity} & \multirow{2}{*}{Organ type} & \multirow{2}{*}{Brainstem} & \multirow{2}{*}{Mandible} & \multicolumn{2}{l}{Parotid gland}               & \multicolumn{2}{l}{Submandibular gland}         & \multicolumn{2}{l}{Optic Nerve}                 & \multirow{2}{*}{Chiasm} \\ \cmidrule(lr){9-14}
                                            &                             &                           &                                    &                           &                             &                            &                           & Left                   & Right                  & Left                   & Right                  & Left                   & Right                  &                         \\ \midrule
Ren {\it{et al.}} \cite{122}                        & 3D CNN                      & 3D CNN                    & HNC (CT) \cite{30}                  & 48                        & 3                           & -                          & -                         & -                      & -                      & -                      & -                      & 0.720                  & 0.700                  & 0.580                   \\
\multirow{2}{*}{Tappeiner {\it{et al.}} \cite{123}} & \multirow{2}{*}{3D CNN}     & \multirow{2}{*}{3D CNN}   & \multirow{2}{*}{HNC (CT) \cite{30}} & \multirow{2}{*}{40}       & \multirow{2}{*}{7}          & \multirow{2}{*}{0.820}     & \multirow{2}{*}{0.910}    & \multirow{2}{*}{0.800} & \multirow{2}{*}{0.810} & \multirow{2}{*}{-}     & \multirow{2}{*}{-}     & \multirow{2}{*}{0.640} & \multirow{2}{*}{0.630} & \multirow{2}{*}{0.420}  \\
                                            &                             &                           &                                    &                           &                             &                            &                           &                        &                        &                        &                        &                        &                        &                         \\
\multirow{2}{*}{Pu {\it{et al.}} \cite{124}}        & \multirow{2}{*}{2.5D U-Net} & \multirow{2}{*}{3D U-Net} & \multirow{2}{*}{HNC (CT) \cite{30}} & \multirow{2}{*}{48}       & \multirow{2}{*}{9}          & \multirow{2}{*}{0.880}     & \multirow{2}{*}{0.940}    & \multirow{2}{*}{0.860} & \multirow{2}{*}{0.865} & \multirow{2}{*}{0.788} & \multirow{2}{*}{0.802} & \multirow{2}{*}{0.743} & \multirow{2}{*}{0.768} & \multirow{2}{*}{0.612}  \\
                                            &                             &                           &                                    &                           &                             &                            &                           &                        &                        &                        &                        &                        &                        &                         \\
\multirow{2}{*}{Ma {\it{et al.}} \cite{125}}       & \multirow{2}{*}{3D U-Net}   & \multirow{2}{*}{}         & \multirow{2}{*}{HNC (CT) \cite{30}} & \multirow{2}{*}{48}       & \multirow{2}{*}{9}          & \multirow{2}{*}{0.879}     & \multirow{2}{*}{0.945}    & \multirow{2}{*}{0.892} & \multirow{2}{*}{0.884} & \multirow{2}{*}{0.829} & \multirow{2}{*}{0.815} & \multirow{2}{*}{0.753} & \multirow{2}{*}{0.747} & \multirow{2}{*}{0.659}  \\
                                            &                             &                           &                                    &                           &                             &                            &                           &                        &                        &                        &                        &                        &                        &                         \\
\multirow{2}{*}{Fang {\it{et al.}} \cite{109}}      & \multirow{2}{*}{2D FCN}     & \multirow{2}{*}{3D U-Net} & HNC (CT) \cite{30}                  & 32                        & 9                           & 0.849                      & 0.924                     & 0.842                  & 0.849                  & 0.734                  & 0.782                  & 0.676                  & 0.684                  & 0.547                   \\
                                            &                             &                           & Private (CT)                       & 56                        & 14                          & 0.863                      & 0.905                     & 0.582                  & 0.687                  & 0.668                  & 0.575                  &                        &                        &                         \\ \bottomrule
\end{tabular}%
}
\end{table*}

\begin{table*}[ht!]
\caption{DSC-Based Summary of the Literature on Multi-Organ Coarse-to-Fine Segmentation Methods for the Abdomen}
\label{tab_S_5}
\resizebox{\textwidth}{!}{%
\begin{tabular}{@{}lllllllllllll@{}}
\toprule
\multirow{2}{*}{Ref}    & \multirow{2}{*}{Coarse} & \multirow{2}{*}{Fine} & \multirow{2}{*}{Datasets}                                           & \multirow{2}{*}{Quantity} & \multirow{2}{*}{Category} & \multirow{2}{*}{Liver} & \multirow{2}{*}{Spleen} & \multicolumn{2}{l}{Kidney} & \multirow{2}{*}{Pancreas} & \multirow{2}{*}{Gallbladder} & \multirow{2}{*}{Stomach} \\ \cmidrule(lr){9-10}
                        &                         &                       &                                                                     &                           &                           &                        &                         & Left         & Right       &                           &                              &                          \\ \midrule
Hu {\it{et al.}} \cite{126}     & 3D FCN                  & Refinement Model      & Private (CT)                                                        & 140                       & 4                         & 0.960                  & 0.942                   & 0.954        &             & -                         & -                            & -                        \\
Roth {\it{et al.}} \cite{127}   & 3D FCN                  & 3D FCN                & Private (CT)                                                        & 331                       & 3                         & 0.932                  & 0.906                   & -            & -           & 0.631                     & 0.706                        & 0.843                    \\
Wang {\it{et al.}} \cite{128}   & 2.5D FCN                & 2.5D FCN              & Private (CT)                                                        & 236                       & 13                        & 0.980                  & 0.971                   & 0.968        & 0.984       & 0.878                     & 0.905                        & 0.952                    \\
Zhang {\it{et al.}} \cite{129}  & 3D V-Net                & 3D V-Net              & BTCV (CT) \cite{29}                                                  & 30                        & 13                        & 0.945                  & 0.915                   & 0.909        & 0.919       & 0.694                     & 0.682                        & 0.784                    \\
Xie {\it{et al.}} \cite{130}    & 2.5D FCN                & 2.5D FCN              & Private (CT)                                                        & 200                       & 16                        & 0.969                  & 0.968                   & 0.962        & 0.960       & 0.877                     & 0.894                        & 0.951                    \\
Zhang {\it{et al.}} \cite{131}  & 3D U-Net                & 3D U-Net              & \begin{tabular}[c]{@{}l@{}}FLARE\\ 2021 (CT) \cite{132}\end{tabular} & 511                       & 4                         & 0.954                  & 0.942                   & \multicolumn{2}{l}{0.936}  & 0.753                     & -                            & -                        \\
Lee {\it{et al.}} \cite{133}    & 3D U-Net                & 3D U-Net              & Private (CT)                                                        & 100                       & 13                        & 0.960                  & 0.965                   & 0.945        & 0.920       & 0.766                     & 0.793                        & 0.833                    \\
Kakeya {\it{et al.}} \cite{134} & 3D U-Net                & 3D U-Net              & Private (CT)                                                        & 47                        & 8                         & 0.971                  & 0.969                   & 0.984        & 0.975       & 0.861                     & 0.918                        & -                        \\ \bottomrule
\end{tabular}%
}
\end{table*}

\begin{table*}[ht!]
\caption{DSC-Based Summary of the Literature on Multi-Organ Coarse-to-Fine Segmentation Methods for the Thorax}
\label{tab_S_6}
\resizebox{\textwidth}{!}{%
\begin{tabular}{@{}lllllllllllll@{}}
\toprule
\multirow{2}{*}{Ref}    & \multirow{2}{*}{Coarse} & \multirow{2}{*}{Fine} & \multirow{2}{*}{Datasets} & \multirow{2}{*}{Quantity} & \multirow{2}{*}{Category} & \multirow{2}{*}{Heart} & \multirow{2}{*}{Esophagus} & \multirow{2}{*}{Trachea} & \multicolumn{2}{l}{Lung} & \multirow{2}{*}{Aorta} & \multirow{2}{*}{Spinal cord} \\ \cmidrule(lr){10-11}
                        &                         &                       &                           &                           &                           &                        &                            &                          & Left       & Right       &                        &                              \\ \midrule
Trullo {\it{et al.}} \cite{135} & 2D FCN                  & 2D FCN                & Private (CT)              & 30                        & 4                         & 0.900                  & 0.690                      & 0.870                    & -          & -           & 0.89                   & -                            \\
Cao {\it{et al.}} \cite{136}    & 2D U-Net                & 2D U-Net              & SegTHOR (CT) \cite{33}     & 50                        & 6                         & 0.945                  & 0.850                      & 0.807                    & 0.97       & 0.966       & -                      & 0.91                         \\
Zhang {\it{et al.}} \cite{129}  & 3D V-Net                & 3D V-Net              & SegTHOR (CT) \cite{33}     & 50                        & 4                         & 0.930                  & 0.785                      & 0.890                    & -          & -           & 0.916                  & -                            \\ \bottomrule
\end{tabular}%
}
\vspace{80pt}
\end{table*}

\begin{table*}[ht!]
\caption{DSC-BASED SUMMARY OF THE LITERATURE ON MULTI-ORGAN LOCALIZATION AND SEGMENTATION METHODS FOR THE HEAD AND NECK}
\label{tab_S_7}
\resizebox{\textwidth}{!}{%
\begin{tabular}{@{}lllllllllllllll@{}}
\toprule
\multirow{2}{*}{Ref}   & \multirow{2}{*}{Localization} & \multirow{2}{*}{Segmentation} & \multirow{2}{*}{Datasets}             & \multirow{2}{*}{Quantity} & \multirow{2}{*}{Category} & \multirow{2}{*}{Brainstem} & \multirow{2}{*}{Mandible} & \multicolumn{2}{l}{Parotid gland} & \multicolumn{2}{l}{Submandibular gland} & \multicolumn{2}{l}{Optic nerve} & \multirow{2}{*}{Chiasm} \\ \cmidrule(lr){9-14}
                       &                               &                               &                                       &                           &                           &                            &                           & Left            & Right           & Left               & Right              & Left           & Right          &                         \\ \midrule
Wang {\it{et al.}} \cite{137}   & 3D U-Net                      & 3D U-Net                      & HNC (CT) \cite{30}                     & 48                        & 9                         & 0.875                      & 0.930                     & 0.864           & 0.848           & 0.758              & 0.733              & 0.737          & 0.736          & 0.451                   \\
Men {\it{et al.}} \cite{138}    & 3D U-Net                      & 3D U-Net                      & TCIA (CT) \cite{56,57}          & 100                       & 7                         & 0.900                      & 0.920                     & 0.860           & 0.860           & -                  & -                  & -              & -              & -                       \\
Tang {\it{et al.}} \cite{139}   & 3D U-Net                      & 3D U-Net                      & Private (CT)                          & 215                       & 28                        & 0.863                      & 0.931                     & 0.849           & 0.849           & 0.807              & 0.825              & 0.757          & 0.761          & 0.642                   \\
Tang {\it{et al.}} \cite{139}   & 3D U-Net                      & 3D U-Net                      & PDDCA (CT) \cite{30}                   & 48                        & 9                         & 0.875                      & 0.95                      & 0.887           & 0.875           & 0.823              & 0.815              & 0.748          & 0.723          & 0.615                   \\
Yang {\it{et al.}} \cite{140}   & 3D CNN                        & 2D U-Net                      & Private (CT)                          & 88                        & 17                        & 0.831                      & 0.875                     & 0.807           & 0.811           & -                  & -                  & 0.638          & 0.675          & -                       \\
Liang {\it{et al.}} \cite{141}  & 2D CNN                        & 2D CNN                        & Private (CT)                          & 185                       & 18                        & 0.896                      & \begin{tabular}[c]{@{}l@{}}left: 0.914;\\ right: 0.912\end{tabular} & 0.852           & 0.85            & -                  & -                  & 0.661          & 0.717          & -                       \\
Gao {\it{et al.}} \cite{142}    & 3D CNN                        & 3D CNN                        & Private (CT)                          & 50                        & 18                        & 0.858                      & -                         & 0.772           & 0.800           & -                  & -                  & 0.639          & 0.617          & 0.638                   \\
Gao {\it{et al.}} \cite{142}   & 3D CNN                        & 3D CNN                        & HNC (CT) \cite{30}                     & 48                        & 9                         & 0.875                      & 0.935                     & 0.863           & 0.879           & 0.798              & 0.801              & 0.735          & 0.744          & 0.596                   \\
Liang {\it{et al.}} \cite{143}  & 2.5D CNN                      & 2.5D CNN                      & HNC (CT) \cite{30}                     & 48                        & 9                         & 0.923                      & 0.941                     & \multicolumn{2}{l}{0.876}         & \multicolumn{2}{l}{0.808}               & \multicolumn{2}{l}{0.736}       & 0.713                   \\
Liang {\it{et al.}} \cite{143} & 2.5D CNN                      & 2.5D CNN                      & Private (CT)                          & 96                        & 11                        & -                          & \begin{tabular}[c]{@{}l@{}}Left: 0.911;\\ right: 0.914\end{tabular} & 0.883           & 0.868           & -                  & -                  & 0.871          & 0.874          & -                       \\
Lei {\it{et al.}} \cite{144}   & 3D CNN                        & 3D U-Net                      & Private (CT)                          & 15                        & 8                         & -                          & 0.850                     & 0.820           & 0.810           & -                  & -                  & -              & -              & -                       \\
Huang {\it{et al.}} \cite{145}  & 3D CNN                        & 3D CNN                        & HNC (CT) \cite{30}                     & 48                        & 9                         & 0.879                      & 0.916                     & 0.884           & 0.878           & 0.801              & 0.776              & 0.677          & 0.706          & 0.643                   \\
Huang {\it{et al.}} \cite{145} & 3D CNN                        & 3D CNN                        & StructSeg (CT) \cite{1}                & 15                        & 7                         & 0.769                      & 0.807                     & 0.802           & 0.802           & -                  & -                  & 0.499          & 0.534          & 0.211                   \\
Huang {\it{et al.}} \cite{145} & 3D CNN                        & 3D CNN                        & Private (CT)                          & 15                        & 9                         & 0.957                      & 0.848                     & 0.962           & 0.946           & 0.846              & 0.808              & 0.824          & 0.843          & 0.434                   \\
Korte {\it{et al.}} \cite{146}  & 3D U-Net                      & 3D U-Net                      & 
\begin{tabular}[c]{@{}l@{}}Public RT-MAC dataset\\ (MRI) \cite{147}\end{tabular} & 31                        & 8                         & -                          & -                         & 0.860           & 0.857           & 0.830              & 0.785              & -              & -              & -                       \\
Korte {\it{et al.}} \cite{146} & 3D U-Net                      & 3D U-Net                      & Private (MRI)                         & 10                        & 8                         & -                          & -                         & 0.730           & 0.775           & 0.537              & 0.435              & -              & -              & -                       \\
Gao {\it{et al.}} \cite{108}    & 3D CNN                        & 3D CNN                        & Private (CT)                          & 1164                      & 22                        & 0.891                      & \begin{tabular}[c]{@{}l@{}}Left: 0.924;\\ right: 0.925\end{tabular} & 0.846           & 0.87            & -                  & -                  & 0.713          & 0.753          & 0.612                   \\
Gao {\it{et al.}} \cite{108}   & 3D CNN                        & 3D CNN                        & HNC (CT) \cite{30}                     & 48                        & 9                         & 0.882                      & 0.947                     & 0.898           & 0.881           & 0.840              & 0.838              & 0.790          & 0.817          & 0.713                   \\ \bottomrule
\end{tabular}%
}
\end{table*}

\begin{table*}[ht!]
\caption{DSC-Based Summary of the Literature on Multi-Organ Localization and Segmentation Methods for the Abdomen}
\label{tab_S_8}
\resizebox{\textwidth}{!}{%
%
%
}
\end{table*}

\begin{table*}[ht!]
\caption{DSC-Based Summary of the Literature on Multi-Organ Localization and Segmentation Methods for the Thorax}
\label{tab_S_9}
\resizebox{\textwidth}{!}{%
\begin{tabular}{@{}lllllllllllll@{}}
\toprule
\multirow{2}{*}{Ref}     & \multirow{2}{*}{Localization} & \multirow{2}{*}{Segmentation} & \multirow{2}{*}{Dataset} & \multirow{2}{*}{Quantity} & \multirow{2}{*}{Category} & \multirow{2}{*}{Heart} & \multirow{2}{*}{Esophagus} & \multirow{2}{*}{Trachea} & \multicolumn{2}{l}{Lung} & \multirow{2}{*}{Aorta} & \multirow{2}{*}{Spinal cord} \\ \cmidrule(lr){10-11}
                         &                               &                               &                          &                           &                           &                        &                            &                          & Left        & Right      &                        &                              \\ \midrule
Francis {\it{et al.}} \cite{136} & 3D U-Net                      & Two 3D U-Net                  & AAPM (CT) \cite{31}       & 60                        & 5                         & 0.941                  & 0.738                      & -                        & 0.979       & 0.973      & -                      & 0.899                        \\
Feng {\it{et al.}} \cite{151}    & 3D U-Net                      & 3D U-Net                      & AAPM (CT) \cite{31}       & 60                        & 5                         & 0.925                  & 0.726                      & -                        & 0.979       & 0.972      & -                      & 0.893                        \\
Feng {\it{et al.}} \cite{151}    & 3D U-Net                      & 3D U-Net                      & Private (CT)             & 30                        & 5                         & 0.86                   & 0.685                      & -                        & 0.976       & 0.977      & -                      & 0.852                        \\ \bottomrule
\end{tabular}%
}
\end{table*}

\begin{table*}[ht!]
\caption{DSC-Based Summary of the Literature on Multi-Organ Single-State Segmentation Methods for the Head and Neck-Supplementary Material}
\label{tab_S_10}
\resizebox{\textwidth}{!}{%
\begin{tabular}{@{}lllllm{0.6\textwidth}@{}}
\toprule
\multirow{2}{*}{Ref}        & \multirow{2}{*}{Backbone} & \multirow{2}{*}{Datasets} & \multirow{2}{*}{Quantity} & \multirow{2}{*}{Organ type} & \multirow{2}{*}{Other organs}                                                                                                                                                                                                                                                                                                                                                                                                                                                                                                             \\
                            &                           &                           &                           &                             &                                                                                                                                                                                                                                                                                                                                                                                                                                                                                                                                           \\
\bottomrule
\end{tabular}%
}
\end{table*}

\begin{table*}[ht!]
\caption{DSC-Based Summary of the Literature on Multi-Organ Single-State Segmentation Methods for the Abdomen-Supplementary Material}
\label{tab_S_11}
\resizebox{\textwidth}{!}{%
\begin{tabular}{@{}lllllm{0.5\textwidth}@{}}
\toprule
\multirow{2}{*}{Ref}        & \multirow{2}{*}{Network} & \multirow{2}{*}{Datasets}                    & \multirow{2}{*}{Quantity} & \multirow{2}{*}{Organ type} & \multirow{2}{*}{Other organs}                                                                                                                                                      \\
                            &                          &                                              &                           &                             &                                                                                                                                                                                    \\ \midrule
Gibson {\it{et al.}} \cite{55}     & 3D CNN                   & TCIA \cite{56,57} \&BTCV \cite{29} (CT) & 72                        & 4                           & Esophagus: 0.730                                                                                                                                                                   \\
Men {\it{et al.}} \cite{58}         & 2D CNN                   & Private (CT)                                 & 278                       & 5                           & Bladder: 0.934; Intestine: 0.653; Left femoral head: 0.921; Right femoral head: 0.923; Colon: 0.618                                                                                \\
Shen {\it{et al.}} \cite{59}        & 3D U-Net                 & Private (CT)                                 & 377                       & 7                           & Artery: 0.892; Vein: 0.793                                                                                                                                                         \\
\bottomrule
\end{tabular}%
}
\end{table*}

\begin{table*}[ht!]
\caption{DSC-Based Summary of the Literature on Multi-Organ Coarse-to-Fine Segmentation Methods for the Head and Neck-Supplementary Material}
\label{tab_S_12}
\resizebox{\textwidth}{!}{%
\begin{tabular}{@{}llllllm{0.6\textwidth}@{}}
\toprule
\multirow{2}{*}{Ref} & \multirow{2}{*}{Coarse} & \multirow{2}{*}{Fine} & \multirow{2}{*}{Datasets} & \multirow{2}{*}{Quantity} & \multirow{2}{*}{Organ type} & \multirow{2}{*}{Other organs}                                                                                                                         \\
                     &                         &                       &                           &                           &                             &                                                                                                                                                       \\ \midrule
Fang {\it{et al.}} \cite{109} & 2D FCN                  & 3D U-Net              & Private (CT)              & 56                        & 14                          & Right eyeball: 0.634; Left eyeball: 0.636; Lips: 0.676; Oral Cavity: 0.829; throat: 0.389; Esophagus: 0.735; Thyroid gland: 0.642; spinal cord: 0.782 \\ \bottomrule
\end{tabular}%
}
\end{table*}

\begin{table*}[ht!]
\caption{DSC-Based Summary of the Literature on Multi-Organ Coarse-to-Fine Segmentation Methods for the Abdomen}
\label{tab_S_13}
\resizebox{\textwidth}{!}{%
\begin{tabular}{@{}llllllm{0.6\textwidth}@{}}
\toprule
\multirow{2}{*}{Ref} & \multirow{2}{*}{Coarse network} & \multirow{2}{*}{Fine network} & \multirow{2}{*}{Dataset} & \multirow{2}{*}{Quantity} & \multirow{2}{*}{Category} & \multirow{2}{*}{Other organs}                                                                                                                                              \\
                     &                                 &                               &                          &                           &                           &                                                                                                                                                                            \\ \midrule
Roth \cite{144}              & 3D FCN                          & 3D FCN                        & Private (CT)             & 331                       & 3                         & Artery: 0.796; vein: 0.731                                                                                                                                                 \\
Wang \cite{145}              & 2.5D FCN                        & 2.5D FCN                      & Private (CT)             & 236                       & 13                        & Aorta: 0.918; colon: 0.830; duodenum: 0.754; Inferior vena cava: 0.870; small intestine: 0.801; vein: 0.807                                                                \\
Zhang \cite{128}             & 3D V-Net                        & 3D V-Net                      & BTCV (CT) \cite{29}              & 30                        & 13                        & Esophagus: 0.691; aorta: 0.877; Inferior vena cava: 0.865; Portal vein and splenic vein: 0.688; right adrenal gland: 0.651; left adrenal gland: 0.619                      \\
\bottomrule
\end{tabular}%
}
\vspace{100pt}
\end{table*}

\begin{table*}[ht!]
\caption{DSC-Based Summary of the Literature on Multi-Organ Localization and Segmentation Methods for the Head and Neck-Supplementary Material}
\label{tab_S_14}
\resizebox{\textwidth}{!}{%
%
 & 43                        & 8                         & Secondary lymph nodes (left): 0.708; secondary lymph nodes (right): 0.715; tertiary lymph nodes (left): 0.561; tertiary lymph nodes (right): 0.573;                                                                                                                                                                                                                                                                                        \\
                                        &                               &                               & Private (MRI)                         & 10                        & 8                         & Secondary lymph nodes (left): 0.553; Secondary lymph nodes (right): 0.525; Tertiary lymph nodes (left): 0.304; Tertiary lymph nodes (right): 0.189;                                                                                                                                                                                                                                                                                        \\
Gao {\it{et al.}} \cite{108}                    & 3D CNN                        & 3D CNN                        & Private (CT)                          & 1164                      & 22                        & Left eye: 0.897; right eye: 0.895; left lens: 0.819; right lens: 0.825; pituitary gland: 0.722; left temporal lobe: 0.877; right temporal lobe: 0.883; spinal cord: 0.831; left inner ear: 0.864; right inner ear: 0.855; left middle ear: 0.857; right middle ear: 0.843; left temporomandibular joint: 0.764; right temporomandibular joint: 0.789;                                                                                      \\ \bottomrule
\end{tabular}%
}
\end{table*}

\begin{table*}[ht!]
\caption{DSC-Based Summary of the Literature on Multi-Organ Localization and Segmentation Methods for the Abdomen-Supplementary Material}
\label{tab_S_15}
\resizebox{\textwidth}{!}{%
\begin{tabular}{@{}llllllm{0.5\textwidth}@{}}
\toprule
Ref                                    & Localization                  & Segmentation              & Dataset                                                    & Quantity & Category & Other organs                                                                                                                                          \\ \midrule
Larsson {\it{et al.}} \cite{148}               & Multi-Atlas                   & 3D FCN                    & BTCV (CT) \cite{29}                                         & 30       & 13       & Esophagus: 0.588; aorta: 0.870; Inferior vena cava: 0.758; Portal vein and splenic vein: 0.715; Right adrenal gland: 0.630; Left adrenal gland: 0.631 \\
\multirow{2}{*}{Zhao {\it{et al.}} \cite{149}} & \multirow{2}{*}{Registration} & \multirow{2}{*}{2D U-Net} & \begin{tabular}[c]{@{}l@{}}VISCERAL challenge dataset Nonenhanced\\ CT (CTwb) \cite{150}\end{tabular} & 20       & 4        & Left adrenal gland: 0.472; Right adrenal gland: 0.390                                                                                                 \\
                                       &                               &                           & \begin{tabular}[c]{@{}l@{}}VISCERAL challenge dataset enhanced\\ CT (CTce) \cite{150}\end{tabular}    & 20       & 4        & Left adrenal gland: 0.403; Right adrenal gland: 0.434                                                                                                 \\ \bottomrule
\end{tabular}%
}
\vspace{300pt}
\end{table*}

}



\begin{thebibliography}{100}
\providecommand{\url}[1]{#1}
\csname url@samestyle\endcsname
\providecommand{\newblock}{\relax}
\providecommand{\bibinfo}[2]{#2}
\providecommand{\BIBentrySTDinterwordspacing}{\spaceskip=0pt\relax}
\providecommand{\BIBentryALTinterwordstretchfactor}{4}
\providecommand{\BIBentryALTinterwordspacing}{\spaceskip=\fontdimen2\font plus
\BIBentryALTinterwordstretchfactor\fontdimen3\font minus
  \fontdimen4\font\relax}
\providecommand{\BIBforeignlanguage}[2]{{%
\expandafter\ifx\csname l@#1\endcsname\relax
\typeout{** WARNING: IEEEtran.bst: No hyphenation pattern has been}%
\typeout{** loaded for the language `#1'. Using the pattern for}%
\typeout{** the default language instead.}%
\else
\language=\csname l@#1\endcsname
\fi
#2}}
\providecommand{\BIBdecl}{\relax}
\BIBdecl

\bibitem{1}
B.~Van~Ginneken, C.~M. Schaefer-Prokop, and M.~Prokop, ``Computer-aided
  diagnosis: how to move from the laboratory to the clinic,'' \emph{Radiology},
  vol. 261, no.~3, pp. 719--732, 2011.

\bibitem{2}
J.~Sykes, ``Reflections on the current status of commercial automated
  segmentation systems in clinical practice,'' pp. 131--134, 2014.

\bibitem{3}
D.~G. Pfister, S.~Spencer, D.~Adelstein, D.~Adkins, Y.~Anzai, D.~M. Brizel,
  J.~Y. Bruce, P.~M. Busse, J.~J. Caudell, A.~J. Cmelak \emph{et~al.}, ``Head
  and neck cancers, version 2.2020, nccn clinical practice guidelines in
  oncology,'' \emph{Journal of the National Comprehensive Cancer Network},
  vol.~18, no.~7, pp. 873--898, 2020.

\bibitem{4}
J.~K. Molitoris, T.~Diwanji, J.~W. Snider~III, S.~Mossahebi, S.~Samanta, S.~N.
  Badiyan, C.~B. Simone, P.~Mohindra \emph{et~al.}, ``Advances in the use of
  motion management and image guidance in radiation therapy treatment for lung
  cancer,'' \emph{Journal of thoracic disease}, vol.~10, no. Suppl 21, pp.
  S2437--S2450, 2018.

\bibitem{5}
M.~A. Vyfhuis, N.~Onyeuku, T.~Diwanji, S.~Mossahebi, N.~P. Amin, S.~N. Badiyan,
  P.~Mohindra, and C.~B. Simone, ``Advances in proton therapy in lung cancer,''
  \emph{Therapeutic advances in respiratory disease}, vol.~12, p.
  1753466618783878, 2018.

\bibitem{6}
C.~W. Hurkmans, J.~H. Borger, B.~R. Pieters, N.~S. Russell, E.~P. Jansen, and
  B.~J. Mijnheer, ``Variability in target volume delineation on ct scans of the
  breast,'' \emph{International Journal of Radiation Oncology Biology Physics},
  vol.~50, no.~5, pp. 1366--1372, 2001.

\bibitem{7}
C.~Rasch, R.~Steenbakkers, and M.~van Herk, ``Target definition in prostate,
  head, and neck,'' in \emph{Seminars in radiation oncology}, vol.~15,
  no.~3.\hskip 1em plus 0.5em minus 0.4em\relax Elsevier, 2005, pp. 136--145.

\bibitem{8}
J.~Van~de Steene, N.~Linthout, J.~De~Mey, V.~Vinh-Hung, C.~Claassens,
  M.~Noppen, A.~Bel, and G.~Storme, ``Definition of gross tumor volume in lung
  cancer: inter-observer variability,'' \emph{Radiotherapy and oncology},
  vol.~62, no.~1, pp. 37--49, 2002.

\bibitem{9}
J.~Breunig, S.~Hernandez, J.~Lin, S.~Alsager, C.~Dumstorf, J.~Price, J.~Steber,
  R.~Garza, S.~Nagda, E.~Melian \emph{et~al.}, ``A system for continual quality
  improvement of normal tissue delineation for radiation therapy treatment
  planning,'' \emph{International Journal of Radiation Oncology Biology
  Physics}, vol.~83, no.~5, pp. e703--e708, 2012.

\bibitem{10}
X.~Chen and L.~Pan, ``A survey of graph cuts/graph search based medical image
  segmentation,'' \emph{IEEE reviews in biomedical engineering}, vol.~11, pp.
  112--124, 2018.

\bibitem{11}
I.~El~Naqa, D.~Yang, A.~Apte, D.~Khullar, S.~Mutic, J.~Zheng, J.~D. Bradley,
  P.~Grigsby, and J.~O. Deasy, ``Concurrent multimodality image segmentation by
  active contours for radiotherapy treatment planning a,'' \emph{Medical
  physics}, vol.~34, no.~12, pp. 4738--4749, 2007.

\bibitem{12}
A.~Pratondo, C.-K. Chui, and S.-H. Ong, ``Robust edge-stop functions for
  edge-based active contour models in medical image segmentation,'' \emph{IEEE
  Signal Processing Letters}, vol.~23, no.~2, pp. 222--226, 2015.

\bibitem{13}
A.~Tsai, A.~Yezzi, W.~Wells, C.~Tempany, D.~Tucker, A.~Fan, W.~E. Grimson, and
  A.~Willsky, ``A shape-based approach to the segmentation of medical imagery
  using level sets,'' \emph{IEEE transactions on medical imaging}, vol.~22,
  no.~2, pp. 137--154, 2003.

\bibitem{14}
A.~M. Saranathan and M.~Parente, ``Threshold based segmentation method for
  hyperspectral images,'' in \emph{2013 5Th workshop on hyperspectral image and
  signal processing: evolution in remote sensing (WHISPERS)}.\hskip 1em plus
  0.5em minus 0.4em\relax Gainesville: IEEE, 2013, pp. 1--4.

\bibitem{15}
J.~Shi and J.~Malik, ``Normalized cuts and image segmentation,'' \emph{IEEE
  Transactions on pattern analysis and machine intelligence}, vol.~22, no.~8,
  pp. 888--905, 2000.

\bibitem{16}
A.~J. Vyavahare and R.~Thool, ``Segmentation using region growing algorithm
  based on clahe for medical images,'' in \emph{Fourth International Conference
  on Advances in Recent Technologies in Communication and Computing
  (ARTCom2012)}.\hskip 1em plus 0.5em minus 0.4em\relax Bangalore, India: IET,
  2012, pp. 182--185.

\bibitem{17}
I.~Isgum, M.~Staring, A.~Rutten, M.~Prokop, M.~A. Viergever, and
  B.~Van~Ginneken, ``Multi-atlas-based segmentation with local decision
  fusion---application to cardiac and aortic segmentation in ct scans,''
  \emph{IEEE transactions on medical imaging}, vol.~28, no.~7, pp. 1000--1010,
  2009.

\bibitem{18}
P.~Aljabar, R.~A. Heckemann, A.~Hammers, J.~V. Hajnal, and D.~Rueckert,
  ``Multi-atlas based segmentation of brain images: atlas selection and its
  effect on accuracy,'' \emph{Neuroimage}, vol.~46, no.~3, pp. 726--738, 2009.

\bibitem{19}
O.~Ecabert, J.~Peters, H.~Schramm, C.~Lorenz, J.~von Berg, M.~J. Walker,
  M.~Vembar, M.~E. Olszewski, K.~Subramanyan, G.~Lavi \emph{et~al.},
  ``Automatic model-based segmentation of the heart in ct images,'' \emph{IEEE
  transactions on medical imaging}, vol.~27, no.~9, pp. 1189--1201, 2008.

\bibitem{20}
A.~A. Qazi, V.~Pekar, J.~Kim, J.~Xie, S.~L. Breen, and D.~A. Jaffray,
  ``Auto-segmentation of normal and target structures in head and neck ct
  images: a feature-driven model-based approach,'' \emph{Medical physics},
  vol.~38, no.~11, pp. 6160--6170, 2011.

\bibitem{21}
E.~A. Smirnov, D.~M. Timoshenko, and S.~N. Andrianov, ``Comparison of
  regularization methods for imagenet classification with deep convolutional
  neural networks,'' \emph{Aasri Procedia}, vol.~6, pp. 89--94, 2014.

\bibitem{22}
A.~Mobiny and H.~Van~Nguyen, ``Fast capsnet for lung cancer screening,'' in
  \emph{Medical Image Computing and Computer Assisted Intervention--MICCAI
  2018: 21st International Conference, Granada, Spain, September 16-20, 2018,
  Proceedings, Part II 11}.\hskip 1em plus 0.5em minus 0.4em\relax Cham:
  Springer, 2018, pp. 741--749.

\bibitem{23}
M.~Z. Alom, C.~Yakopcic, M.~Hasan, T.~M. Taha, and V.~K. Asari, ``Recurrent
  residual u-net for medical image segmentation,'' \emph{Journal of Medical
  Imaging}, vol.~6, no.~1, p. 014006, 2019.

\bibitem{24}
R.~Wang, T.~Lei, R.~Cui, B.~Zhang, H.~Meng, and A.~K. Nandi, ``Medical image
  segmentation using deep learning: A survey,'' \emph{IET Image Processing},
  vol.~16, no.~5, pp. 1243--1267, 2022.

\bibitem{25}
B.~Huang, F.~Yang, M.~Yin, X.~Mo, and C.~Zhong, ``A review of multimodal
  medical image fusion techniques,'' \emph{Computational and mathematical
  methods in medicine}, vol. 2020, p. 8279342, 2020.

\bibitem{26}
Y.~Fu, Y.~Lei, T.~Wang, W.~J. Curran, T.~Liu, and X.~Yang, ``Deep learning in
  medical image registration: a review,'' \emph{Physics in Medicine \&
  Biology}, vol.~65, no.~20, p. 20TR01, 2020.

\bibitem{27}
Y.~Lei, Y.~Fu, T.~Wang, R.~L. Qiu, W.~J. Curran, T.~Liu, and X.~Yang, ``Deep
  learning in multi-organ segmentation,'' \emph{arXiv preprint
  arXiv:2001.10619}, 2020.

\bibitem{28}
Y.~Fu, Y.~Lei, T.~Wang, W.~J. Curran, T.~Liu, and X.~Yang, ``A review of deep
  learning based methods for medical image multi-organ segmentation,''
  \emph{Physica Medica}, vol.~85, pp. 107--122, 2021.

\bibitem{29}
B.~Landman, Z.~Xu, J.~E. Igelsias, M.~Styner, T.~Langerak, and A.~Klein,
  ``Segmentation outside the cranial vault challenge,'' \emph{Synapse}, 2015.

\bibitem{30}
P.~F. Raudaschl, P.~Zaffino, G.~C. Sharp, M.~F. Spadea, A.~Chen, B.~M. Dawant,
  T.~Albrecht, T.~Gass, C.~Langguth, M.~L{\"u}thi \emph{et~al.}, ``Evaluation
  of segmentation methods on head and neck ct: auto-segmentation challenge
  2015,'' \emph{Medical physics}, vol.~44, no.~5, pp. 2020--2036, 2017.

\bibitem{31}
J.~Yang, H.~Veeraraghavan, S.~G. Armato~III, K.~Farahani, J.~S. Kirby,
  J.~Kalpathy-Kramer, W.~van Elmpt, A.~Dekker, X.~Han, X.~Feng \emph{et~al.},
  ``Autosegmentation for thoracic radiation treatment planning: a grand
  challenge at aapm 2017,'' \emph{Medical physics}, vol.~45, no.~10, pp.
  4568--4581, 2018.

\bibitem{32}
A.~E. Kavur, N.~S. Gezer, M.~Bar{\i}{\c{s}}, S.~Aslan, P.-H. Conze, V.~Groza,
  D.~D. Pham, S.~Chatterjee, P.~Ernst, S.~{\"O}zkan \emph{et~al.}, ``Chaos
  challenge-combined (ct-mr) healthy abdominal organ segmentation,''
  \emph{Medical Image Analysis}, vol.~69, p. 101950, 2021.

\bibitem{33}
X.~Feng, K.~Qing, N.~J. Tustison, C.~H. Meyer, and Q.~Chen, ``Deep
  convolutional neural network for segmentation of thoracic organs-at-risk
  using cropped 3d images,'' \emph{Medical physics}, vol.~46, no.~5, pp.
  2169--2180, 2019.

\bibitem{34}
X.~Xu, F.~Zhou, B.~Liu, and X.~Bai, ``Annotations for body organ localization
  based on miccai lits dataset,'' \emph{IEEE Dataport}, 2018.

\bibitem{35}
A.~Babier, B.~Zhang, R.~Mahmood, K.~L. Moore, T.~G. Purdie, A.~L. McNiven, and
  T.~C. Chan, ``Openkbp: the open-access knowledge-based planning grand
  challenge and dataset,'' \emph{Medical Physics}, vol.~48, no.~9, pp.
  5549--5561, 2021.

\bibitem{36}
J.~Ma, Y.~Zhang, S.~Gu, C.~Zhu, C.~Ge, Y.~Zhang, X.~An, C.~Wang, Q.~Wang,
  X.~Liu \emph{et~al.}, ``Abdomenct-1k: Is abdominal organ segmentation a
  solved problem?'' \emph{IEEE Transactions on Pattern Analysis and Machine
  Intelligence}, vol.~44, no.~10, pp. 6695--6714, 2021.

\bibitem{92}
Y.~LeCun, B.~Boser, J.~S. Denker, D.~Henderson, R.~E. Howard, W.~Hubbard, and
  L.~D. Jackel, ``Backpropagation applied to handwritten zip code
  recognition,'' \emph{Neural computation}, vol.~1, no.~4, pp. 541--551, 1989.

\bibitem{93}
R.~Karthik, R.~Menaka, A.~Johnson, and S.~Anand, ``Neuroimaging and deep
  learning for brain stroke detection-a review of recent advancements and
  future prospects,'' \emph{Computer Methods and Programs in Biomedicine}, vol.
  197, p. 105728, 2020.

\bibitem{94}
X.~Zhao, K.~Chen, G.~Wu, G.~Zhang, X.~Zhou, C.~Lv, S.~Wu, Y.~Chen, G.~Xie, and
  Z.~Yao, ``Deep learning shows good reliability for automatic segmentation and
  volume measurement of brain hemorrhage, intraventricular extension, and
  peripheral edema,'' \emph{European radiology}, vol.~31, no.~7, pp.
  5012--5020, 2021.

\bibitem{55}
E.~Gibson, F.~Giganti, Y.~Hu, E.~Bonmati, S.~Bandula, K.~Gurusamy, B.~R.
  Davidson, S.~P. Pereira, M.~J. Clarkson, and D.~C. Barratt, ``Towards
  image-guided pancreas and biliary endoscopy: automatic multi-organ
  segmentation on abdominal ct with dense dilated networks,'' in \emph{Medical
  Image Computing and Computer Assisted Intervention- MICCAI 2017: 20th
  International Conference, Quebec City, QC, Canada, September 11-13, 2017,
  Proceedings, Part I 20}.\hskip 1em plus 0.5em minus 0.4em\relax Cham,
  Switzerland: Springer, 2017, pp. 728--736.

\bibitem{37}
B.~Ibragimov and L.~Xing, ``Segmentation of organs-at-risks in head and neck ct
  images using convolutional neural networks,'' \emph{Medical physics},
  vol.~44, no.~2, pp. 547--557, 2017.

\bibitem{38}
K.~Fritscher, P.~Raudaschl, P.~Zaffino, M.~F. Spadea, G.~C. Sharp, and
  R.~Schubert, ``Deep neural networks for fast segmentation of 3d medical
  images,'' in \emph{Medical Image Computing and Computer-Assisted
  Intervention--MICCAI 2016: 19th International Conference, Athens, Greece,
  October 17-21, 2016, Proceedings, Part II 19}.\hskip 1em plus 0.5em minus
  0.4em\relax Cham, Switzerland: Springer, 2016, pp. 158--165.

\bibitem{95}
P.~Moeskops, J.~M. Wolterink, B.~H. Van Der~Velden, K.~G. Gilhuijs, T.~Leiner,
  M.~A. Viergever, and I.~I{\v{s}}gum, ``Deep learning for multi-task medical
  image segmentation in multiple modalities,'' in \emph{Medical Image Computing
  and Computer-Assisted Intervention--MICCAI 2016: 19th International
  Conference, Athens, Greece, October 17-21, 2016, Proceedings, Part II
  19}.\hskip 1em plus 0.5em minus 0.4em\relax Cham, Switzerland: Springer,
  2016, pp. 478--486.

\bibitem{96}
Long, Jonathan, Shelhamer, Evan, Darrell, and Trevor, ``Fully convolutional
  networks for semantic segmentation,'' \emph{IEEE Transactions on Pattern
  Analysis \& Machine Intelligence}, vol.~39, no.~4, pp. 640--651, 2017.

\bibitem{97}
Y.~Wang, Y.~Zhou, P.~Tang, W.~Shen, E.~K. Fishman, and A.~L. Yuille, ``Training
  multi-organ segmentation networks with sample selection by relaxed upper
  confident bound,'' in \emph{Medical Image Computing and Computer Assisted
  Intervention--MICCAI 2018: 21st International Conference, Granada, Spain,
  September 16-20, 2018, Proceedings, Part IV 11}.\hskip 1em plus 0.5em minus
  0.4em\relax Cham, Switzerland: Springer, 2018, pp. 434--442.

\bibitem{98}
P.~O. Pinheiro, T.-Y. Lin, R.~Collobert, and P.~Doll{\'a}r, ``Learning to
  refine object segments,'' in \emph{Computer Vision--ECCV 2016: 14th European
  Conference, Amsterdam, The Netherlands, October 11--14, 2016, Proceedings,
  Part I 14}.\hskip 1em plus 0.5em minus 0.4em\relax Cham, Switzerland:
  Springer, 2016, pp. 75--91.

\bibitem{99}
O.~Ronneberger, P.~Fischer, and T.~Brox, ``U-net: Convolutional networks for
  biomedical image segmentation,'' in \emph{Medical Image Computing and
  Computer-Assisted Intervention--MICCAI 2015: 18th International Conference,
  Munich, Germany, October 5-9, 2015, Proceedings, Part III 18}.\hskip 1em plus
  0.5em minus 0.4em\relax Cham, Switzerland: Springer, 2015, pp. 234--241.

\bibitem{39}
W.~Zhu, Y.~Huang, L.~Zeng, X.~Chen, Y.~Liu, Z.~Qian, N.~Du, W.~Fan, and X.~Xie,
  ``Anatomynet: deep learning for fast and fully automated whole-volume
  segmentation of head and neck anatomy,'' \emph{Medical physics}, vol.~46,
  no.~2, pp. 576--589, 2019.

\bibitem{40}
W.~van Rooij, M.~Dahele, H.~R. Brandao, A.~R. Delaney, B.~J. Slotman, and W.~F.
  Verbakel, ``Deep learning-based delineation of head and neck organs at risk:
  geometric and dosimetric evaluation,'' \emph{International Journal of
  Radiation Oncology Biology Physics}, vol. 104, no.~3, pp. 677--684, 2019.

\bibitem{42}
S.~Gou, N.~Tong, S.~Qi, S.~Yang, R.~Chin, and K.~Sheng,
  ``Self-channel-and-spatial-attention neural network for automated multi-organ
  segmentation on head and neck ct images,'' \emph{Physics in Medicine \&
  Biology}, vol.~65, no.~24, p. 245034, 2020.

\bibitem{48}
Z.~Zhang, T.~Zhao, H.~Gay, W.~Zhang, and B.~Sun, ``Weaving attention u-net: A
  novel hybrid cnn and attention-based method for organs-at-risk segmentation
  in head and neck ct images,'' \emph{Medical physics}, vol.~48, no.~11, pp.
  7052--7062, 2021.

\bibitem{61}
H.~R. Roth, C.~Shen, H.~Oda, M.~Oda, Y.~Hayashi, K.~Misawa, and K.~Mori, ``Deep
  learning and its application to medical image segmentation,'' \emph{Medical
  Imaging Technology}, vol.~36, no.~2, pp. 63--71, 2018.

\bibitem{69}
C.~Jia and J.~Wei, ``Amo-net: abdominal multi-organ segmentation in mri with a
  extend unet,'' in \emph{2021 IEEE 4th Advanced Information Management,
  Communicates, Electronic and Automation Control Conference (IMCEC)},
  vol.~4.\hskip 1em plus 0.5em minus 0.4em\relax Chongqing, China: IEEE, 2021,
  pp. 1770--1775.

\bibitem{78}
R.~Berzoini, A.~A. Colombo, S.~Bardini, A.~Conelli, E.~D'Arnese, and M.~D.
  Santambrogio, ``An optimized u-net for unbalanced multi-organ segmentation,''
  in \emph{2022 44th Annual International Conference of the IEEE Engineering in
  Medicine \& Biology Society (EMBC)}.\hskip 1em plus 0.5em minus 0.4em\relax
  Glasgow, Scotland: IEEE, 2022, pp. 3764--3767.

\bibitem{86}
Z.~Lambert, C.~Petitjean, B.~Dubray, and S.~Kuan, ``Segthor: Segmentation of
  thoracic organs at risk in ct images,'' in \emph{2020 Tenth International
  Conference on Image Processing Theory, Tools and Applications (IPTA)}.\hskip
  1em plus 0.5em minus 0.4em\relax Paris, France: IEEE, 2020, pp. 1--6.

\bibitem{100}
F.~Milletari, N.~Navab, and S.-A. Ahmadi, ``V-net: Fully convolutional neural
  networks for volumetric medical image segmentation,'' in \emph{2016 fourth
  international conference on 3D vision (3DV)}.\hskip 1em plus 0.5em minus
  0.4em\relax Stanford, CA: IEEE, 2016, pp. 565--571.

\bibitem{60}
E.~Gibson, F.~Giganti, Y.~Hu, E.~Bonmati, S.~Bandula, K.~Gurusamy, B.~Davidson,
  S.~P. Pereira, M.~J. Clarkson, and D.~C. Barratt, ``Automatic multi-organ
  segmentation on abdominal ct with dense v-networks,'' \emph{IEEE transactions
  on medical imaging}, vol.~37, no.~8, pp. 1822--1834, 2018.

\bibitem{77}
M.~Xu, H.~Guo, J.~Zhang, K.~Yan, and L.~Lu, ``A new probabilistic v-net model
  with hierarchical spatial feature transform for efficient abdominal
  multi-organ segmentation,'' \emph{arXiv preprint arXiv:2208.01382}, 2022.

\bibitem{50}
G.~Podobnik, B.~Ibragimov, P.~Strojan, P.~Peterlin, and T.~Vrtovec,
  ``Segmentation of organs-at-risk from ct and mr images of the head and neck:
  Baseline results,'' in \emph{2022 IEEE 19th International Symposium on
  Biomedical Imaging (ISBI)}.\hskip 1em plus 0.5em minus 0.4em\relax Kolkata,
  India: IEEE, 2022, pp. 1--4.

\bibitem{102}
F.~Isensee, P.~F. J{\"a}ger, P.~M. Full, P.~Vollmuth, and K.~H. Maier-Hein,
  ``nnu-net for brain tumor segmentation,'' in \emph{Brainlesion: Glioma,
  Multiple Sclerosis, Stroke and Traumatic Brain Injuries: 6th International
  Workshop, BrainLes 2020, Held in Conjunction with MICCAI 2020, Lima, Peru,
  October 4, 2020, Revised Selected Papers, Part II 6}.\hskip 1em plus 0.5em
  minus 0.4em\relax Cham, Switzerland: Springer, 2021, pp. 118--132.

\bibitem{103}
G.~Zhang, Z.~Yang, B.~Huo, S.~Chai, and S.~Jiang, ``Multiorgan segmentation
  from partially labeled datasets with conditional nnu-net,'' \emph{Computers
  in Biology and Medicine}, vol. 136, p. 104658, 2021.

\bibitem{104}
N.~Altini, A.~Brunetti, V.~P. Napoletano, F.~Girardi, E.~Allegretti, S.~M.
  Hussain, G.~Brunetti, V.~Triggiani, V.~Bevilacqua, and D.~Buongiorno, ``A
  fusion biopsy framework for prostate cancer based on deformable superellipses
  and nnu-net,'' \emph{Bioengineering}, vol.~9, no.~8, p. 343, 2022.

\bibitem{105}
I.~Goodfellow, J.~Pouget-Abadie, M.~Mirza, B.~Xu, D.~Warde-Farley, S.~Ozair,
  A.~Courville, and Y.~Bengio, ``Generative adversarial networks,'' \emph{arXiv
  preprint arXiv:1406.2661}, 2014.

\bibitem{41}
N.~Tong, S.~Gou, S.~Yang, M.~Cao, and K.~Sheng, ``Shape constrained fully
  convolutional densenet with adversarial training for multiorgan segmentation
  on head and neck ct and low-field mr images,'' \emph{Medical physics},
  vol.~46, no.~6, pp. 2669--2682, 2019.

\bibitem{62}
J.~Cai, Y.~Xia, D.~Yang, D.~Xu, L.~Yang, and H.~Roth, ``End-to-end adversarial
  shape learning for abdomen organ deep segmentation,'' in \emph{Machine
  Learning in Medical Imaging: 10th International Workshop, MLMI 2019, Held in
  Conjunction with MICCAI 2019, Shenzhen, China, October 13, 2019, Proceedings
  10}.\hskip 1em plus 0.5em minus 0.4em\relax Cham, Switzerland: Springer,
  2019, pp. 124--132.

\bibitem{84}
X.~Dong, Y.~Lei, T.~Wang, M.~Thomas, L.~Tang, W.~J. Curran, T.~Liu, and
  X.~Yang, ``Automatic multiorgan segmentation in thorax ct images using
  u-net-gan,'' \emph{Medical physics}, vol.~46, no.~5, pp. 2157--2168, 2019.

\bibitem{106}
R.~Trullo, C.~Petitjean, B.~Dubray, and S.~Ruan, ``Multiorgan segmentation
  using distance-aware adversarial networks,'' \emph{Journal of Medical
  Imaging}, vol.~6, no.~1, p. 014001, 2019.

\bibitem{107}
F.~Mahmood, D.~Borders, R.~J. Chen, G.~N. McKay, K.~J. Salimian, A.~Baras, and
  N.~J. Durr, ``Deep adversarial training for multi-organ nuclei segmentation
  in histopathology images,'' \emph{IEEE transactions on medical imaging},
  vol.~39, no.~11, pp. 3257--3267, 2019.

\bibitem{108}
Y.~Gao, R.~Huang, Y.~Yang, J.~Zhang, K.~Shao, C.~Tao, Y.~Chen, D.~N. Metaxas,
  H.~Li, and M.~Chen, ``Focusnetv2: Imbalanced large and small organ
  segmentation with adversarial shape constraint for head and neck ct images,''
  \emph{Medical Image Analysis}, vol.~67, p. 101831, 2021.

\bibitem{109}
H.~Fang, Y.~Fang, and X.~Yang, ``Multi-organ segmentation network with
  adversarial performance validator,'' \emph{arXiv preprint arXiv:2204.07850},
  2022.

\bibitem{110}
A.~Vaswani, N.~Shazeer, and N.~Parmar, ``Attention is all uou need,''
  \emph{arXiv:170603762}, 2021.

\bibitem{111}
I.~Bello, ``Lambdanetworks: Modeling long-range interactions without
  attention,'' \emph{arXiv preprint arXiv:2102.08602}, 2021.

\bibitem{112}
Y.~Gao, M.~Zhou, and D.~N. Metaxas, ``Utnet: a hybrid transformer architecture
  for medical image segmentation,'' in \emph{Medical Image Computing and
  Computer Assisted Intervention--MICCAI 2021: 24th International Conference,
  Strasbourg, France, September 27--October 1, 2021, Proceedings, Part III
  24}.\hskip 1em plus 0.5em minus 0.4em\relax Cham, Switzerland: Springer,
  2021, pp. 61--71.

\bibitem{113}
C.~Yao, M.~Hu, G.~Zhai, and X.~Zhang, ``Transclaw u-net: Claw u-net with
  transformers for medical image segmentation,'' \emph{arXiv preprint
  arXiv:2107.05188}, 2021.

\bibitem{114}
J.~M.~J. Valanarasu, P.~Oza, I.~Hacihaliloglu, and V.~M. Patel, ``Medical
  transformer: Gated axial-attention for medical image segmentation,'' in
  \emph{Medical Image Computing and Computer Assisted Intervention--MICCAI
  2021: 24th International Conference, Strasbourg, France, September
  27--October 1, 2021, Proceedings, Part I 24}.\hskip 1em plus 0.5em minus
  0.4em\relax Springer, 2021, pp. 36--46.

\bibitem{115}
S.~Pan, Y.~Lei, T.~Wang, J.~Wynne, C.-W. Chang, J.~Roper, A.~B. Jani, P.~Patel,
  J.~D. Bradley, T.~Liu \emph{et~al.}, ``Male pelvic multi-organ segmentation
  using token-based transformer vnet,'' \emph{Physics in Medicine \& Biology},
  vol.~67, no.~20, p. 205012, 2022.

\bibitem{71}
H.~Cao, Y.~Wang, J.~Chen, D.~Jiang, X.~Zhang, Q.~Tian, and M.~Wang,
  ``Swin-unet: Unet-like pure transformer for medical image segmentation,''
  \emph{arXiv preprint arXiv:2105.05537}, 2021.

\bibitem{75}
X.~Huang, Z.~Deng, D.~Li, and X.~Yuan, ``Missformer: An effective medical image
  segmentation transformer,'' \emph{arXiv preprint arXiv:2109.07162}, 2021.

\bibitem{82}
Y.~Xie, J.~Zhang, C.~Shen, and Y.~Xia, ``Cotr: Efficiently bridging cnn and
  transformer for 3d medical image segmentation,'' in \emph{Medical Image
  Computing and Computer Assisted Intervention--MICCAI 2021: 24th International
  Conference, Strasbourg, France, September 27--October 1, 2021, Proceedings,
  Part III 24}.\hskip 1em plus 0.5em minus 0.4em\relax Cham, Switzerland:
  Springer, 2021, pp. 171--180.

\bibitem{116}
H.~Wang, P.~Cao, J.~Wang, and O.~R. Zaiane, ``Uctransnet: rethinking the skip
  connections in u-net from a channel-wise perspective with transformer,'' in
  \emph{Proceedings of the AAAI conference on artificial intelligence},
  vol.~36, no.~3, 2022, pp. 2441--2449.

\bibitem{117}
H.~Wang, S.~Xie, L.~Lin, Y.~Iwamoto, X.-H. Han, Y.-W. Chen, and R.~Tong,
  ``Mixed transformer u-net for medical image segmentation,'' in \emph{ICASSP
  2022-2022 IEEE International Conference on Acoustics, Speech and Signal
  Processing (ICASSP)}.\hskip 1em plus 0.5em minus 0.4em\relax Singapore: IEEE,
  2022, pp. 2390--2394.

\bibitem{118}
G.~Xu, X.~Wu, X.~Zhang, and X.~He, ``Levit-unet: Make faster encoders with
  transformer for medical image segmentation,'' \emph{arXiv preprint
  arXiv:2107.08623}, 2021.

\bibitem{119}
Y.~Zhang, H.~Liu, and Q.~Hu, ``Transfuse: Fusing transformers and cnns for
  medical image segmentation,'' in \emph{Medical Image Computing and Computer
  Assisted Intervention--MICCAI 2021: 24th International Conference,
  Strasbourg, France, September 27--October 1, 2021, Proceedings, Part I
  24}.\hskip 1em plus 0.5em minus 0.4em\relax Cham, Switzerland: Springer,
  2021, pp. 14--24.

\bibitem{76}
C.~Suo, X.~Li, D.~Tan, Y.~Zhang, and X.~Gao, ``I2-net: Intra-and inter-scale
  collaborative learning network for abdominal multi-organ segmentation,'' in
  \emph{Proceedings of the 2022 International Conference on Multimedia
  Retrieval}, New York, NY, 2022, pp. 654--660.

\bibitem{51}
H.~Kan, J.~Shi, M.~Zhao, Z.~Wang, W.~Han, H.~An, Z.~Wang, and S.~Wang,
  ``Itunet: Integration of transformers and unet for organs-at-risk
  segmentation,'' in \emph{2022 44th Annual International Conference of the
  IEEE Engineering in Medicine \& Biology Society (EMBC)}.\hskip 1em plus 0.5em
  minus 0.4em\relax IEEE, 2022, pp. 2123--2127.

\bibitem{72}
J.~Chen, Y.~Lu, Q.~Yu, X.~Luo, E.~Adeli, Y.~Wang, L.~Lu, A.~L. Yuille, and
  Y.~Zhou, ``Transunet: Transformers make strong encoders for medical image
  segmentation,'' \emph{arXiv preprint arXiv:2102.04306}, 2021.

\bibitem{66}
A.~Hatamizadeh, Y.~Tang, V.~Nath, D.~Yang, A.~Myronenko, B.~Landman, H.~R.
  Roth, and D.~Xu, ``Unetr: Transformers for 3d medical image segmentation,''
  in \emph{Proceedings of the IEEE/CVF winter conference on applications of
  computer vision}, Waikoloa, HI, 2022, pp. 574--584.

\bibitem{120}
P.-H. Chen, C.-H. Huang, S.-K. Hung, L.-C. Chen, H.-L. Hsieh, W.-Y. Chiou,
  M.-S. Lee, H.-Y. Lin, and W.-M. Liu, ``Attention-lstm fused u-net
  architecture for organ segmentation in ct images,'' in \emph{2020
  International Symposium on Computer, Consumer and Control (IS3C)}.\hskip 1em
  plus 0.5em minus 0.4em\relax Taichung City, Taiwan: IEEE, 2020, pp. 304--307.

\bibitem{121}
A.~Chakravarty and J.~Sivaswamy, ``Race-net: a recurrent neural network for
  biomedical image segmentation,'' \emph{IEEE journal of biomedical and health
  informatics}, vol.~23, no.~3, pp. 1151--1162, 2018.

\bibitem{123}
E.~Tappeiner, S.~Pr{\"o}ll, M.~H{\"o}nig, P.~F. Raudaschl, P.~Zaffino, M.~F.
  Spadea, G.~C. Sharp, R.~Schubert, and K.~Fritscher, ``Multi-organ
  segmentation of the head and neck area: an efficient hierarchical neural
  networks approach,'' \emph{International journal of computer assisted
  radiology and surgery}, vol.~14, no.~5, pp. 745--754, 2019.

\bibitem{124}
Y.~Pu, S.-I. Kamata, and Y.~Wang, ``A coarse to fine framework for multi-organ
  segmentation in head and neck images,'' in \emph{2020 Joint 9th International
  Conference on Informatics, Electronics \& Vision (ICIEV) and 2020 4th
  International Conference on Imaging, Vision \& Pattern Recognition
  (icIVPR)}.\hskip 1em plus 0.5em minus 0.4em\relax Kitakyushu, Japan: IEEE,
  2020, pp. 1--6.

\bibitem{125}
Q.~Ma, C.~Zu, X.~Wu, J.~Zhou, and Y.~Wang, ``Coarse-to-fine segmentation of
  organs at risk in nasopharyngeal carcinoma radiotherapy,'' in \emph{Medical
  Image Computing and Computer Assisted Intervention--MICCAI 2021: 24th
  International Conference, Strasbourg, France, September 27--October 1, 2021,
  Proceedings, Part I 24}.\hskip 1em plus 0.5em minus 0.4em\relax Cham,
  Switzerland: Springer, 2021, pp. 358--368.

\bibitem{126}
P.~Hu, F.~Wu, J.~Peng, Y.~Bao, F.~Chen, and D.~Kong, ``Automatic abdominal
  multi-organ segmentation using deep convolutional neural network and
  time-implicit level sets,'' \emph{International journal of computer assisted
  radiology and surgery}, vol.~12, no.~3, pp. 399--411, 2017.

\bibitem{129}
L.~Zhang, J.~Zhang, P.~Shen, G.~Zhu, P.~Li, X.~Lu, H.~Zhang, S.~A. Shah, and
  M.~Bennamoun, ``Block level skip connections across cascaded v-net for
  multi-organ segmentation,'' \emph{IEEE Transactions on Medical Imaging},
  vol.~39, no.~9, pp. 2782--2793, 2020.

\bibitem{130}
L.~Xie, Q.~Yu, Y.~Zhou, Y.~Wang, E.~K. Fishman, and A.~L. Yuille, ``Recurrent
  saliency transformation network for tiny target segmentation in abdominal ct
  scans,'' \emph{IEEE transactions on medical imaging}, vol.~39, no.~2, pp.
  514--525, 2019.

\bibitem{133}
H.~H. Lee, Y.~Tang, S.~Bao, R.~G. Abramson, Y.~Huo, and B.~A. Landman,
  ``Rap-net: Coarse-to-fine multi-organ segmentation with single random
  anatomical prior,'' in \emph{2021 IEEE 18th International Symposium on
  Biomedical Imaging (ISBI)}.\hskip 1em plus 0.5em minus 0.4em\relax Nice,
  France: IEEE, 2021, pp. 1491--1494.

\bibitem{152}
P.~F. Christ, M.~E.~A. Elshaer, F.~Ettlinger, S.~Tatavarty, M.~Bickel,
  P.~Bilic, M.~Rempfler, M.~Armbruster, F.~Hofmann, M.~D'Anastasi
  \emph{et~al.}, ``Automatic liver and lesion segmentation in ct using cascaded
  fully convolutional neural networks and 3d conditional random fields,'' in
  \emph{International conference on medical image computing and
  computer-assisted intervention}.\hskip 1em plus 0.5em minus 0.4em\relax Cham,
  Switzerland: Springer, 2016, pp. 415--423.

\bibitem{153}
D.~Lachinov, E.~Vasiliev, and V.~Turlapov, ``Glioma segmentation with cascaded
  unet,'' in \emph{Brainlesion: Glioma, Multiple Sclerosis, Stroke and
  Traumatic Brain Injuries: 4th International Workshop, BrainLes 2018, Held in
  Conjunction with MICCAI 2018, Granada, Spain, September 16, 2018, Revised
  Selected Papers, Part II 4}.\hskip 1em plus 0.5em minus 0.4em\relax Cham,
  Switzerland: Springer, 2019, pp. 189--198.

\bibitem{154}
S.~Li, Y.~Chen, S.~Yang, and W.~Luo, ``Cascade dense-unet for prostate
  segmentation in mr images,'' in \emph{Intelligent Computing Theories and
  Application: 15th International Conference, ICIC 2019, Nanchang, China,
  August 3--6, 2019, Proceedings, Part I 15}.\hskip 1em plus 0.5em minus
  0.4em\relax Cham, Switzerland: Springer, 2019, pp. 481--490.

\bibitem{83}
R.~Trullo, C.~Petitjean, S.~Ruan, B.~Dubray, D.~Nie, and D.~Shen,
  ``Segmentation of organs at risk in thoracic ct images using a sharpmask
  architecture and conditional random fields,'' in \emph{2017 IEEE 14th
  international symposium on biomedical imaging (ISBI 2017)}.\hskip 1em plus
  0.5em minus 0.4em\relax Melbourne, Australia: IEEE, 2017, pp. 1003--1006.

\bibitem{137}
Y.~Wang, L.~Zhao, M.~Wang, and Z.~Song, ``Organ at risk segmentation in head
  and neck ct images using a two-stage segmentation framework based on 3d
  u-net,'' \emph{IEEE Access}, vol.~7, pp. 144\,591--144\,602, 2019.

\bibitem{138}
K.~Men, H.~Geng, C.~Cheng, H.~Zhong, M.~Huang, Y.~Fan, J.~P. Plastaras, A.~Lin,
  and Y.~Xiao, ``More accurate and efficient segmentation of organs-at-risk in
  radiotherapy with convolutional neural networks cascades,'' \emph{Medical
  physics}, vol.~46, no.~1, pp. 286--292, 2019.

\bibitem{144}
Y.~Lei, J.~Zhou, X.~Dong, T.~Wang, H.~Mao, M.~McDonald, W.~J. Curran, T.~Liu,
  and X.~Yang, ``Multi-organ segmentation in head and neck mri using
  u-faster-rcnn,'' in \emph{Medical Imaging 2020: Image Processing}, vol.
  113133A.\hskip 1em plus 0.5em minus 0.4em\relax Houston, TX: SPIE, 2020, pp.
  826--831.

\bibitem{151}
S.~Francis, P.~Jayaraj, P.~Pournami, M.~Thomas, A.~T. Jose, A.~J. Binu, and
  N.~Puzhakkal, ``Thoraxnet: a 3d u-net based two-stage framework for oar
  segmentation on thoracic ct images,'' \emph{Physical and Engineering Sciences
  in Medicine}, vol.~45, no.~1, pp. 189--203, 2022.

\bibitem{139}
H.~Tang, X.~Chen, Y.~Liu, Z.~Lu, J.~You, M.~Yang, S.~Yao, G.~Zhao, Y.~Xu,
  T.~Chen \emph{et~al.}, ``Clinically applicable deep learning framework for
  organs at risk delineation in ct images,'' \emph{Nature Machine
  Intelligence}, vol.~1, no.~10, pp. 480--491, 2019.

\bibitem{146}
J.~C. Korte, N.~Hardcastle, S.~P. Ng, B.~Clark, T.~Kron, and P.~Jackson,
  ``Cascaded deep learning-based auto-segmentation for head and neck cancer
  patients: Organs at risk on t2-weighted magnetic resonance imaging,''
  \emph{Medical physics}, vol.~48, no.~12, pp. 7757--7772, 2021.

\bibitem{142}
Y.~Gao, R.~Huang, M.~Chen, Z.~Wang, J.~Deng, Y.~Chen, Y.~Yang, J.~Zhang,
  C.~Tao, and H.~Li, ``Focusnet: Imbalanced large and small organ segmentation
  with an end-to-end deep neural network for head and neck ct images,'' in
  \emph{Medical Image Computing and Computer Assisted Intervention--MICCAI
  2019: 22nd International Conference, Shenzhen, China, October 13--17, 2019,
  Proceedings, Part III 22}.\hskip 1em plus 0.5em minus 0.4em\relax Cham,
  Switzerland: Springer, 2019, pp. 829--838.

\bibitem{148}
M.~Larsson, Y.~Zhang, and F.~Kahl, ``Robust abdominal organ segmentation using
  regional convolutional neural networks,'' \emph{Applied Soft Computing},
  vol.~70, pp. 465--471, 2018.

\bibitem{149}
Y.~Zhao, H.~Li, S.~Wan, A.~Sekuboyina, X.~Hu, G.~Tetteh, M.~Piraud, and
  B.~Menze, ``Knowledge-aided convolutional neural network for small organ
  segmentation,'' \emph{IEEE journal of biomedical and health informatics},
  vol.~23, no.~4, pp. 1363--1373, 2019.

\bibitem{122}
X.~Ren, L.~Xiang, D.~Nie, Y.~Shao, H.~Zhang, D.~Shen, and Q.~Wang,
  ``Interleaved 3d-cnn s for joint segmentation of small-volume structures in
  head and neck ct images,'' \emph{Medical physics}, vol.~45, no.~5, pp.
  2063--2075, 2018.

\bibitem{145}
B.~Huang, Y.~Ye, Z.~Xu, Z.~Cai, Y.~He, Z.~Zhong, L.~Liu, X.~Chen, H.~Chen, and
  B.~Huang, ``3d lightweight network for simultaneous registration and
  segmentation of organs-at-risk in ct images of head and neck cancer,''
  \emph{IEEE Transactions on Medical Imaging}, vol.~41, no.~4, pp. 951--964,
  2021.

\bibitem{141}
S.~Liang, F.~Tang, X.~Huang, K.~Yang, T.~Zhong, R.~Hu, S.~Liu, X.~Yuan, and
  Y.~Zhang, ``Deep-learning-based detection and segmentation of organs at risk
  in nasopharyngeal carcinoma computed tomographic images for radiotherapy
  planning,'' \emph{European radiology}, vol.~29, no.~4, pp. 1961--1967, 2019.

\bibitem{155}
N.~Tong, S.~Gou, S.~Yang, D.~Ruan, and K.~Sheng, ``Fully automatic multi-organ
  segmentation for head and neck cancer radiotherapy using shape representation
  model constrained fully convolutional neural networks,'' \emph{Medical
  physics}, vol.~45, no.~10, pp. 4558--4567, 2018.

\bibitem{127}
H.~R. Roth, H.~Oda, Y.~Hayashi, M.~Oda, N.~Shimizu, M.~Fujiwara, K.~Misawa, and
  K.~Mori, ``Hierarchical 3d fully convolutional networks for multi-organ
  segmentation,'' \emph{arXiv preprint arXiv:1704.06382}, 2017.

\bibitem{58}
K.~Men, J.~Dai, and Y.~Li, ``Automatic segmentation of the clinical target
  volume and organs at risk in the planning ct for rectal cancer using deep
  dilated convolutional neural networks,'' \emph{Medical physics}, vol.~44,
  no.~12, pp. 6377--6389, 2017.

\bibitem{44}
Z.~Chen, C.~Li, J.~He, J.~Ye, D.~Song, S.~Wang, L.~Gu, and Y.~Qiao, ``A novel
  hybrid convolutional neural network for accurate organ segmentation in 3d
  head and neck ct images,'' in \emph{Medical Image Computing and Computer
  Assisted Intervention--MICCAI 2021: 24th International Conference,
  Strasbourg, France, September 27--October 1, 2021, Proceedings, Part I
  24}.\hskip 1em plus 0.5em minus 0.4em\relax Cham, Switzerland: Springer,
  2021, pp. 569--578.

\bibitem{67}
Y.~Chen, D.~Ruan, J.~Xiao, L.~Wang, B.~Sun, R.~Saouaf, W.~Yang, D.~Li, and
  Z.~Fan, ``Fully automated multiorgan segmentation in abdominal magnetic
  resonance imaging with deep neural networks,'' \emph{Medical physics},
  vol.~47, no.~10, pp. 4971--4982, 2020.

\bibitem{156}
R.~Jain, A.~Sutradhar, A.~K. Dash, and S.~Das, ``Automatic multi-organ
  segmentation on abdominal ct scans using deep u-net model,'' in \emph{2021
  19th OITS International Conference on Information Technology (OCIT)}.\hskip
  1em plus 0.5em minus 0.4em\relax Bhubaneswar, India: IEEE, 2021, pp. 48--53.

\bibitem{64}
Y.~Ahn, J.~S. Yoon, S.~S. Lee, H.-I. Suk, J.~H. Son, Y.~S. Sung, Y.~Lee, B.-K.
  Kang, and H.~S. Kim, ``Deep learning algorithm for automated segmentation and
  volume measurement of the liver and spleen using portal venous phase computed
  tomography images,'' \emph{Korean journal of radiology}, vol.~21, no.~8, pp.
  987--997, 2020.

\bibitem{89}
J.~Shi, K.~Wen, X.~Hao, X.~Xue, H.~An, and H.~Zhang, ``A novel u-like network
  for the segmentation of thoracic organs,'' in \emph{2020 IEEE 17th
  International Symposium on Biomedical Imaging Workshops (ISBI
  Workshops)}.\hskip 1em plus 0.5em minus 0.4em\relax Iowa City, IA: IEEE,
  2020, pp. 1--4.

\bibitem{143}
S.~Liang, K.-H. Thung, D.~Nie, Y.~Zhang, and D.~Shen, ``Multi-view spatial
  aggregation framework for joint localization and segmentation of organs at
  risk in head and neck ct images,'' \emph{IEEE Transactions on Medical
  Imaging}, vol.~39, no.~9, pp. 2794--2805, 2020.

\bibitem{157}
X.~Zhou, R.~Takayama, S.~Wang, T.~Hara, and H.~Fujita, ``Deep learning of the
  sectional appearances of 3d ct images for anatomical structure segmentation
  based on an fcn voting method,'' \emph{Medical physics}, vol.~44, no.~10, pp.
  5221--5233, 2017.

\bibitem{128}
Y.~Wang, Y.~Zhou, W.~Shen, S.~Park, E.~K. Fishman, and A.~L. Yuille,
  ``Abdominal multi-organ segmentation with organ-attention networks and
  statistical fusion,'' \emph{Medical image analysis}, vol.~55, pp. 88--102,
  2019.

\bibitem{68}
H.~Tang, X.~Liu, K.~Han, X.~Xie, X.~Chen, H.~Qian, Y.~Liu, S.~Sun, and N.~Bai,
  ``Spatial context-aware self-attention model for multi-organ segmentation,''
  in \emph{Proceedings of the IEEE/CVF winter conference on applications of
  computer vision}, Waikoloa, HI, 2021, pp. 939--949.

\bibitem{158}
T.~Qu, X.~Wang, C.~Fang, L.~Mao, J.~Li, P.~Li, J.~Qu, X.~Li, H.~Xue, Y.~Yu
  \emph{et~al.}, ``M$^3$net: A multi-scale multi-view framework for multi-phase
  pancreas segmentation based on cross-phase non-local attention,''
  \emph{Medical image analysis}, vol.~75, p. 102232, 2022.

\bibitem{159}
Y.~Ding, W.~Zheng, J.~Geng, Z.~Qin, K.-K.~R. Choo, Z.~Qin, and X.~Hou,
  ``Mvfusfra: a multi-view dynamic fusion framework for multimodal brain tumor
  segmentation,'' \emph{IEEE Journal of Biomedical and Health Informatics},
  vol.~26, no.~4, pp. 1570--1581, 2021.

\bibitem{160}
A.~Ouaknine, A.~Newson, P.~P{\'e}rez, F.~Tupin, and J.~Rebut, ``Multi-view
  radar semantic segmentation,'' in \emph{Proceedings of the IEEE/CVF
  International Conference on Computer Vision}, Montreal, QC, 2021, pp.
  15\,671--15\,680.

\bibitem{184}
Z.~S. Cheng, T.~Y. Zeng, S.~J. Huang, and X.~Yang, ``A novel hybrid network for
  h\&n organs at risk segmentation,'' in \emph{Proceedings of the 5th
  International Conference on Biomedical Signal and Image Processing}, Suzhou,
  China, 2020, pp. 7--13.

\bibitem{161}
Y.~Zhou, Z.~Li, S.~Bai, C.~Wang, X.~Chen, M.~Han, E.~Fishman, and A.~L. Yuille,
  ``Prior-aware neural network for partially-supervised multi-organ
  segmentation,'' in \emph{Proceedings of the IEEE/CVF international conference
  on computer vision}, Seoul, South Korea, 2019, pp. 10\,672--10\,681.

\bibitem{162}
S.~Lian, L.~Li, Z.~Luo, Z.~Zhong, B.~Wang, and S.~Li, ``Learning multi-organ
  segmentation via partial-and mutual-prior from single-organ datasets,''
  \emph{Biomedical Signal Processing and Control}, vol.~80, p. 104339, 2023.

\bibitem{163}
O.~Oktay, E.~Ferrante, K.~Kamnitsas, M.~Heinrich, W.~Bai, J.~Caballero, S.~A.
  Cook, A.~De~Marvao, T.~Dawes, D.~P. O‘Regan \emph{et~al.}, ``Anatomically
  constrained neural networks (acnns): application to cardiac image enhancement
  and segmentation,'' \emph{IEEE transactions on medical imaging}, vol.~37,
  no.~2, pp. 384--395, 2017.

\bibitem{164}
J.~Ho, A.~Jain, and P.~Abbeel, ``Denoising diffusion probabilistic models,''
  \emph{Advances in Neural Information Processing Systems}, vol.~33, pp.
  6840--6851, 2020.

\bibitem{165}
J.~Song, C.~Meng, and S.~Ermon, ``Denoising diffusion implicit models,''
  \emph{arXiv preprint arXiv:2010.02502}, 2020.

\bibitem{54}
I.~Isler, C.~Lisle, J.~Rineer, P.~Kelly, D.~Turgut, J.~Ricci, and U.~Bagci,
  ``Enhancing organ at risk segmentation with improved deep neural networks,''
  in \emph{Medical Imaging 2022: Image Processing}, vol. 12032.\hskip 1em plus
  0.5em minus 0.4em\relax San Diego, CA: SPIE, 2022, pp. 814--820.

\bibitem{88}
S.~Vesal, N.~Ravikumar, and A.~Maier, ``A 2d dilated residual u-net for
  multi-organ segmentation in thoracic ct,'' \emph{arXiv preprint
  arXiv:1905.07710}, 2019.

\bibitem{166}
T.-Y. Lin, P.~Doll{\'a}r, R.~Girshick, K.~He, B.~Hariharan, and S.~Belongie,
  ``Feature pyramid networks for object detection,'' in \emph{Proceedings of
  the IEEE conference on computer vision and pattern recognition}, Honolulu,
  HI, 2017, pp. 2117--2125.

\bibitem{167}
L.-C. Chen, G.~Papandreou, I.~Kokkinos, K.~Murphy, and A.~L. Yuille, ``Deeplab:
  Semantic image segmentation with deep convolutional nets, atrous convolution,
  and fully connected crfs,'' \emph{IEEE transactions on pattern analysis and
  machine intelligence}, vol.~40, no.~4, pp. 834--848, 2017.

\bibitem{49}
A.~Srivastava, D.~Jha, E.~Keles, B.~Aydogan, M.~Abazeed, and U.~Bagci, ``An
  efficient multi-scale fusion network for 3d organ at risk (oar)
  segmentation,'' \emph{arXiv preprint arXiv:2208.07417}, 2022.

\bibitem{168}
O.~Oktay, J.~Schlemper, L.~L. Folgoc, M.~Lee, M.~Heinrich, K.~Misawa, K.~Mori,
  S.~McDonagh, N.~Y. Hammerla, B.~Kainz \emph{et~al.}, ``Attention u-net:
  Learning where to look for the pancreas,'' \emph{arXiv preprint
  arXiv:1804.03999}, 2018.

\bibitem{169}
S.~Hu \emph{et~al.}, ``Hu j., shen l., albanie s., sun g., wu e,''
  \emph{Squeeze-and-excitation networks, IEEE Transactions on Pattern Analysis
  and Machine Intelligence}, vol.~42, no.~8, pp. 2011--2023, 2019.

\bibitem{45}
Z.~Liu, H.~Wang, W.~Lei, and G.~Wang, ``Csaf-cnn: cross-layer spatial attention
  map fusion network for organ-at-risk segmentation in head and neck ct
  images,'' in \emph{2020 IEEE 17th International Symposium on Biomedical
  Imaging (ISBI)}.\hskip 1em plus 0.5em minus 0.4em\relax Iowa City, IA: IEEE,
  2020, pp. 1522--1525.

\bibitem{70}
H.~Lin, Z.~Li, Z.~Yang, and Y.~Wang, ``Variance-aware attention u-net for
  multi-organ segmentation,'' \emph{Medical Physics}, vol.~48, no.~12, pp.
  7864--7876, 2021.

\bibitem{170}
Z.~Zhou, M.~M. Rahman~Siddiquee, N.~Tajbakhsh, and J.~Liang, ``Unet++: A nested
  u-net architecture for medical image segmentation,'' in \emph{Deep Learning
  in Medical Image Analysis and Multimodal Learning for Clinical Decision
  Support: 4th International Workshop, DLMIA 2018, and 8th International
  Workshop, ML-CDS 2018, Held in Conjunction with MICCAI 2018, Granada, Spain,
  September 20, 2018, Proceedings 4}.\hskip 1em plus 0.5em minus 0.4em\relax
  Cham, Switzerland: Springer, 2018, pp. 3--11.

\bibitem{171}
G.~Huang, Z.~Liu, L.~Van Der~Maaten, and K.~Q. Weinberger, ``Densely connected
  convolutional networks,'' in \emph{Proceedings of the IEEE conference on
  computer vision and pattern recognition}, Honolulu, HI, 2017, pp. 4700--4708.

\bibitem{172}
K.~He, X.~Zhang, S.~Ren, and J.~Sun, ``Deep residual learning for image
  recognition,'' in \emph{Proceedings of the IEEE conference on computer vision
  and pattern recognition}, San Juan, PR, 2016, pp. 770--778.

\bibitem{173}
J.~Dai, H.~Qi, Y.~Xiong, Y.~Li, G.~Zhang, H.~Hu, and Y.~Wei, ``Deformable
  convolutional networks,'' in \emph{Proceedings of the IEEE international
  conference on computer vision}, Venice, Italy, 2017, pp. 764--773.

\bibitem{63}
M.~P. Heinrich, O.~Oktay, and N.~Bouteldja, ``Obelisk-net: Fewer layers to
  solve 3d multi-organ segmentation with sparse deformable convolutions,''
  \emph{Medical image analysis}, vol.~54, pp. 1--9, 2019.

\bibitem{80}
N.~Shen, Z.~Wang, J.~Li, H.~Gao, W.~Lu, P.~Hu, and L.~Feng, ``Multi-organ
  segmentation network for abdominal ct images based on spatial attention and
  deformable convolution,'' \emph{Expert Systems with Applications}, vol. 211,
  p. 118625, 2023.

\bibitem{174}
Q.~Hou, L.~Zhang, M.-M. Cheng, and J.~Feng, ``Strip pooling: Rethinking spatial
  pooling for scene parsing,'' in \emph{Proceedings of the IEEE/CVF conference
  on computer vision and pattern recognition}, Seattle, WA, 2020, pp.
  4003--4012.

\bibitem{131}
F.~Zhang, Y.~Wang, and H.~Yang, ``Efficient context-aware network for abdominal
  multi-organ segmentation,'' \emph{arXiv preprint arXiv:2109.10601}, 2021.

\bibitem{175}
S.~Jadon, ``A survey of loss functions for semantic segmentation,'' in
  \emph{2020 IEEE conference on computational intelligence in bioinformatics
  and computational biology (CIBCB)}.\hskip 1em plus 0.5em minus 0.4em\relax
  Via del Mar, Chile: IEEE, 2020, pp. 1--7.

\bibitem{176}
M.~Yi-de, L.~Qing, and Q.~Zhi-Bai, ``Automated image segmentation using
  improved pcnn model based on cross-entropy,'' in \emph{Proceedings of 2004
  International Symposium on Intelligent Multimedia, Video and Speech
  Processing, 2004.}\hskip 1em plus 0.5em minus 0.4em\relax Hong Kong, China:
  IEEE, 2004, pp. 743--746.

\bibitem{177}
C.~H. Sudre, W.~Li, T.~Vercauteren, S.~Ourselin, and M.~Jorge~Cardoso,
  ``Generalised dice overlap as a deep learning loss function for highly
  unbalanced segmentations,'' in \emph{Deep Learning in Medical Image Analysis
  and Multimodal Learning for Clinical Decision Support: Third International
  Workshop, DLMIA 2017, and 7th International Workshop, ML-CDS 2017, Held in
  Conjunction with MICCAI 2017, Qu{\'e}bec City, QC, Canada, September 14,
  Proceedings 3}.\hskip 1em plus 0.5em minus 0.4em\relax Cham, Switzerland:
  Springer, 2017, pp. 240--248.

\bibitem{178}
S.~S.~M. Salehi, D.~Erdogmus, and A.~Gholipour, ``Tversky loss function for
  image segmentation using 3d fully convolutional deep networks,'' in
  \emph{Machine Learning in Medical Imaging: 8th International Workshop, MLMI
  2017, Held in Conjunction with MICCAI 2017, Quebec City, QC, Canada,
  September 10, 2017, Proceedings 8}.\hskip 1em plus 0.5em minus 0.4em\relax
  Cham, Switzerland: Springer, 2017, pp. 379--387.

\bibitem{179}
T.-Y. Lin, P.~Goyal, R.~Girshick, K.~He, and P.~Doll{\'a}r, ``Focal loss for
  dense object detection,'' in \emph{Proceedings of the IEEE international
  conference on computer vision}, Venice, Italy, 2017, pp. 2980--2988.

\bibitem{180}
V.~Pihur, S.~Datta, and S.~Datta, ``Weighted rank aggregation of cluster
  validation measures: a monte carlo cross-entropy approach,''
  \emph{Bioinformatics}, vol.~23, no.~13, pp. 1607--1615, 2007.

\bibitem{59}
C.~Shen, H.~R. Roth, H.~Oda, M.~Oda, Y.~Hayashi, K.~Misawa, and K.~Mori, ``On
  the influence of dice loss function in multi-class organ segmentation of
  abdominal ct using 3d fully convolutional networks,'' \emph{arXiv preprint
  arXiv:1801.05912}, 2018.

\bibitem{181}
E.~Tappeiner, M.~Welk, and R.~Schubert, ``Tackling the class imbalance problem
  of deep learning-based head and neck organ segmentation,''
  \emph{International Journal of Computer Assisted Radiology and Surgery},
  vol.~17, no.~11, pp. 2103--2111, 2022.

\bibitem{182}
N.~Abraham and N.~M. Khan, ``A novel focal tversky loss function with improved
  attention u-net for lesion segmentation,'' in \emph{2019 IEEE 16th
  international symposium on biomedical imaging (ISBI 2019)}.\hskip 1em plus
  0.5em minus 0.4em\relax Venice, Italy: IEEE, 2019, pp. 683--687.

\bibitem{183}
K.~C. Wong, M.~Moradi, H.~Tang, and T.~Syeda-Mahmood, ``3d segmentation with
  exponential logarithmic loss for highly unbalanced object sizes,'' in
  \emph{Medical Image Computing and Computer Assisted Intervention--MICCAI
  2018: 21st International Conference, Granada, Spain, September 16-20, 2018,
  Proceedings, Part III 11}.\hskip 1em plus 0.5em minus 0.4em\relax Cham,
  Switzerland: Springer, 2018, pp. 612--619.

\bibitem{73}
J.~Song, X.~Chen, Q.~Zhu, F.~Shi, D.~Xiang, Z.~Chen, Y.~Fan, L.~Pan, and
  W.~Zhu, ``Global and local feature reconstruction for medical image
  segmentation,'' \emph{IEEE Transactions on Medical Imaging}, vol.~41, no.~9,
  pp. 2273--2284, 2022.

\bibitem{47}
W.~Lei, H.~Mei, Z.~Sun, S.~Ye, R.~Gu, H.~Wang, R.~Huang, S.~Zhang, S.~Zhang,
  and G.~Wang, ``Automatic segmentation of organs-at-risk from head-and-neck ct
  using separable convolutional neural network with hard-region-weighted
  loss,'' \emph{Neurocomputing}, vol. 442, pp. 184--199, 2021.

\bibitem{185}
P.~Bilic, P.~Christ, H.~B. Li, E.~Vorontsov, A.~Ben-Cohen, G.~Kaissis,
  A.~Szeskin, C.~Jacobs, G.~E.~H. Mamani, G.~Chartrand \emph{et~al.}, ``The
  liver tumor segmentation benchmark (lits),'' \emph{Medical Image Analysis},
  vol.~84, p. 102680, 2023.

\bibitem{186}
N.~Heller, N.~Sathianathen, A.~Kalapara, E.~Walczak, K.~Moore, H.~Kaluzniak,
  J.~Rosenberg, P.~Blake, Z.~Rengel, M.~Oestreich \emph{et~al.}, ``The kits19
  challenge data: 300 kidney tumor cases with clinical context, ct semantic
  segmentations, and surgical outcomes,'' \emph{arXiv preprint
  arXiv:1904.00445}, 2019.

\bibitem{187}
A.~L. Simpson, M.~Antonelli, S.~Bakas, M.~Bilello, K.~Farahani,
  B.~Van~Ginneken, A.~Kopp-Schneider, B.~A. Landman, G.~Litjens, B.~Menze
  \emph{et~al.}, ``A large annotated medical image dataset for the development
  and evaluation of segmentation algorithms,'' \emph{arXiv preprint
  arXiv:1902.09063}, 2019.

\bibitem{188}
S.~Chen, K.~Ma, and Y.~Zheng, ``Med3d: Transfer learning for 3d medical image
  analysis,'' \emph{arXiv preprint arXiv:1904.00625}, 2019.

\bibitem{189}
K.~Dmitriev and A.~E. Kaufman, ``Learning multi-class segmentations from
  single-class datasets,'' in \emph{Proceedings of the IEEE/CVF Conference on
  Computer Vision and Pattern Recognition}, Long Beach, CA, 2019, pp.
  9501--9511.

\bibitem{190}
J.~Zhang, Y.~Xie, Y.~Xia, and C.~Shen, ``Dodnet: Learning to segment
  multi-organ and tumors from multiple partially labeled datasets,'' in
  \emph{Proceedings of the IEEE/CVF conference on computer vision and pattern
  recognition}, Nashville, TN, 2021, pp. 1195--1204.

\bibitem{191}
Y.~Xie, J.~Zhang, Y.~Xia, and C.~Shen, ``Learning from partially labeled data
  for multi-organ and tumor segmentation,'' \emph{arXiv preprint
  arXiv:2211.06894}, 2022.

\bibitem{101}
F.~Isensee, P.~F. Jaeger, S.~A. Kohl, J.~Petersen, and K.~H. Maier-Hein,
  ``nnu-net: a self-configuring method for deep learning-based biomedical image
  segmentation,'' \emph{Nature methods}, vol.~18, no.~2, pp. 203--211, 2021.

\bibitem{192}
H.~Wu, S.~Pang, and A.~Sowmya, ``Tgnet: A task-guided network architecture for
  multi-organ and tumour segmentation from partially labelled datasets,'' in
  \emph{2022 IEEE 19th International Symposium on Biomedical Imaging
  (ISBI)}.\hskip 1em plus 0.5em minus 0.4em\relax Kolkata, India: IEEE, 2022,
  pp. 1--5.

\bibitem{193}
P.~Liu, L.~Xiao, and S.~K. Zhou, ``Incremental learning for multi-organ
  segmentation with partially labeled datasets,'' \emph{arXiv preprint
  arXiv:2103.04526}, 2021.

\bibitem{194}
X.~Xu and P.~Yan, ``Federated multi-organ segmentation with partially labeled
  data,'' \emph{arXiv preprint arXiv:2206.07156}, 2022.

\bibitem{195}
X.~Fang and P.~Yan, ``Multi-organ segmentation over partially labeled datasets
  with multi-scale feature abstraction,'' \emph{IEEE Transactions on Medical
  Imaging}, vol.~39, no.~11, pp. 3619--3629, 2020.

\bibitem{196}
G.~Shi, L.~Xiao, Y.~Chen, and S.~K. Zhou, ``Marginal loss and exclusion loss
  for partially supervised multi-organ segmentation,'' \emph{Medical Image
  Analysis}, vol.~70, p. 101979, 2021.

\bibitem{197}
R.~Huang, Y.~Zheng, Z.~Hu, S.~Zhang, and H.~Li, ``Multi-organ segmentation via
  co-training weight-averaged models from few-organ datasets,'' in
  \emph{Medical Image Computing and Computer Assisted Intervention--MICCAI
  2020: 23rd International Conference, Lima, Peru, October 4--8, 2020,
  Proceedings, Part IV 23}.\hskip 1em plus 0.5em minus 0.4em\relax Cham,
  Switzerland: Springer, 2020, pp. 146--155.

\bibitem{198}
L.~Zhang, S.~Feng, Y.~Wang, Y.~Wang, Y.~Zhang, X.~Chen, and Q.~Tian,
  ``Unsupervised ensemble distillation for multi-organ segmentation,'' in
  \emph{2022 IEEE 19th International Symposium on Biomedical Imaging
  (ISBI)}.\hskip 1em plus 0.5em minus 0.4em\relax Kolkata, India: IEEE, 2022,
  pp. 1--5.

\bibitem{199}
Q.~Wu, Y.~Chen, N.~Huang, and X.~Yue, ``Weakly-supervised cerebrovascular
  segmentation network with shape prior and model indicator,'' in
  \emph{Proceedings of the 2022 International Conference on Multimedia
  Retrieval}, Newark, NJ, 2022, pp. 668--676.

\bibitem{200}
F.~Kanavati, K.~Misawa, M.~Fujiwara, K.~Mori, D.~Rueckert, and B.~Glocker,
  ``Joint supervoxel classification forest for weakly-supervised organ
  segmentation,'' in \emph{Machine Learning in Medical Imaging: 8th
  International Workshop, MLMI 2017, Held in Conjunction with MICCAI 2017,
  Quebec City, QC, Canada, September 10, 2017, Proceedings 8}.\hskip 1em plus
  0.5em minus 0.4em\relax Cham, Switzerland: Springer, 2017, pp. 79--87.

\bibitem{201}
W.~Bai, O.~Oktay, M.~Sinclair, H.~Suzuki, M.~Rajchl, G.~Tarroni, B.~Glocker,
  A.~King, P.~M. Matthews, and D.~Rueckert, ``Semi-supervised learning for
  network-based cardiac mr image segmentation,'' in \emph{Medical Image
  Computing and Computer-Assisted Intervention- MICCAI 2017: 20th International
  Conference, Quebec City, QC, Canada, September 11-13, 2017, Proceedings, Part
  II 20}.\hskip 1em plus 0.5em minus 0.4em\relax Cham, Switzerland: Springer,
  2017, pp. 253--260.

\bibitem{202}
X.~Luo, M.~Hu, T.~Song, G.~Wang, and S.~Zhang, ``Semi-supervised medical image
  segmentation via cross teaching between cnn and transformer,'' in
  \emph{International Conference on Medical Imaging with Deep Learning}.\hskip
  1em plus 0.5em minus 0.4em\relax Zurich, Switzerland: PMLR, 2022, pp.
  820--833.

\bibitem{203}
J.~Chen, J.~Zhang, K.~Debattista, and J.~Han, ``Semi-supervised unpaired
  medical image segmentation through task-affinity consistency,'' \emph{IEEE
  Transactions on Medical Imaging}, 2022.

\bibitem{204}
Y.~Wu, Z.~Ge, D.~Zhang, M.~Xu, L.~Zhang, Y.~Xia, and J.~Cai, ``Mutual
  consistency learning for semi-supervised medical image segmentation,''
  \emph{Medical Image Analysis}, vol.~81, p. 102530, 2022.

\bibitem{205}
X.~Luo, W.~Liao, J.~Chen, T.~Song, Y.~Chen, S.~Zhang, N.~Chen, G.~Wang, and
  S.~Zhang, ``Efficient semi-supervised gross target volume of nasopharyngeal
  carcinoma segmentation via uncertainty rectified pyramid consistency,'' in
  \emph{Medical Image Computing and Computer Assisted Intervention--MICCAI
  2021: 24th International Conference, Strasbourg, France, September
  27--October 1, 2021, Proceedings, Part II 24}.\hskip 1em plus 0.5em minus
  0.4em\relax Cham, Switzerland: Springer, 2021, pp. 318--329.

\bibitem{206}
R.~Jiao, Y.~Zhang, L.~Ding, R.~Cai, and J.~Zhang, ``Learning with limited
  annotations: a survey on deep semi-supervised learning for medical image
  segmentation,'' \emph{arXiv preprint arXiv:2207.14191}, 2022.

\bibitem{207}
Y.~Zhou, Y.~Wang, P.~Tang, S.~Bai, W.~Shen, E.~Fishman, and A.~Yuille,
  ``Semi-supervised 3d abdominal multi-organ segmentation via deep multi-planar
  co-training,'' in \emph{2019 IEEE Winter Conference on Applications of
  Computer Vision (WACV)}.\hskip 1em plus 0.5em minus 0.4em\relax Waikoloa, HI:
  IEEE, 2019, pp. 121--140.

\bibitem{208}
Y.~Xia, D.~Yang, Z.~Yu, F.~Liu, J.~Cai, L.~Yu, Z.~Zhu, D.~Xu, A.~Yuille, and
  H.~Roth, ``Uncertainty-aware multi-view co-training for semi-supervised
  medical image segmentation and domain adaptation,'' \emph{Medical image
  analysis}, vol.~65, p. 101766, 2020.

\bibitem{209}
H.~Lai, T.~Wang, and S.~Zhou, ``Dlunet: Semi-supervised learning based
  dual-light unet for multi-organ segmentation,'' in \emph{Fast and
  Low-Resource Semi-supervised Abdominal Organ Segmentation: MICCAI 2022
  Challenge, FLARE 2022, Held in Conjunction with MICCAI 2022, Singapore,
  September 22, 2022, Proceedings}.\hskip 1em plus 0.5em minus 0.4em\relax
  Springer, 2023, pp. 64--73.

\bibitem{210}
H.~H. Lee, Y.~Tang, O.~Tang, Y.~Xu, Y.~Chen, D.~Gao, S.~Han, R.~Gao, M.~R.
  Savona, R.~G. Abramson \emph{et~al.}, ``Semi-supervised multi-organ
  segmentation through quality assurance supervision,'' in \emph{Medical
  Imaging 2020: Image Processing}, vol. 11313.\hskip 1em plus 0.5em minus
  0.4em\relax Houston, TX: SPIE, 2020, pp. 363--369.

\bibitem{211}
A.~Raju, C.-T. Cheng, Y.~Huo, J.~Cai, J.~Huang, J.~Xiao, L.~Lu, C.~Liao, and
  A.~P. Harrison, ``Co-heterogeneous and adaptive segmentation from
  multi-source and multi-phase ct imaging data: A study on pathological liver
  and lesion segmentation,'' in \emph{Computer Vision--ECCV 2020: 16th European
  Conference, Glasgow, UK, August 23--28, 2020, Proceedings, Part XXIII}.\hskip
  1em plus 0.5em minus 0.4em\relax Cham, Switzerland: Springer, 2020, pp.
  448--465.

\bibitem{212}
D.~Guo, D.~Jin, Z.~Zhu, T.-Y. Ho, A.~P. Harrison, C.-H. Chao, J.~Xiao, and
  L.~Lu, ``Organ at risk segmentation for head and neck cancer using stratified
  learning and neural architecture search,'' in \emph{Proceedings of the
  IEEE/CVF Conference on Computer Vision and Pattern Recognition}, Seattle, WA,
  2020, pp. 4223--4232.

\bibitem{213}
N.~Tajbakhsh, L.~Jeyaseelan, Q.~Li, J.~N. Chiang, Z.~Wu, and X.~Ding,
  ``Embracing imperfect datasets: A review of deep learning solutions for
  medical image segmentation,'' \emph{Medical Image Analysis}, vol.~63, p.
  101693, 2020.

\bibitem{214}
L.~Qu, S.~Liu, X.~Liu, M.~Wang, and Z.~Song, ``Towards label-efficient
  automatic diagnosis and analysis: a comprehensive survey of advanced deep
  learning-based weakly-supervised, semi-supervised and self-supervised
  techniques in histopathological image analysis,'' \emph{Physics in Medicine
  \& Biology}, vol.~67, no.~20, p. 20TR01, 2022.

\bibitem{65}
S.~Fu, Y.~Lu, Y.~Wang, Y.~Zhou, W.~Shen, E.~Fishman, and A.~Yuille, ``Domain
  adaptive relational reasoning for 3d multi-organ segmentation,'' in
  \emph{Medical Image Computing and Computer Assisted Intervention--MICCAI
  2020: 23rd International Conference, Lima, Peru, October 4--8, 2020,
  Proceedings, Part I 23}.\hskip 1em plus 0.5em minus 0.4em\relax Cham,
  Switzerland: Springer, 2020, pp. 656--666.

\bibitem{81}
J.~Hong, Y.-D. Zhang, and W.~Chen, ``Source-free unsupervised domain adaptation
  for cross-modality abdominal multi-organ segmentation,''
  \emph{Knowledge-Based Systems}, vol. 250, p. 109155, 2022.

\bibitem{43}
Y.~Liu, Y.~Lei, Y.~Fu, T.~Wang, J.~Zhou, X.~Jiang, M.~McDonald, J.~J. Beitler,
  W.~J. Curran, T.~Liu \emph{et~al.}, ``Head and neck multi-organ
  auto-segmentation on ct images aided by synthetic mri,'' \emph{Medical
  physics}, vol.~47, no.~9, pp. 4294--4302, 2020.

\bibitem{46}
S.~Cros, E.~Vorontsov, and S.~Kadoury, ``Managing class imbalance in
  multi-organ ct segmentation in head and neck cancer patients,'' in \emph{2021
  IEEE 18th International Symposium on Biomedical Imaging (ISBI)}.\hskip 1em
  plus 0.5em minus 0.4em\relax Nice, France: IEEE, 2021, pp. 1360--1364.

\bibitem{52}
J.~Jiang, S.~Elguindi, S.~L. Berry, I.~Onochie, L.~Cervino, J.~O. Deasy, and
  H.~Veeraraghavan, ``Nested block self-attention multiple resolution residual
  network for multiorgan segmentation from ct,'' \emph{Medical Physics},
  vol.~49, no.~8, pp. 5244--5257, 2022.

\bibitem{53}
S.~Francis, G.~Pooloth, S.~B.~S. Singam, N.~Puzhakkal,
  P.~Pulinthanathu~Narayanan, and J.~Pottekkattuvalappil~Balakrishnan,
  ``Sabos-net: Self-supervised attention based network for automatic organ
  segmentation of head and neck ct images,'' \emph{International Journal of
  Imaging Systems and Technology}, vol.~33, no.~1, pp. 175--191, 2023.

\bibitem{56}
K.~Clark, B.~Vendt, K.~Smith, J.~Freymann, J.~Kirby, P.~Koppel, S.~Moore,
  S.~Phillips, D.~Maffitt, M.~Pringle \emph{et~al.}, ``The cancer imaging
  archive (tcia): maintaining and operating a public information repository,''
  \emph{Journal of digital imaging}, vol.~26, no.~6, pp. 1045--1057, 2013.

\bibitem{57}
H.~R. Roth, L.~Lu, A.~Farag, H.-C. Shin, J.~Liu, E.~B. Turkbey, and R.~M.
  Summers, ``Deeporgan: Multi-level deep convolutional networks for automated
  pancreas segmentation,'' in \emph{Medical Image Computing and
  Computer-Assisted Intervention--MICCAI 2015: 18th International Conference,
  Munich, Germany, October 5-9, 2015, Proceedings, Part I 18}.\hskip 1em plus
  0.5em minus 0.4em\relax Cham, Switzerland: Springer, 2015, pp. 556--564.

\bibitem{74}
V.~Kumar, M.~K. Sharma, R.~Jehadeesan, B.~Venkatraman, and D.~Sheet,
  ``Adversarial training of deep convolutional neural network for multi-organ
  segmentation from multi-sequence mri of the abdomen,'' in \emph{2021
  International Conference on Intelligent Technologies (CONIT)}.\hskip 1em plus
  0.5em minus 0.4em\relax Hubli, India: IEEE, 2021, pp. 1--6.

\bibitem{79}
B.~Rister, D.~Yi, K.~Shivakumar, T.~Nobashi, and D.~L. Rubin, ``Ct-org, a new
  dataset for multiple organ segmentation in computed tomography,''
  \emph{Scientific Data}, vol.~7, no.~1, p. 381, 2020.

\bibitem{85}
C.~C. Vu, Z.~A. Siddiqui, L.~Zamdborg, A.~B. Thompson, T.~J. Quinn,
  E.~Castillo, and T.~M. Guerrero, ``Deep convolutional neural networks for
  automatic segmentation of thoracic organs-at-risk in radiation oncology--use
  of non-domain transfer learning,'' \emph{Journal of Applied Clinical Medical
  Physics}, vol.~21, no.~6, pp. 108--113, 2020.

\bibitem{87}
M.~S.~K. Gali, N.~Garg, S.~Vasamsetti \emph{et~al.}, ``Dilated u-net based
  segmentation of organs at risk in thoracic ct images,'' in \emph{SegTHOR@
  ISBI}, 2019.

\bibitem{90}
H.~Mahmood, S.~M.~S. Islam, J.~Hill, and G.~Tay, ``Rapid segmentation of
  thoracic organs using u-net architecture,'' in \emph{2021 Digital Image
  Computing: Techniques and Applications (DICTA)}.\hskip 1em plus 0.5em minus
  0.4em\relax Gold Coast, Australia: IEEE, 2021, pp. 1--6.

\bibitem{91}
F.~Zhang, Q.~Wang, A.~Yang, N.~Lu, H.~Jiang, D.~Chen, Y.~Yu, and Y.~Wang,
  ``Geometric and dosimetric evaluation of the automatic delineation of organs
  at risk (oars) in non-small-cell lung cancer radiotherapy based on a modified
  densenet deep learning network,'' \emph{Frontiers in Oncology}, vol.~12, p.
  861857, 2022.

\bibitem{132}
J.~Ma, Y.~Zhang, S.~Gu, X.~An, Z.~Wang, C.~Ge, C.~Wang, F.~Zhang, Y.~Wang,
  Y.~Xu \emph{et~al.}, ``Fast and low-gpu-memory abdomen ct organ segmentation:
  the flare challenge,'' \emph{Medical Image Analysis}, vol.~82, p. 102616,
  2022.

\bibitem{134}
H.~Kakeya, T.~Okada, and Y.~Oshiro, ``3d u-japa-net: mixture of convolutional
  networks for abdominal multi-organ ct segmentation,'' in \emph{Medical Image
  Computing and Computer Assisted Intervention--MICCAI 2018: 21st International
  Conference, Granada, Spain, September 16-20, 2018, Proceedings, Part IV
  11}.\hskip 1em plus 0.5em minus 0.4em\relax Cham, Switzerland: Springer,
  2018, pp. 426--433.

\bibitem{135}
R.~Trullo, C.~Petitjean, D.~Nie, D.~Shen, and S.~Ruan, ``Joint segmentation of
  multiple thoracic organs in ct images with two collaborative deep
  architectures,'' in \emph{Deep Learning in Medical Image Analysis and
  Multimodal Learning for Clinical Decision Support: Third International
  Workshop, DLMIA 2017, and 7th International Workshop, ML-CDS 2017, Held in
  Conjunction with MICCAI 2017, Qu{\'e}bec City, QC, Canada, September 14,
  Proceedings 3}, vol. 10553.\hskip 1em plus 0.5em minus 0.4em\relax Springer,
  2017, pp. 21--29.

\bibitem{136}
Z.~Cao, B.~Yu, B.~Lei, H.~Ying, X.~Zhang, D.~Z. Chen, and J.~Wu, ``Cascaded
  se-resunet for segmentation of thoracic organs at risk,''
  \emph{Neurocomputing}, vol. 453, pp. 357--368, 2021.

\bibitem{140}
Q.~Yang, S.~Zhang, X.~Sun, J.~Sun, and K.~Yuan, ``Automatic segmentation of
  head-neck organs by multi-mode cnns for radiation therapy,'' in \emph{2019
  International Conference on Medical Imaging Physics and Engineering
  (ICMIPE)}.\hskip 1em plus 0.5em minus 0.4em\relax Shenzhen, China: IEEE,
  2019, pp. 1--5.

\bibitem{147}
C.~E. Cardenas, A.~S. Mohamed, J.~Yang, M.~Gooding, H.~Veeraraghavan,
  J.~Kalpathy-Cramer, S.~P. Ng, Y.~Ding, J.~Wang, S.~Y. Lai \emph{et~al.},
  ``Head and neck cancer patient images for determining auto-segmentation
  accuracy in t2-weighted magnetic resonance imaging through expert manual
  segmentations,'' \emph{Medical physics}, vol.~47, no.~5, pp. 2317--2322,
  2020.

\bibitem{150}
O.~Jimenez-del Toro, H.~M{\"u}ller, M.~Krenn, K.~Gruenberg, A.~A. Taha,
  M.~Winterstein, I.~Eggel, A.~Foncubierta-Rodr{\'\i}guez, O.~Goksel, A.~Jakab
  \emph{et~al.}, ``Cloud-based evaluation of anatomical structure segmentation
  and landmark detection algorithms: Visceral anatomy benchmarks,'' \emph{IEEE
  transactions on medical imaging}, vol.~35, no.~11, pp. 2459--2475, 2016.

\end{thebibliography}
\end{document}